\documentclass[11pt]{article}
\usepackage{color, amsmath, amssymb, amsbsy, amsthm, graphicx, bbm, amsfonts, bm,dsfont,enumitem,booktabs}
\usepackage{caption}   %
\usepackage{booktabs}  %
\usepackage{setspace, comment, graphicx, algorithm, latexsym,titlesec}
\usepackage{multirow}
\usepackage[top=1in, bottom=1in, left=1in, right=1in]{geometry}
\usepackage[flushleft]{threeparttable}
\usepackage[colorlinks,allcolors=blue]{hyperref}
\usepackage[square,numbers]{natbib}  
\usepackage{titlesec} 
\allowdisplaybreaks
\usepackage{algorithm,algpseudocode}
\usepackage{wrapfig,lipsum,booktabs}
\newtheoremstyle{theoremstyle} 
{\topsep} 
{\topsep} 
{\itshape} 
{} 
{} 
{} 
{.5em} 
{\color{black}\ifthenelse{\equal{#3}{}}{{\bfseries #1 #2}}{{\bfseries #1 #2 (#3)}}}
\newtheoremstyle{examplestyle}
{\topsep} 
{\topsep} 
{} 
{} 
{} 
{} 
{.5em} 
{\color{black}\ifthenelse{\equal{#3}{}}{{\bfseries #1 #2}}{{\bfseries #1 #2 (#3)}}}
\theoremstyle{theoremstyle}\newtheorem{theorem}{Theorem}
\theoremstyle{theoremstyle}     
\theoremstyle{theoremstyle}  
\theoremstyle{theoremstyle}\newtheorem{corollary}{Corollary}        
\theoremstyle{theoremstyle}
\theoremstyle{theoremstyle}\newtheorem{assumption}{Assumption}
\theoremstyle{theoremstyle}

\theoremstyle{examplestyle}\newtheorem{Step}{Step}

\theoremstyle{examplestyle}\newtheorem{StepDiscrete}{Step}

\theoremstyle{examplestyle}

\def \hat{\widehat}
\def \tilde{\widetilde}

\def \cov{ \mathbb{C}\mathrm{ov} }
\def \corr{ \mathbb{C}\mathrm{orr} }

\def \diff{ \mathrm{d} }

\def \Probability{ \mathbb{P} }
\def \Expectation{ \mathbb{E} }
\def \Variance{ \mathbb{V} }
\def \Indicator{ \mathds{1} }
\def \bbeta{ \boldsymbol{\beta} }
\def \bomega{ \boldsymbol{\omega} }
\def \transpose{ \intercal }
\def \bSigma{ \boldsymbol{\Sigma} }
\def \bsigma{ \boldsymbol{\sigma} }
\begin{document} 

\title{\Large Can Language Models Boost the Power of Randomized Experiments Without Statistical Bias?
}
	
\author{Xinrui Ruan$^1$ \and Xinwei Ma$^2$ \and Yingfei Wang$^3$ \and Waverly Wei$^4$ \and Jingshen Wang$^1$}

\date{
    $^1${\small Division of Biostatistics, University of California, Berkeley}\\
    $^2${\small Department of Economics, University of California, San Diego}\\
    $^3${\small Michael G. Foster School of Business, University of Washington}\\
    $^4${\small Department of Data Sciences and Operations, University of Southern California}\\
    \ \\
    {\small October 7, 2025 \\ Revised: \today}
}

\maketitle

\begin{abstract}
\noindent Randomized controlled trials (RCTs) are widely adopted for causal inference, yet cost and sample-size constraints limit power.  We introduce \textbf{CALM} (\textit{Causal Analysis leveraging Language Models}), a statistical framework that integrates insights generated by large language models (LLMs) into the analysis of RCTs using established causal estimators to increase precision while preserving statistical validity. In particular, CALM treats LLM-generated outputs as auxiliary prognostic information and corrects their potential bias via a heterogeneous calibration step that residualizes and optimally reweights predictions. We prove that CALM remains consistent even when LLM predictions are biased and achieves efficiency gains over augmented inverse probability weighting estimators for various causal estimands. In particular, CALM develops a few-shot variant that aggregates predictions across randomly sampled demonstration sets. The resulting U-statistic-like predictor restores i.i.d. structure and also mitigates prompt-selection variability. Empirically, in simulations calibrated to a mobile-app depression RCT, CALM delivers lower variance relative to other benchmarking methods, is effective in zero- and few-shot settings, and remains stable across prompt designs. By principled use of LLMs to harness unstructured data and external knowledge learned during pretraining, CALM provides a practical path to more precise causal analyses.

\end{abstract}

\thispagestyle{empty}
\clearpage 

\thispagestyle{plain}\setcounter{page}{1}

\doublespacing
\section{Introduction}\label{section 1:Introduction}

Randomized controlled trials (RCTs) are widely adopted to evaluate the effectiveness of socioeconomic, business, biomedical, and behavioral interventions. This is achieved either by estimating an average effect or by performing subgroup analyses to explore treatment effect heterogeneity and thereby facilitate targeted interventions. Despite their prevalence, most RCT datasets are analyzed in isolation without incorporating external information. Additionally, regression analyses remain the standard practice with a set of hand-picked pre-intervention covariates, primarily due to the simplicity of such models. This approach, however, inevitably leaves considerable baseline information unexploited, including multimodal variables such as open-ended survey responses, product images, clinical notes, facial and gestural cues.

In this paper, we introduce a versatile framework for analyzing randomized controlled trials, accommodating pre-intervention covariates of multimodal nature without requiring specialized computational architectures. As such, our methods are simple to implement and remain accessible to a broad audience of applied researchers. Specifically, the first step in our framework leverages modern large language models (LLMs) to generate predictions of the potential outcomes using design information and baseline covariates that can be tabular or unstructured. LLMs are self-supervised models trained on broad, heterogeneous corpora, yielding general-purpose representations adaptable to many downstream tasks. Integrating LLM-generated predictions into causal inference methods therefore offers opportunities to boost the power of RCTs by leveraging both knowledge acquired during pretraining and information extracted from baseline covariates. 

Despite the popularity of language models in prediction tasks, it remains a critical challenge to develop rigorous and statistically valid causal inference methods incorporating LLM-generated predictions. This is because these predictions are typically biased for the true potential outcomes, and moreover, their accuracy can exhibit substantial heterogeneity depending on individual characteristics. As a result, naively plugging LLM-based predictions into causal estimators can induce bias and lead to problematic statistical inference. Our framework, \textbf{C}ausal \textbf{A}nalysis leveraging \textbf{L}anguage \textbf{M}odels (CALM), tackles these issues via three key components. First, we employ few-shot learning with demonstrative examples resampling to improve prediction relevance and precision. We then introduce a residualization step coupled with heterogeneous calibration weighting to avoid introducing statistical bias and deliver more efficient treatment effect estimation. We demonstrate such efficiency gains both through rigorous theoretical investigations and via an array of numerical evidence, including a carefully crafted simulation study based on calibrated data and a re-evaluation of a behavioral trial. Finally, the estimator takes a Neyman-orthogonal form so that it admits an asymptotically normal distribution, thereby enabling valid statistical inference. In the remainder of this introduction, we elaborate on these features of our methods, further situating our contribution within the literature. 

\textit{Few-shot Learning with Demonstrative Examples Resampling}. A key strength and novelty of CALM is its ability to incorporate few-shot learning (a.k.a. in-context learning) to improve prediction relevance and precision. Specifically, the idea is to include a few demonstrative examples in the prompt, which has been well-documented in the literature due to its simplicity and effectiveness for rapid task adaptation \citep{brown2020language,gao2020making,min2022rethinking,schick2020exploiting}. However, few-shot learning is not without its limitations: the generated predictions can be sensitive to the examples and their ordering \cite{zhao2021calibrate}. Such dependence and lack of invariance may introduce unnecessary noise and bias into the predictions, threatening downstream treatment effect estimation and statistical inference. Another challenge of incorporating few-shot learning is that the demonstrative examples have to be chosen in-sample, and as a result, this induces correlations among the predicted potential outcomes. Unfortunately, it is generally not possible to quantify such correlations, rendering any post hoc adjustment to the statistical inference procedure infeasible. 

Our proposed CALM tackles these challenges by augmenting few-shot learning with resampling-based prediction aggregation. On the practical side, the method remains easy to implement: one only needs to regenerate predictions a few times, each with randomly sampled demonstrative examples. This can easily be done through a chat-based platform for small datasets, or otherwise via a standard API. Theoretically, we show that averaging predictions across multiple small, randomly drawn sets of demonstrative samples yields an LLM-based predictor with a U-statistic-like structure, which is not only robust to demonstrative sample selection but also behaves in an iid manner in large samples. As a result, both our treatment effect estimator and its companion inferential method remain straightforward, making them accessible to practitioners.

\textit{Residualization with Heterogeneous Calibration Weights}. Despite their popularity, language model predictions are by no means perfect: they can be imprecise and even worse, biased in ways that are difficult to characterize. To remain agnostic about the nature of such bias, our proposed CALM incorporates the residualized version of the LLM predictions, where tabular covariates are partialled out from the predicted potential outcomes via nonparametric regressions. A natural concern is whether residualization leads to information loss. Fortunately, this is not the case. Our final estimator builds on the classical augmented inverse probability weighting method, which already incorporates the tabular covariate information efficiently. Residualization therefore preserves the predictive information captured by the unstructured covariates while simultaneously removing any potential bias in the language model predictions.

Incorporating mean-zero quantities into an estimator does not, by itself, guarantee efficiency gains. To ensure that our approach improves upon classical methods, we recognize that prediction accuracy from language models may be heterogeneous, varying across treatment arms and covariate values. We therefore introduce an additional calibration weighting scheme. The idea is simple yet powerful: by assigning weights that are both treatment- and covariate-specific, the estimator fully exploits the predictive power of the unstructured covariates and the underlying language model, leading to improved estimation efficiency and statistical power. This is borne out in both our theoretical analysis and our numerical evidence.

\textit{Neyman Orthogonal Estimating Equation}. Our proposed CALM requires the estimation of several nuisance functions, including the conditional first and second moments of the true and predicted potential outcomes. Following the literature, we recommend the use of sample splitting together with cross-fitting, which is enabled by our carefully crafted estimating equation that is fully Neyman orthogonal with respect to the nuisance functions. It is worth noting that sample splitting and cross-fitting may not be necessary, depending on the specific methods employed for first-step  estimation. Although we do not pursue such directions, see \cite{cattaneo2010efficient,cattaneo2018kernel,cattaneo2019two,hirano2003efficient,robinson1988root} for relevant examples.

In addition to its methodological and theoretical advances, our work provides systematic empirical evidence on the practical utility of CALM. Using synthetic data calibrated to the BRIGHTEN study with both structured and unstructured information, we demonstrate that CALM yields more accurate estimates of causal effects, achieving lower bias and reduced variance compared to benchmark methods. We further demonstrate that CALM remains effective in both zero- and few-shot settings without domain-specific fine-tuning, adapts well to heterogeneous covariate strata, and avoids the under-coverage issues that arise when LLM predictions are naively incorporated as covariates in benchmark methods. In the BRIGHTEN case study, CALM not only yields tighter confidence intervals than benchmark methods, but also detects subgroup treatment effects (e.g., among female Hispanic participants) that benchmark methods miss, indicating its potential to enhance the discovery of treatment heterogeneity. Finally, robustness checks across multiple LLMs and diverse prompt-engineering strategies confirm that CALM's empirical performance is stable to design choices, thereby reducing practical barriers to applying LLMs in real-world experiment analysis.

Before closing this introduction, we briefly discuss several related strands of literature to which this paper contributes. To start, CALM is motivated by recent advances in LLMs, which have demonstrated strong in-context learning abilities, encompassing zero-shot and few-shot prediction, that make them adaptable to new tasks with minimal supervision \citep{brown2020language,driess2023palme,touvron2023llama, ruan2026hero}. In scientific and biomedical domains, domain-adapted models further improve relevance and accuracy \citep{luo2022biogpt,taylor2022galactica}, while multimodal extensions enable joint reasoning over text, images, and audio \citep{alayrac2022flamingo,li2023blip,liu2023llava,openai2024gpt4,radford2021clip}. These capabilities make LLMs well-suited for extracting rich information from the unstructured data increasingly collected in modern RCTs, including clinical notes, medical images, and participant narratives, now standard across oncology, mental health \citep{pratap2022real,ramos2024assessing}, dementia \citep{guterman2023care}, and healthcare delivery \citep{bryan2025developing,hanvey2024investigating}. Yet most LLM-based prediction pipelines offer little uncertainty quantification or statistical guarantees, which are precisely what causal inference in high-stakes biomedical settings demands.

As CALM incorporates LLM-based predictions in a principled manner without introducing additional bias, it also contributes to the growing literature on causal inference with potentially biased auxiliary information and prediction-powered inference (PPI) \citep{angelopoulos2023prediction,arbour2026ai,carlson2025unifying,chen2008semiparametric,egami2023using,ludwig2025large,zhang2025agentic}. Originally developed for settings with missing labels, PPI leverages black-box predictions of unobserved outcomes to improve estimation efficiency, and has since been extended to randomized experiments \citep{DeBartolomeis2025Efficient,poulet2025prediction}, with refinements including optimally tuned parameters \citep{angelopoulos2023ppi++}, stratified tuning \citep{fisch2024stratified}, and cross-fitted machine learning predictions \citep{zrnic2024cross}. Unlike CALM, however, these methods do not guarantee efficiency gains and may even incur efficiency loss when auxiliary predictions are poor; moreover, they are not designed to handle the correlated predictions that arise from LLM few-shot learning.

Our work also relates to the literature on improving the efficiency of randomized experiments through covariate adjustment. Classical approaches such as regression adjustment and stratification are well known to reduce variance without compromising statistical validity \citep{freedman2008regression,lin2013agnostic}, but they are limited in their ability to utilize unstructured data and thus may not achieve full efficiency. More recently, double/debiased machine learning \citep{chernozhukov2018double} leverages flexible machine learning together with cross-fitting. However, existing flexible machine learning tools, such as random forests and boosting, are primarily limited to structured data and are not directly applicable to unstructured data \citep{coravos2019modernizing,guterman2023care,pratap2022real}. In addition, the inclusion of many covariates may lead to biases in the asymptotic normal distribution and therefore renders standard normal-based inference invalid \cite{cattaneo2019two}.

The rest of the paper unfolds as follows. Section~\ref{section 2:Setup and Methodology} introduces our theoretical framework and notation before presenting the proposed method. To keep the exposition clear, the section focuses on a single potential outcome, allowing us to highlight the key components of our approach without cumbersome notation. Section~\ref{section 3:Extensions} expands on this setup to cover zero-shot prediction with CALM, and the estimation of conditional and unconditional average treatment effects. Section~\ref{section 4:Theoretical investigations} presents our theoretical results alongside their formal assumptions. Finally, Sections~\ref{section 5:Simulation studies} and~\ref{section 6:case Study} showcase our numerical results, which include a carefully designed simulation study with synthetic data and a re-analysis of a randomized controlled trial.

\section{Setup and Methodology}\label{section 2:Setup and Methodology}

We discuss in Subsection~\ref{subsection 2-1:Motivation and Efficiency Gains of CALM} the heuristic motivation for CALM, introduce the residualized, calibration-weighted correction term that incorporates LLM-based counterfactual predictions, and characterize the resulting efficiency gain over AIPW. Next, Subsection~\ref{subsection 2-2:CALM with Few-Shot Learning: Implementation Details} details the step-by-step implementation of CALM under few-shot learning. Finally, Subsection~\ref{subsection 2-3:Misspecified Outcome Models and Calibration} extends this framework to settings where outcome models are parsimonious and potentially misspecified, showing that CALM retains a guaranteed efficiency gain over AIPW even when calibration is restricted to a constrained class. We then specialize this result to the practically important case of covariate discretization.

Consider a randomized controlled trial with $n$ units, labeled by $i=1,2,\dots,n$. For each participant, we observe their treatment status, $T_i\in \{1,2,\dots,k\}$, a scalar outcome, $Y_i\in \mathbb{R}$, and pre-intervention covariates, including both structured $X_i \in \mathcal{X} \subseteq \mathbb{R}^p$ (such as age, educational background, etc.) and unstructured $Z_i\in\mathcal{Z}$ (e.g., open-ended surveys). As such, the information observed by the analyst is $O_i = (Y_i, T_i, X_i, Z_i)$. Following the standard potential outcomes framework, the observed outcome is $Y_i = \sum_{t=1}^k \Indicator(T_i=t) Y_i(t)$, and $(Y_i(1),\dots,Y_i(k))$ denote the $k$ potential outcomes for unit $i$. To facilitate further discussion, we also introduce the propensity scores and the conditional expectations of the potential outcomes:
\begin{align*}
  e_t(X_i) = \Probability[T_i=t|X_i]\quad \text{and}\quad \mu_t(X_i) = \Expectation[Y_i(t)|X_i]\quad \text{for }t=1,2,\dots,k.
\end{align*}
Since the data arise from a randomized trial, we assume that the propensity scores are known. We note, however, that our proposed method readily extends to settings with unknown $e_t(X_i)$ with minor changes to the underlying technical conditions. The framework accommodates both complete randomization, in which the treatment probability is constant (say, $e_t(X_i) = 1/k$), and stratified experiments, where units with different pre-intervention covariates may receive different treatment probabilities. For expositional clarity, the formal analysis below adopts an i.i.d. superpopulation formulation, while the CALM construction itself remains applicable under fixed-allocation complete or stratified designs. In addition, our methodology allows some of the unstructured covariates to be missing, thanks to the versatility of language models in handling partially observed inputs. Such flexibility is particularly relevant, as it is common for participants not to complete a survey.

This paper develops estimation strategies for both unconditional and conditional average treatment effects that incorporate structured and unstructured information by leveraging language models. The parameters of interest are
\begin{align*}
  \tau_{t,t'}=\Expectation[Y_i(t)-Y_i(t')]\quad \text{and}\quad \tau_{t,t'}(x)=\Expectation[Y_i(t)-Y_i(t')|X_i=x],
\end{align*}
for two treatment levels $t$ and $t'$. However, we focus in this section on the simpler estimand $\mu_t = \Expectation[Y_i(t)]$ to keep the exposition clear. Once the main idea and implementation details have been introduced, the same strategy carries over naturally to the estimation of $\tau_{t,t'}$ (see Section \ref{section 3:Extensions}).

\subsection{Motivation and Efficiency Gains of CALM}\label{subsection 2-1:Motivation and Efficiency Gains of CALM}

To provide some heuristic justification for our proposed CALM, first consider $n^{-1}\sum_{i=1}^n Y_i(t)$. This ``estimator'' is ideal as it employs the sample average to estimate a population expectation. However, it is clearly infeasible: $Y_i(t)$ is only observed for individuals who are assigned to treatment arm $t$. Now rewrite it into $n^{-1}\sum_{i=1}^n \{Y_i(t) - \mu_t(X_i) + \mu_t(X_i)\}$. Importantly, $\mu_t(X_i)$ is estimable for each unit, so it is $Y_i(t) - \mu_t(X_i)$ that is subject to missingness. This is the motivation behind the augmented inverse probability weighting (AIPW) approach
\begin{align*}
  \overline{\mu}_{t,\mathtt{AIPW}} = \frac{1}{n}\sum_{i=1}^n \frac{\Indicator(T_i=t)}{e_t(X_i)}\Big(Y_i - \mu_t(X_i)\Big) + \mu_t(X_i),
\end{align*} 
which is widely adopted in practice. Strictly speaking, the above is still infeasible as it involves the unknown nuisance function $\mu_t(X_i)$. See below for implementation details of CALM, which also applies to AIPW. In the absence of unstructured covariates, the AIPW estimator is known to be efficient \cite{cattaneo2010efficient,hahn1998role}.

As the discussion so far has made clear, the information in $Y_i(t)$ can be partitioned into the part captured by the structured covariates through $\mu_t(X_i)$ and the part that cannot, and it is the latter that is only available for units in treatment arm $t$. Our proposal is to employ language models to predict such information for units in other treatment arms via unstructured data $Z_i$. For now, assume we are able to generate LLM-based predictions of the potential outcomes, denoted $Y^\dagger_i(t)$ for each $i=1,2,\dots,n$ (see Step \ref{step2:few-shot} below for implementation details, including our novel few-shot learning prediction with demonstrative examples resampling). The proposed estimator takes the form (strictly speaking, below is the asymptotically linear representation of our CALM estimator)
\begin{align}\label{eq:CALM asymptotic rep}
  \overline{\mu}_{t,\mathtt{CALM}} = \frac{1}{n}\sum_{i=1}^n \underbrace{\frac{\Indicator(T_i=t)}{e_t(X_i)}\Big(Y_i - \mu_t(X_i)\Big) + \mu_t(X_i)}_{\text{AIPW}} + \underbrace{\Big(1-\frac{\Indicator(T_i=t)}{e_t(X_i)}\Big)\Big(Y_i^\dagger(t) - \mu^\dagger_t(X_i)\Big)\omega_t(X_i)}_{\text{CALM improvement}}.
\end{align} 

We start with the middle term, $Y_i^\dagger(t) - \mu^\dagger_t(X_i)$, which incorporates LLM-based counterfactual predictions without introducing statistical bias. Importantly, our method does not require the predicted counterfactuals $Y_i^\dagger(t)$ to be ``consistent'' or ``unbiased'' for the true $Y_i(t)$, as such conditions are difficult to justify in practice. Such robustness is achieved by residualizing the prediction against its own conditional expectation,
\begin{align*}
  \mu^\dagger_t(X_i) = \Expectation[Y_i^\dagger(t)|X_i],
\end{align*}
so that this residual is mean zero by construction, avoiding bias in the estimating equation regardless of how $Y_i^\dagger(t)$ relates to $Y_i(t)$. This construction is also motivated by our earlier discussion: $Y_i(t) - \mu_t(X_i)$ is exactly what is missing for units not in treatment arm $t$, and our approach leverages language models and unstructured information to ``recover'' it.

The residualized LLM predictions are followed by the calibration weights $\omega_t(X_i)$, which deliver guaranteed efficiency gains over the AIPW approach. The term $\omega_t(X_i)$ takes the form
\begin{align*}
  \omega_t(X_i) = \frac{\cov[Y_i(t),Y_i^\dagger(t)|X_i]}{\Variance[Y_i^\dagger(t)|X_i]} = \frac{\cov[Y_i,Y_i^\dagger(t)|T_i=t,X_i]}{\Variance[Y_i^\dagger(t)|X_i]},
\end{align*}
assigning higher weights when the LLM predictions are more strongly correlated with the true potential outcomes. Together with this weighting scheme, $(Y_i^\dagger(t) - \mu^\dagger_t(X_i))\omega_t(X_i)$ acts as a linear predictor of $Y_i(t) - \mu_t(X_i)$, although this is done nonparametrically by conditioning on the structured covariates. This allows the predictive precision of the language model to vary across units, a feature likely to arise in practice. We remark that this heterogeneous weighting is crucial: without it, incorporating a mean-zero term into the estimating equation can lead to efficiency loss when $Y_i^\dagger(t)$ provides little information about $Y_i(t)$ beyond what is already captured by $X_i$.

Another distinguishing feature of CALM is its principled incorporation of few-shot LLM predictions through demonstrative example resampling and aggregation. To compare, a naive approach would generate $Y_i^\dagger(t)$ using an LLM with $X_i, Z_i$ and a fixed set of demonstrative examples as input. This, however, not only turns out to be suboptimal from a methodological perspective but also leads to nontrivial issues for our theoretical analysis. Specifically, few-shot predictions are known to be sensitive to the composition and ordering of the demonstrative examples in the prompt. Moreover, as those examples are drawn from the same sample, they induce correlations among the predicted outcomes that are difficult to characterize. The former leads to imprecise predictions, undermining the quality and overall performance of CALM; the latter makes asymptotic analysis very challenging. The novel demonstrative examples resampling step we introduce successfully tackles these two issues. See Step~\ref{step2:few-shot} below and the theoretical results in Section~\ref{section 4:Theoretical investigations}.

Finally, $1-\Indicator(T_i=t)/e_t(X_i)$ is introduced so that the estimating equation is Neyman orthogonal to the nuisance functions, a property that ensures valid inference even when those functions are estimated from data. The current form of $\overline{\mu}_{t,\mathtt{CALM}}$ is nonetheless infeasible, since it depends on the unknown nuisance functions $\mu_t(\cdot)$, $\mu^\dagger_t(\cdot)$, and $\omega_t(\cdot)$. As we establish in Section~\ref{section 4:Theoretical investigations}, Neyman orthogonality combined with sample splitting permits a wide range of first-step estimators for these nuisance functions under very mild regularity conditions, including standard kernel- or sieve-based nonparametric methods and more complex machine learning tools (see Steps \ref{step1:sample split}, \ref{step3:nuisance} and \ref{step4:estimation and inference}).

Before providing the detailed implementation, we briefly discuss the efficiency gains of CALM. Under mild regularity conditions, one can show that the formulas we provide, $\overline{\mu}_{t,\mathtt{AIPW}}$ and $\overline{\mu}_{t,\mathtt{CALM}}$, correspond to the asymptotically linear representations of the AIPW approach and our CALM, respectively. Thus, their variances allow a straightforward comparison of statistical efficiency:
\begin{align}
  \nonumber&\mathsf{V}_{t,\mathtt{AIPW}} = n\cdot\Variance[\overline{\mu}_{t,\mathtt{AIPW}}] =  \Expectation\Big[ \frac{\Variance[Y_i(t)|X_i]}{e_t(X_i)} + (\mu_t(X_i) - \mu_t)^2 \Big],\\
  \label{eq:CALM asy Variance}\text{while}\quad &\mathsf{V}_{t,\mathtt{CALM}} = n\cdot\Variance[\overline{\mu}_{t,\mathtt{CALM}}] = \mathsf{V}_{t,\mathtt{AIPW}} -  \Expectation\Big[ \frac{\Variance[Y_i(t)|X_i]}{e_t(X_i)}(1-e_t(X_i))\rho^2_t(X_i) \Big],
\end{align}
and $\rho_t(X_i) = \corr[Y_i(t),Y_i^\dagger(t)|X_i]$. Thus, CALM delivers nontrivial efficiency gains over AIPW except in the very rare case in which the LLM-based predictions turn out to be random noise for all $X_i$. 

\subsection{CALM with Few-Shot Learning: Implementation Details}\label{subsection 2-2:CALM with Few-Shot Learning: Implementation Details}

We now describe the step-by-step procedure for estimating $\mu_t = \Expectation[Y_i(t)]$ using CALM with few-shot learning. Alongside the detailed steps, we also provide heuristics and guidance for practical use. The first step is a three-way sample split:

\begin{Step}[Three-way sample splitting]\label{step1:sample split}
Randomly split the units $i=1,2,\dots,n$ into three non-overlapping subsamples $\mathcal{I}_1$, $\mathcal{I}_2$, and $\mathcal{I}_3$.
\end{Step}

For now, one may think of the first fold $\mathcal{I}_1$ as a reservoir of demonstrative examples for few-shot learning, the second, $\mathcal{I}_2$, as the fold used to estimate the nuisance functions $\mu_t(\cdot)$, $\mu_t^\dagger(\cdot)$, and $\omega_t(\cdot)$, and the last $\mathcal{I}_3$ for forming estimating equations. Later we will employ cross-fitting (Step \ref{step4:estimation and inference}), in which the roles of the three folds are rotated.

The next step generates predictions of the potential outcomes via few-shot learning with demonstrative-example resampling.

\begin{Step}[Few-shot predictions with demonstrative examples resampling]\label{step2:few-shot}
Fix some (small) $m$. Let $\mathcal{S}^* = \big\{(X_{j}^{*}, Z_{j}^{*}, T^{*}_{j}, Y_{j}^{*}), j=1, \ldots,m \big\}$ 
be a collection of ordered observations, randomly drawn without replacement from the arm-$t$ observations in $\mathcal{I}_1$. Let $i$ denote a unit from $\mathcal{I}_2$ or $\mathcal{I}_3$. Then instruct the LLM to predict the outcome for unit $i$ as if enrolled in treatment $t$, given a context describing the RCT setting, the serialized covariates $(X_i, Z_i)$, and the serialized demonstrative examples $\mathcal{S}^*$. By repeating the above procedure $B$ times with independently resampled examples $\mathcal{S}_b^{*}$ for $b=1, 2,\dots, B$, we obtain the aggregated prediction: 
\begin{align*}
  X_i,Z_i,\mathcal{S}^{*}
  \ \ \xrightarrow[\text{LLM few-shot}]
               {\text{Serialization}}
  \ \ 
  Y^{\dagger}_{i}(t; \mathcal{S}^{*})
  \ \ 
  \xrightarrow[\text{Aggregation}]
               {\text{Resampling }\mathcal{S}^{*}}
  \ \ 
  Y^{\dagger}_{i}(t; \mathcal{I}_{1}) = \frac1{B}\sum_{b=1}^{B}   Y^{\dagger}_{i}(t; \mathcal{S}^{*}_b). 
\end{align*}
\end{Step}

The above step involves constructing a prompt with four components to enable few-shot prediction from a pre-trained LLM. To start, we provide an instruction describing the task: predicting the potential outcome for unit $i$ as if they were assigned to treatment $t$. To ensure the language model correctly interprets the data, the instruction also includes natural language descriptions of the input variables and additional contextual knowledge, such as a description of the treatment arms. Next, we provide the demonstrative examples $\mathcal{S}^{*}$, where each example contains the unit's covariates, treatment assignment, and their observed outcome. Third, we provide the input data $X_i$ and $Z_i$ for unit $i$, where structured covariates are serialized into natural language and unstructured textual data (e.g., free-text responses) are presented directly. Other modalities, such as images, can be uploaded separately if supported by the model interface. Finally, we specify the desired output by requesting the counterfactual outcome prediction in a fixed format. A concrete example of our few-shot prompt design is provided in the Supplementary Materials.

Step 2 simultaneously tackles the two statistical challenges of integrating few-shot predictions into CALM discussed earlier. For the first challenge, it repeatedly resamples small, random sets of demonstrative examples and aggregates the resulting few-shot predictions. This averaging procedure effectively neutralizes the variability introduced by the random composition and ordering of demonstrative examples, which would otherwise persist even as the sample size grows. This step also addresses the second challenge: a naive use of demonstrative examples can induce correlations among the predicted outcomes. More precisely, a naive implementation would stop at $Y^{\dagger}_{i}(t; \mathcal{S}^{*})$ without further resampling or aggregation. This, however, leads to correlated predictions for units $i$ and $i'$ whenever their demonstrative examples overlap. Such correlations are, unfortunately, very difficult to characterize. By contrast, the novel resampling and aggregation we introduce give the final prediction $Y^{\dagger}_{i}(t; \mathcal{I}_{1})$ a U-statistic structure. Together with sample splitting, this ensures that the predictions for different units $i$ and $i'$ are asymptotically independent, a key property we exploit in our theoretical analysis.

Now that predicted potential outcomes have been generated for all units in $\mathcal{I}_2$ and $\mathcal{I}_3$, the next step involves estimating the nuisance functions using subsample $\mathcal{I}_2$ and constructing ``estimating equations'' for units in $\mathcal{I}_3$.

\begin{Step}[Nuisance function and parameter estimation]\label{step3:nuisance}
Estimate $\mu_t(\cdot)$, $\mu_t^{\dagger}(\cdot)$, and $\omega_t(\cdot)$ using subsample $\mathcal{I}_{2}$. This can be done via sieve-based or kernel-based nonparametric procedures, or other machine learning methods. Denote the resulting estimates as $\hat{\mu}_{t}(\cdot;\mathcal{I}_{2})$, $\hat{\mu}^{\dagger}_{t}(\cdot;\mathcal{I}_{2})$, and $\hat{\omega}_{t}(\cdot;\mathcal{I}_{2})$, respectively. Form the estimating equation for observations $i$ in $\mathcal{I}_{3}$:
\begin{align*}
\hat{\varphi}_{t}(i;\mathcal{I}_{1},\mathcal{I}_{2}) &= \frac{\Indicator(T_i=t)}{e_t(X_i)}\Big(Y_i - \hat{\mu}_{t}(X_i;\mathcal{I}_{2})\Big) + \hat{\mu}_{t}(X_i;\mathcal{I}_{2}) 
\\
&\qquad\qquad+ \Big(1-\frac{\Indicator(T_i=t)}{e_t(X_i)}\Big)\Big(Y^{\dagger}_{i}(t; \mathcal{I}_{1}) - \hat{\mu}^{\dagger}_{t}(X_i;\mathcal{I}_{2})\Big)\hat{\omega}_{t}(X_i;\mathcal{I}_{2}).
\end{align*}
\end{Step}

As the notation suggests, $\hat{\varphi}_{t}(i;\mathcal{I}_{1},\mathcal{I}_{2})$ depends on all three data folds: $\mathcal{I}_1$ serves as a pool of demonstrative examples, $\mathcal{I}_2$ provides the nuisance function estimates, and $\mathcal{I}_3$ provides the unit's own observations. Without loss of generality, we will assume that all calibration weights we construct are bounded. 

The final step rotates the roles of the three data folds, a technique known as cross-fitting. We also provide a valid asymptotic variance estimator and the corresponding confidence interval construction.

\begin{Step}[CALM-based estimation and statistical inference]\label{step4:estimation and inference}
Rotate the roles of $\mathcal{I}_1$, $\mathcal{I}_2$, and $\mathcal{I}_3$ by repeating Steps \ref{step2:few-shot}--\ref{step3:nuisance}, and form the final CALM estimator of $\mu_t$:
\begin{align*}
  \hat{\mu}_{t,\mathtt{CALM}} = \frac{1}{n}\Bigg\{\sum_{i\in\mathcal{I}_3} \hat{\varphi}_{t}(i;\mathcal{I}_{1},\mathcal{I}_{2}) + \sum_{i\in\mathcal{I}_1} \hat{\varphi}_{t}(i;\mathcal{I}_{2},\mathcal{I}_{3}) + \sum_{i\in\mathcal{I}_2} \hat{\varphi}_{t}(i;\mathcal{I}_{3},\mathcal{I}_{1})\Bigg\}.
\end{align*}
Construct the asymptotic variance estimator
\begin{align*}
  \hat{\mathsf{V}}_{t,\mathtt{CALM}} = \frac{1}{n}\Bigg\{\sum_{i\in\mathcal{I}_3} \Big(\hat{\varphi}_{t}(i;\mathcal{I}_{1},\mathcal{I}_{2}) - \hat{\mu}_{t,\mathtt{CALM}}\Big)^2 + \sum_{i\in\mathcal{I}_1} \Big(\hat{\varphi}_{t}(i;\mathcal{I}_{2},\mathcal{I}_{3})- \hat{\mu}_{t,\mathtt{CALM}}\Big)^2 + \sum_{i\in\mathcal{I}_2} \Big(\hat{\varphi}_{t}(i;\mathcal{I}_{3},\mathcal{I}_{1})- \hat{\mu}_{t,\mathtt{CALM}}\Big)^2\Bigg\},
\end{align*}
leading to the $1-\alpha$ confidence interval $[\hat{\mu}_{t,\mathtt{CALM}} \pm z_{\alpha/2}\hat{\mathsf{V}}_{t,\mathtt{CALM}}^{1/2}/\sqrt{n}]$ where $z_{\alpha/2}$ denotes the upper $\alpha/2$-quantile of the standard normal distribution.
\end{Step}

\subsection{Misspecified Outcome Models and Constrained Calibration}\label{subsection 2-3:Misspecified Outcome Models and Calibration} 

Step~\ref{step3:nuisance} allows for the estimation of nuisance functions using general nonparametric methods. Empirical practice, however, may favor parsimonious outcome models, such as an additively separable functional form or linear specifications based on finite-dimensional basis expansions. A prominent example of the latter is covariate discretization, also known as coarsening or binning. Two primary considerations drive these modeling choices. First, a parsimonious outcome model greatly simplifies estimation, producing results that are straightforward to compute, report, and interpret, which may be particularly relevant in data-scarce applications. Second, in the specific case of discretization, many randomized controlled trials employ stratified or blocked randomization, making the treatment assignment probability $e_t(X_i)$ piecewise constant across a finite set of strata. Aligning covariate adjustment with the granularity of the randomization design ensures internal consistency. The tradeoff behind parsimonious outcome models is that they are likely misspecified. 

Therefore, in this subsection, we discuss how outcome model misspecification affects our proposed CALM framework; that is, we consider a scenario where $\hat{\mu}_t(\cdot;\mathcal{I}_2)$ and $\hat{\mu}^\dagger_t(\cdot;\mathcal{I}_2)$ converge to $\tilde{\mu}_t(\cdot)$ and $\tilde{\mu}^\dagger_t(\cdot)$, where these limits may differ from the true conditional expectations. In line with parsimonious modeling choices, we also allow the calibration weights to be estimated within a constrained class $\mathcal{W}$. Depending on the specific application and the analyst's modeling preferences, this class can range from fully nonparametric to, at the other extreme, parametric functions of the covariates depending on a finite-dimensional calibration parameter.

We first present a general result on the guaranteed efficiency gain relative to the AIPW approach, even if the outcome models are misspecified and calibration is constrained. To keep the exposition accessible, the discussion in this subsection is heuristic, and formal results are presented in Section~\ref{section 4:Theoretical investigations}. We again start with the asymptotically linear representation: under very mild regularity conditions, our CALM estimator is equivalent to 
\begin{align}\label{eq:CALM with misspecification}
  \overline{\mu}_{t,\mathtt{CALM}} = \frac{1}{n}\sum_{i=1}^n \underbrace{\frac{\Indicator(T_i=t)}{e_t(X_i)}\Big(Y_i - \tilde{\mu}_t(X_i)\Big) + \tilde{\mu}_t(X_i)}_{\text{AIPW with misspecification}} + \underbrace{\Big(1-\frac{\Indicator(T_i=t)}{e_t(X_i)}\Big)\Big(Y_i^\dagger(t) - \tilde{\mu}^\dagger_t(X_i)\Big)\omega_t(X_i)}_{\omega_t(\cdot)\in\mathcal{W}:\ \text{constrained calibration}},
\end{align} 
for some calibration weight that may not be the optimal one. Asymptotic variances of the CALM and AIPW approaches can be found from the above representation. Let $\lambda_t(X_i) = (1-e_t(X_i))/e_t(X_i)$. Then 
\begin{align}
  \nonumber \mathsf{V}_{t,\mathtt{AIPW}} 
  &= \Expectation\Big[ \frac{\Variance[Y_i(t)|X_i]}{e_t(X_i)} + \big(\mu_t(X_i) - \mu_t\big)^2 + \lambda_t(X_i)\big(\tilde{\mu}_{t}(X_i)-\mu_t(X_i)\big)^2\Big],\\
  \label{eq:CALM asy Variance misspecified} \mathsf{V}_{t,\mathtt{CALM}} 
  \nonumber&=  \mathsf{V}_{t,\mathtt{AIPW}} + 
  \Expectation\Big[\omega_{t}^2(X_i)\lambda_t(X_i)\Expectation\big[\big(Y_i^\dagger(t) - \tilde{\mu}^\dagger_{t}(X_i)\big)^2\big|X_i\big] \\
    &\qquad\qquad\qquad\qquad\qquad- 2\omega_{t}(X_i)\lambda_t(X_i)\Expectation\big[\big(Y_i(t) - \tilde{\mu}_{t}(X_i)\big)\big(Y_i^\dagger(t) - \tilde{\mu}^\dagger_{t}(X_i))\big|X_i\big]\Big].
\end{align}
If $\omega_t(\cdot)$ is chosen to minimize the last term over $\mathcal{W}$, then, because the AIPW approach corresponds to $\omega_t(\cdot)=0$, the optimally calibrated CALM estimator is weakly more efficient than AIPW whenever $0\in\mathcal{W}$. This is an extremely mild requirement, and is satisfied, for example, when $\mathcal{W}$ consists of linear models of the covariates, or basis transformations thereof. 

We now discuss a few implementation details and some associated technical considerations. First, a potential change is to employ only two data folds, where $\mathcal{I}_1$ continues to serve as the reservoir for demonstrative examples (i.e., Step \ref{step2:few-shot} remains the same), and $\mathcal{I}_2$ is used for nuisance function estimation, calibration weights construction, and forming an estimator for $\mu_t$. In this way, cross-fitting only needs to be done twice by flipping the roles of $\mathcal{I}_1$ and $\mathcal{I}_2$, thus significantly reducing computation. Formally,

\setcounter{StepDiscrete}{3} 
\begin{StepDiscrete}\label{step 4 (reduced cross-fit)}
Flip the roles of $\mathcal{I}_1$ and $\mathcal{I}_2$ by repeating Steps \ref{step2:few-shot}--\ref{step3:nuisance}, and form the CALM estimator of $\mu_t$:
\begin{align*}
  \hat{\mu}_{t,\mathtt{CALM}} = \frac{1}{n}\Bigg\{\sum_{i\in\mathcal{I}_2} \hat{\varphi}_{t}(i;\mathcal{I}_{1},\mathcal{I}_{2}) + \sum_{i\in\mathcal{I}_1} \hat{\varphi}_{t}(i;\mathcal{I}_{2},\mathcal{I}_{1})\Bigg\}.
\end{align*}
Construct the asymptotic variance estimator
\begin{align*}
  \hat{\mathsf{V}}_{t,\mathtt{CALM}} = \frac{1}{n}\Bigg\{\sum_{i\in\mathcal{I}_2} \Big(\hat{\varphi}_{t}(i;\mathcal{I}_{1},\mathcal{I}_{2}) - \hat{\mu}_{t,\mathtt{CALM}}\Big)^2 + \sum_{i\in\mathcal{I}_1} \Big(\hat{\varphi}_{t}(i;\mathcal{I}_{2},\mathcal{I}_{1})- \hat{\mu}_{t,\mathtt{CALM}}\Big)^2 \Bigg\},
\end{align*}
leading to the $1-\alpha$ confidence interval $[\hat{\mu}_{t,\mathtt{CALM}} \pm z_{\alpha/2}\hat{\mathsf{V}}_{t,\mathtt{CALM}}^{1/2}/\sqrt{n}]$ where $z_{\alpha/2}$ denotes the upper $\alpha/2$-quantile of the standard normal distribution.\label{Step:4'}
\end{StepDiscrete}

Removing one data fold, however, comes at a cost: its validity typically requires that nuisance function estimation is conducted in function classes that are simple or parsimonious (hence Donsker-type), precisely the main motivating scenario of this subsection. We adopt this alternative asymptotic reasoning in Section \ref{section 4:Theoretical investigations} when providing sufficient conditions for CALM under misspecification with constrained calibration. When such a restriction on the nuisance function classes is not plausible, the three-way sample split of Section \ref{subsection 2-2:CALM with Few-Shot Learning: Implementation Details} remains the more robust choice: our CALM estimator then admits an asymptotically linear representation (and hence is asymptotically normally distributed) under very mild conditions on the nuisance function estimates.

We now consider the issue of estimating the calibration weight. The appropriate approach depends on the class $\mathcal{W}$. In the case of unconstrained $\mathcal{W}$ consisting of all measurable functions, the optimal calibration weight is the ratio of $\Expectation[(Y_i(t) - \tilde{\mu}_{t}(X_i))(Y_i^\dagger(t) - \tilde{\mu}^\dagger_{t}(X_i))|X_i]$ and $\Expectation[(Y_i^\dagger(t) - \tilde{\mu}^\dagger_{t}(X_i))^2|X_i]$. On the other hand, one may specify a flexible, parametric function class for the calibration weights through a parameter vector, which may exhibit more stable finite-sample performance in data-scarce environments; that is, $\omega_t(\cdot,\bbeta)$. A convenient estimator for $\bbeta$ can be found by optimizing the estimated asymptotic variance: using the same notation developed above and in Section \ref{subsection 2-2:CALM with Few-Shot Learning: Implementation Details}, define
\begin{align*}
\hat{\bbeta} = \underset{\bbeta}{\mathrm{argmin}} \frac{1}{|\mathcal{I}_2|}\sum_{i\in\mathcal{I}_2} \Big(\hat{\varphi}_{t}(i;\mathcal{I}_{1},\mathcal{I}_{2};\bbeta) - \overline{\varphi}_{t}(\mathcal{I}_{1},\mathcal{I}_{2};\bbeta)\Big)^2,
\end{align*}
where, with a slight abuse of notation, 
\begin{align*}
\hat{\varphi}_{t}(i;\mathcal{I}_{1},\mathcal{I}_{2};\bbeta) &= \frac{\Indicator(T_i=t)}{e_t(X_i)}\Big(Y_i - \hat{\mu}_{t}(X_i;\mathcal{I}_{2})\Big) + \hat{\mu}_{t}(X_i;\mathcal{I}_{2}) 
\\
&\qquad\qquad+ \Big(1-\frac{\Indicator(T_i=t)}{e_t(X_i)}\Big)\Big(Y^{\dagger}_{i}(t; \mathcal{I}_{1}) - \hat{\mu}^{\dagger}_{t}(X_i;\mathcal{I}_{2})\Big){\omega}_{t}(X_i;{\bbeta}),
\end{align*}
and $\overline{\varphi}_{t}(\mathcal{I}_{1},\mathcal{I}_{2};\bbeta)$ denotes the sample average of $\hat{\varphi}_{t}(i;\mathcal{I}_{1},\mathcal{I}_{2};\bbeta) $ in data fold $\mathcal{I}_2$.

In the remainder of this section, we focus on the special case of covariate discretization (a.k.a. coarsening or binning). This approach is particularly attractive due to its simplicity: nuisance functions are estimated as block averages. It also leads to an estimator that is more aligned with the sampling scheme when the underlying trial is conducted in a stratified manner. Building on this insight, we assume there is a pre-specified discretization that partitions the covariate space into $J$ blocks $\mathcal{X}_1,\dots,\mathcal{X}_J$, and that the propensity score is piecewise constant: $e_t(X_i) = \sum_{j=1}^J \Indicator(X_i \in \mathcal{X}_j)e_{t,j}$. Then the two conditional expectations can be estimated via simple block averages 
\begin{align*}
  \hat{\mu}_{t}(X_i;\mathcal{I}_2) = \sum_{j=1}^J \Indicator(X_i \in \mathcal{X}_j) \hat{\mu}_{t,j}(\mathcal{I}_2),\qquad \hat{\mu}_{t,j}(\mathcal{I}_2) = \frac{1}{n_{t,j}(\mathcal{I}_2)}\sum_{i \in \mathcal{I}_{2}} \Indicator(T_i=t,X_i \in \mathcal{X}_j)Y_i,\\
  \hat{\mu}^\dagger_{t}(X_i;\mathcal{I}_2) = \sum_{j=1}^J \Indicator(X_i \in \mathcal{X}_j) \hat{\mu}_{t,j}^\dagger(\mathcal{I}_2),\qquad \hat{\mu}_{t,j}^\dagger(\mathcal{I}_2) = \frac{1}{n_{j}(\mathcal{I}_2)}\sum_{i \in \mathcal{I}_{2}} \Indicator(X_i \in \mathcal{X}_j)Y_i^\dagger(t;\mathcal{I}_1).
\end{align*}
In the above, $n_{t,j}(\mathcal{I}_2) = \sum_{i \in \mathcal{I}_{2}} \Indicator(T_i=t,X_i \in \mathcal{X}_j)$ counts the units in data fold $\mathcal{I}_2$ that belong to block $j$ and are assigned to treatment $t$. Analogously, $n_{j}(\mathcal{I}_2)=\sum_{i \in \mathcal{I}_{2}} \Indicator(X_i \in \mathcal{X}_j)$ is the number of units in $\mathcal{I}_2$ that fall into block $j$. 

Due to discretization, large-sample limits of the estimated nuisance functions are also of piecewise-constant form. Building on the general result established earlier for CALM with misspecified outcomes, it is routine to show that the optimal calibration also takes a piecewise constant structure, and can be estimated via the simple formula:
\begin{align*}
&\hat{\omega}_{t}(X_i;\mathcal{I}_2)
= \sum_{j=1}^J \Indicator(X_i \in \mathcal{X}_j)\hat{\omega}_{t,j}(\mathcal{I}_2),\\ 
&\qquad\text{with }\hat{\omega}_{t,j}(\mathcal{I}_2) = 
\frac{
  n_{t,j}^{-1}(\mathcal{I}_2)\sum\limits_{i \in \mathcal{I}_{2}} \Indicator(T_i=t,X_i \in \mathcal{X}_j)
   \big(Y_i-\hat{\mu}_{t,j}(\mathcal{I}_2)\big)
   \big(Y_{i}^{\dagger}(t;\mathcal{I}_1)-\hat{\mu}_{t,j}^{\dagger}(\mathcal{I}_2)\big)
}{
   n_{j}^{-1}(\mathcal{I}_2)\sum\limits_{i \in \mathcal{I}_{2}} \Indicator(X_i \in \mathcal{X}_j)
   \big(Y_{i}^{\dagger}(t;\mathcal{I}_1)-\hat\mu_{t,j}^{\dagger}(\mathcal{I}_2)\big)^2
}.
\end{align*}
Notably, the weight $\lambda_t(X_i)$ from the general result above does not appear in $\hat{\omega}_{t,j}(\mathcal{I}_2)$: since the propensity score is piecewise constant, $\lambda_t(X_i)=\lambda_{t,j}=(1-e_{t,j})/e_{t,j}$ for all $i$ with $X_i\in\mathcal{X}_j$.

As a final remark, the granularity of $\mathcal{X}_1,\dots,\mathcal{X}_J$ trades off robustness against the size of the efficiency gain: coarser discretizations average over larger cells, which may stabilize the estimates in finite samples but can dilute genuine heterogeneity in the true calibration weight, weakening the correlation between the true and LLM-generated potential outcomes. On the other hand, finer discretizations do the opposite, at the cost of noisier within-cell estimates when cells contain few observations.
 
\section{Extensions}\label{section 3:Extensions}

We now discuss special cases and extensions of the CALM framework introduced in Section \ref{section 2:Setup and Methodology}. These include counterfactual predictions from zero-shot learning and the estimation of both unconditional and conditional average treatment effects.

\subsection{CALM with Zero-Shot Predictions}\label{subsection 3-1:CALM with Zero-Shot Predictions}

A special case of CALM is $m=0$, where the LLM-generated counterfactual predictions are produced without any demonstrative examples. This is commonly referred to as zero-shot learning \citep{brown2020language}. Throughout, we use the subscript $\mathtt{ZS}$ to denote objects specific to this zero-shot special case, in parallel to the subscript-free notation used for the few-shot CALM estimator of Section~\ref{subsection 2-2:CALM with Few-Shot Learning: Implementation Details}. The zero-shot prediction is obtained via
\begin{align*}
  X_i,Z_i \ \ \xrightarrow[\text{LLM zero-shot}]{\text{Serialization}} \ \ Y_{i,\mathtt{ZS}}^\dagger(t), \quad \text{for }1\leq i\leq n,
\end{align*}
using a prompt with three components: an instruction describing the counterfactual prediction task and the treatment arms, the serialized input $(X_i,Z_i)$, and the desired output format. No demonstrative examples $\mathcal{S}^*$ are drawn, and consequently no resampling or aggregation step is needed: a single LLM query per unit and treatment arm suffices to obtain $Y_{i,\mathtt{ZS}}^\dagger(t)$.

Because zero-shot predictions no longer depend on a shared pool of demonstrative examples, the reservoir fold $\mathcal{I}_1$ in Step~\ref{step1:sample split} becomes unnecessary. The sample splitting therefore simplifies to a two-way split into folds $\mathcal{I}_1$ and $\mathcal{I}_2$, and Steps~\ref{step3:nuisance}--\ref{step4:estimation and inference} carry over with $Y_i^\dagger(t;\mathcal{I}_1)$ replaced by $Y_{i,\mathtt{ZS}}^\dagger(t)$ and the corresponding nuisance functions
\begin{align*}
  \mu_{t,\mathtt{ZS}}^\dagger(X_i) = \Expectation[Y_{i,\mathtt{ZS}}^\dagger(t)|X_i], \qquad
  \omega_{t,\mathtt{ZS}}(X_i) = \frac{\cov[Y_i(t),Y_{i,\mathtt{ZS}}^\dagger(t)|X_i]}{\Variance[Y_{i,\mathtt{ZS}}^\dagger(t)|X_i]}.
\end{align*}
Estimating $\mu_t(\cdot)$, $\mu_{t,\mathtt{ZS}}^\dagger(\cdot)$, and $\omega_{t,\mathtt{ZS}}(\cdot)$ on one fold and forming the estimating equation on the other, then rotating the two folds, yields the zero-shot CALM estimator
\begin{align*}
  \hat{\mu}_{t,\mathtt{ZS}} = \frac{1}{n}\Bigg\{\sum_{i\in\mathcal{I}_2} \hat{\varphi}_{t,\mathtt{ZS}}(i;\mathcal{I}_1) + \sum_{i\in\mathcal{I}_1} \hat{\varphi}_{t,\mathtt{ZS}}(i;\mathcal{I}_2)\Bigg\},
\end{align*}
with $\hat\varphi_{t,\mathtt{ZS}}(i;\mathcal{I}_1)$ defined exactly as $\hat\varphi_t(i;\mathcal{I}_1,\mathcal{I}_2)$ in Step~\ref{step3:nuisance}, but with $Y_{i,\mathtt{ZS}}^\dagger(t)$, $\hat\mu_{t}(X_i;\mathcal{I}_1)$, $\hat\mu^{\dagger}_{t,\mathtt{ZS}}(X_i;\mathcal{I}_1)$, and $\hat\omega_{t,\mathtt{ZS}}(X_i;\mathcal{I}_1)$ in place of $Y_i^\dagger(t;\mathcal{I}_1)$, $\hat\mu_t(X_i;\mathcal{I}_2)$, $\hat\mu_t^\dagger(X_i;\mathcal{I}_2)$, and $\hat\omega_t(X_i;\mathcal{I}_2)$. Asymptotic variance estimation and confidence interval construction proceed analogously.

The appeal of the zero-shot special case is largely computational. Because no demonstrative examples are resampled, obtaining $\{Y_{i,\mathtt{ZS}}^\dagger(t)\}_{i=1}^n$ requires only $n$ LLM queries, as opposed to the $nB$ queries needed to aggregate few-shot predictions over $B$ resamples in Step~\ref{step2:few-shot}. This reduction in queries translates directly into lower latency and monetary cost, both of which scale with the number of API calls to the LLM in practice. Dropping the demonstrative-example block from the prompt also shortens each query, easing the token-budget constraints imposed by the LLM's context window. Sample splitting is likewise simpler: with no fold reserved purely as a donor pool of examples, the two remaining folds are larger than any two folds under the three-way split, and the resulting estimating equations no longer require the U-statistic argument used in Section~\ref{section 4:Theoretical investigations} to establish asymptotic independence of the predictions across units, as $Y_{i,\mathtt{ZS}}^\dagger(t)$ depends only on unit $i$'s own covariates.

This computational simplicity, however, typically comes at the cost of a smaller efficiency gain over AIPW. Recall from Section~\ref{subsection 2-1:Motivation and Efficiency Gains of CALM} that the variance reduction delivered by CALM is governed by the strength of the conditional correlation between the true and predicted potential outcomes, $\rho_t(X_i)=\corr[Y_i(t),Y_i^\dagger(t)|X_i]$: the corresponding zero-shot analogue $\rho_{t,\mathtt{ZS}}(X_i)=\corr[Y_i(t),Y_{i,\mathtt{ZS}}^\dagger(t)|X_i]$ enters the variance formula in exactly the same way,
\begin{align*}
  \mathsf{V}_{t,\mathtt{ZS}} = \mathsf{V}_{t,\mathtt{AIPW}} - \Expectation\Big[ \frac{\Variance[Y_i(t)|X_i]}{e_t(X_i)}(1-e_t(X_i))\rho^2_{t,\mathtt{ZS}}(X_i) \Big].
\end{align*}
Because a zero-shot prediction is generated from the LLM's pretrained knowledge alone, without ever observing a single realized $(X,Z,Y)$ triple from the trial, it has no opportunity to adapt to the specific empirical relationship between covariates and outcomes in the study population: idiosyncrasies of the outcome scale, site-specific practices, and subtle features of the intervention protocol are all invisible to a zero-shot prompt. Few-shot examples, by contrast, implicitly reveal this relationship to the LLM, which can be viewed as a form of in-context calibration to the trial at hand. We therefore expect $\rho_{t,\mathtt{ZS}}(x)$ to be weaker than its few-shot counterpart in most applications, so that zero-shot CALM delivers a smaller (though still nontrivial) variance reduction relative to AIPW. Crucially, this loss is one of efficiency, not validity: because the correction term remains residualized and optimally weighted, $\hat\mu_{t,\mathtt{ZS}}$ is consistent and no less efficient than AIPW regardless of how weak $\rho_{t,\mathtt{ZS}}(x)$ turns out to be. Zero-shot CALM is thus a computationally attractive default that is always safe to use, while the few-shot approach of Section~\ref{subsection 2-2:CALM with Few-Shot Learning: Implementation Details} is preferable whenever the added query cost is acceptable and reliable in-sample examples are available.

\subsection{Average Treatment Effect Estimation}\label{subsection 3-2:CALM for estimating ATE}

Fixing two treatments $t$ and $t'$, we now turn to the average treatment effect (ATE) $\tau_{t,t'}= \mu_t - \mu_{t'} = \Expectation[Y_i(t)-Y_i(t')] $. A natural candidate combines the two CALM mean estimators of Section~\ref{subsection 2-2:CALM with Few-Shot Learning: Implementation Details}, applied separately to arms $t$ and $t'$: $ \hat\mu_{t,\mathtt{CALM}} - \hat\mu_{t',\mathtt{CALM}}$. This difference inherits consistency, asymptotic normality, and the arm-by-arm efficiency gain over AIPW established in Section~\ref{section 2:Setup and Methodology}. The only ingredient that is not already efficient for the difference is the calibration weight: $\omega_t(X_i)$ and $\omega_{t'}(X_i)$ are each optimal for predicting their own arm's missing outcome, but treating them separately ignores that $Y_i^\dagger(t)$ and $Y_i^\dagger(t')$ are themselves correlated given $X_i$, as both counterfactual predictions are generated by employing the same unit-level pre-intervention information $(X_i,Z_i)$. In addition, it is possible that the counterfactual prediction for arm $t$ is predictive for the $t'$-potential outcome. We now derive the calibration weight that is optimal for the difference.

Substituting the asymptotic linear representation \eqref{eq:CALM asymptotic rep} for each arm, but leaving the calibration weights generic rather than fixing them at their Section~\ref{subsection 2-1:Motivation and Efficiency Gains of CALM} values, gives
\begin{align}
\nonumber&\overline{\tau}_{t,t',\mathtt{CALM}} = \frac1n\sum_{i=1}^n\ \bigg[\frac{\Indicator(T_i=t)}{e_t(X_i)}\big(Y_i-\mu_t(X_i)\big)+\mu_t(X_i)\bigg] - \bigg[\frac{\Indicator(T_i=t')}{e_{t'}(X_i)}\big(Y_i-\mu_{t'}(X_i)\big)+\mu_{t'}(X_i)\bigg]\\
\label{eq:CALM ATE asymptotic rep} &\qquad+\Big(1-\frac{\Indicator(T_i=t)}{e_t(X_i)}\Big)\big(Y_i^\dagger(t)-\mu_t^\dagger(X_i)\big)\omega_t(X_i) - \Big(1-\frac{\Indicator(T_i=t')}{e_{t'}(X_i)}\Big)\big(Y_i^\dagger(t')-\mu_{t'}^\dagger(X_i)\big)\omega_{t'}(X_i).
\end{align}
Importantly, the variance of $\overline\tau_{t,t',\mathtt{CALM}}$ provides insights on how the asymptotic variance of the ATE estimator depends on the calibration weights. To be specific, define
\begin{align}\label{eq:ATE asym variance}
  \mathsf{V}_{t,t',\mathtt{CALM}} = n\cdot\Variance[\overline{\tau}_{t,t',\mathtt{CALM}}] &= n\cdot\Variance[\overline{\tau}_{t,t',\mathtt{AIPW}}] + \Expectation\Big[\bomega(X_i)^\transpose \bSigma^\dagger(X_i) \bomega(X_i) - 2\bomega(X_i)^\transpose \bsigma(X_i)]\Big],
\end{align}
where $\bomega(X_i) = (\omega_t(X_i),\omega_{t'}(X_i))$, and $\bSigma^\dagger(X_i)$ and $\bsigma(X_i)$ are weighted conditional covariance matrices:
\begin{align*}
  \bSigma^\dagger(X_i) &= \begin{bmatrix}
    \lambda_t(X_i)\Variance[Y_i^\dagger(t)|X_i] & &   \cov[Y_i^\dagger(t),Y_i^\dagger(t')|X_i] \\
    \cov[Y_i^\dagger(t),Y_i^\dagger(t')|X_i] & & \lambda_{t'}(X_i)\Variance[Y_i^\dagger(t')|X_i],
  \end{bmatrix}\\  
  \bsigma(X_i) &= \begin{bmatrix}
    \lambda_t(X_i)\cov[Y_i(t),Y_i^\dagger(t)|X_i] \ \ +\ \  \cov[Y_i(t'),Y_i^\dagger(t)|X_i]\\
    \cov[Y_i(t),Y_i^\dagger(t')|X_i] \ \ +\ \  \lambda_{t'}(X_i)\cov[Y_i(t'),Y_i^\dagger(t')|X_i]
  \end{bmatrix},
\end{align*}
with $\lambda_t(X_i) = (1-e_t(X_i))/e_t(X_i)$. As such, the optimal calibration weights are
\begin{align}\label{eq:optimal weights for ATE}
  \bomega_{\mathtt{ATE}}(X_i) = \Big(\omega_{t,\mathtt{ATE}}(X_i),\ \omega_{t',\mathtt{ATE}}(X_i)\Big) = \bSigma^\dagger(X_i)^{-1} \bsigma(X_i).
\end{align}

The weights above clearly reduce to the arm-specific weights $\big(\omega_t(X_i),\omega_{t'}(X_i)\big)$ of Section~\ref{section 2:Setup and Methodology} whenever the two predictions are conditionally uncorrelated, and each prediction is informative only about its own arm's true outcome. In that case $\bSigma^\dagger(X_i)$ is diagonal and each component of $\bsigma(X_i)$ retains only its own-arm term, so $\bomega_{\mathtt{ATE}}(X_i)$ solves two decoupled univariate problems: joint calibration is a strict generalization of, rather than a departure from, the mean-estimation case. 

Estimation replaces the population conditional moments in \eqref{eq:optimal weights for ATE} by their nonparametric estimates, exactly as in Step~\ref{step3:nuisance}, using $Y_i^\dagger(t;\mathcal{I}_1)$ and $Y_i^\dagger(t';\mathcal{I}_1)$ in place of $Y_i^\dagger(t)$ and $Y_i^\dagger(t')$; call the resulting estimates $\hat\omega_{t,\mathtt{ATE}}(\cdot)$ and $\hat\omega_{t',\mathtt{ATE}}(\cdot)$. Substituting them, together with $\hat\mu_t(\cdot)$, $\hat\mu_{t'}(\cdot)$, $\hat\mu_t^\dagger(\cdot)$, $\hat\mu_{t'}^\dagger(\cdot)$, into the estimating equation of Step~\ref{step3:nuisance}, with the arm-specific calibration weight there replaced by its ATE-specific counterpart, defines $\hat{\varphi}_{t,\mathtt{ATE}}(i;\mathcal{I}_1,\mathcal{I}_2)$ and $\hat\varphi_{t',\mathtt{ATE}}(i;\mathcal{I}_1,\mathcal{I}_2)$; the notation again emphasizes that $\mathcal{I}_1$ provides demonstrative examples for counterfactual prediction using few-shot learning, while the nuisance functions are estimated with units from data fold $\mathcal{I}_2$. The final ATE estimator, $\hat\tau_{t,t',\mathtt{CALM}}$, together with standard error and confidence intervals, can be constructed employing the same cross-fitting technique detailed in Step \ref{step4:estimation and inference}.

Before closing this subsection, we remark on the special case of discretized covariates discussed in Section~\ref{subsection 2-3:Misspecified Outcome Models and Calibration}, a popular choice both for experimental design and for post hoc analysis. As argued there, discretizing $X_i$ into $J$ blocks $\mathcal{X}_1,\dots,\mathcal{X}_J$ lets the conditional expectations $\hat\mu_t(\cdot)$, $\hat\mu_{t'}(\cdot)$, $\hat\mu_t^\dagger(\cdot)$, $\hat\mu_{t'}^\dagger(\cdot)$ be estimated as simple block averages. Building on this, one can drop the data fold $\mathcal{I}_3$ (or, more precisely, combine $\mathcal{I}_2$ and $\mathcal{I}_3$) and avoid the extra round of cross-fitting, since the relevant Donsker-type (stochastic equicontinuity) condition is easily verified in this setting; see Step~\ref{step 4 (reduced cross-fit)}, Assumption~\ref{assumption: Misspecified outcome models with constrained calibration}, and Corollary \ref{coro:Covariate discretization} for details. Under covariate discretization based on a stratified randomization, the optimal calibration weights inherit the same simplification: like the nuisance functions, they are piecewise constant, and within each block their estimation reduces to an elementary two-variable regression rather than separate estimation of the underlying variances and covariances. To be specific, fix a block $j$. Using observations from this block in data fold $\mathcal{I}_2$, $\hat\omega_{t,j,\mathtt{ATE}}(\mathcal{I}_2)$ and $\hat\omega_{t',j,\mathtt{ATE}}(\mathcal{I}_2)$ are simply the coefficients from 
\begin{align*}
  &\ \text{regressing}\quad - \frac{\Indicator(T_i=t)}{e_{t,j}}\big(Y_i-\hat{\mu}_{t,j}(\mathcal{I}_2)\big) + \frac{\Indicator(T_i=t')}{e_{t',j}}\big(Y_i-\hat{\mu}_{t',j}(\mathcal{I}_2)\big) \\
  & \text{on}\quad \Big(1-\frac{\Indicator(T_i=t)}{e_{t,j}}\Big)\big(Y_i^\dagger(t;\mathcal{I}_1)-\hat{\mu}_{t,j}^\dagger(\mathcal{I}_2)\big)\quad \text{and}\quad - \Big(1-\frac{\Indicator(T_i=t')}{e_{t',j}}\Big)\big(Y_i^\dagger(t';\mathcal{I}_1)-\hat{\mu}_{t',j}^\dagger(\mathcal{I}_2)\big).
\end{align*}

\subsection{Conditional Average Treatment Effect}\label{subsection:CALM for estimating CATE}

Building on the average treatment effect estimator of Section~\ref{subsection 3-2:CALM for estimating ATE}, we now turn to the conditional average treatment effect (CATE) estimand, $\tau_{t,t'}(x) = \Expectation[Y_i(t)-Y_i(t')|X_i=x]$. Recall $\hat{\varphi}_{t,\mathtt{ATE}}(i;\mathcal{I}_1,\mathcal{I}_2)$ and $\hat\varphi_{t',\mathtt{ATE}}(i;\mathcal{I}_1,\mathcal{I}_2)$ from the previous subsection, and define
\begin{align*}
\hat{\varphi}_{t,t'}(i;\mathcal{I}_1,\mathcal{I}_2) = \hat{\varphi}_{t,\mathtt{ATE}}(i;\mathcal{I}_1,\mathcal{I}_2) - \hat\varphi_{t',\mathtt{ATE}}(i;\mathcal{I}_1,\mathcal{I}_2),
\end{align*}
as well as the kernel-based smoothing weights:
\begin{align*}
  \kappa_i(x;\mathcal{I}_3) = \frac{k_h(X_i-x)}{\sum_{i \in \mathcal{I}_3} k_h( X_i-x)},\qquad k_h(X_i - x) = \frac{1}{h^p}k\Big(\frac{X_i-x}{h}\Big),
\end{align*}
for some kernel function $k$, bandwidth sequence $h$, and covariate dimension $p$. A CALM-based CATE estimator can then be constructed as a cross-fitted local average:
\begin{align*}
\hat\tau_{t,t',\mathtt{CALM}}(x)
&=
\frac{1}{3}\Bigg\{\sum_{i\in \mathcal{I}_3}
\kappa_i(x;\mathcal{I}_3)
\hat{\varphi}_{t,t'}(i;\mathcal{I}_1,\mathcal{I}_2)\\
&\qquad\qquad\qquad+ \sum_{i\in \mathcal{I}_2}
\kappa_i(x;\mathcal{I}_2)
\hat{\varphi}_{t,t'}(i;\mathcal{I}_3,\mathcal{I}_1)
 + \sum_{i\in \mathcal{I}_1}
\kappa_i(x;\mathcal{I}_1)
\hat{\varphi}_{t,t'}(i;\mathcal{I}_2,\mathcal{I}_3)
\Bigg\}.
\end{align*}
This concrete structure allows us to explicitly compute the asymptotic variance of the CATE estimator; in particular, we show in Section \ref{section 4:Theoretical investigations} that CALM again delivers an efficiency gain over the AIPW approach, provided that the language-model-based predictions are correlated with the true potential outcomes. 

To gain some intuition, again consider the asymptotic linear representation. Let $\kappa_i(x) = k_h(X_i-x)/\sum_{i=1}^n k_h(X_i-x)$ denote the local weights constructed using the full sample; then
\begin{align}
\nonumber\overline{\tau}_{t,t',\mathtt{CALM}}(x) = \sum_{i=1}^n \kappa_i(x) &\Bigg\{\frac{\Indicator(T_i=t)}{e_t(X_i)}\big(Y_i-\mu_t(X_i)\big)+\mu_t(X_i) - \frac{\Indicator(T_i=t')}{e_{t'}(X_i)}\big(Y_i-\mu_{t'}(X_i)\big)-\mu_{t'}(X_i)\\
\nonumber &\qquad\qquad+\Big(1-\frac{\Indicator(T_i=t)}{e_t(X_i)}\Big)\big(Y_i^\dagger(t)-\mu_t^\dagger(X_i)\big)\omega_{t,\mathtt{ATE}}(X_i)\\
\label{eq:CALM CATE asymptotic rep}&\qquad\qquad - \Big(1-\frac{\Indicator(T_i=t')}{e_{t'}(X_i)}\Big)\big(Y_i^\dagger(t')-\mu_{t'}^\dagger(X_i)\big)\omega_{t',\mathtt{ATE}}(X_i)\Bigg\}.
\end{align}
Then its conditional variance serves as a natural benchmark for efficiency comparison: define
\begin{align}
  \nonumber&\mathsf{V}_{t,t',\mathtt{CALM}}(x) = nh^p\cdot\Variance[\overline{\tau}_{t,t',\mathtt{CALM}}(x)|\mathbf{X}] \\
  \label{eq:CATE asym variance}&\quad = nh^p\cdot\Variance[\overline{\tau}_{t,t',\mathtt{AIPW}}(x)|\mathbf{X}] + \frac{\kappa_2}{f(x)}\Big(\bomega_{\mathtt{ATE}}(x)^\transpose \bSigma^\dagger(x) \bomega_{\mathtt{ATE}}(x) - 2\bomega_{\mathtt{ATE}}(x)^\transpose \bsigma(x)\Big) + o_p(1).
\end{align}
In the above, $\kappa_2 = \int k(u)^2\diff u$ depends only on the kernel function, $f(x)$ is the density of $X_i$ at the evaluation point $x$, and $\Expectation[\cdot|\mathbf{X}]$ represents expectation conditional on all structured covariates $X_1,X_2,\dots,X_n$. Importantly, because the oracle calibration weights we derived in Section~\ref{subsection 3-2:CALM for estimating ATE} are already specific to each value of the structured covariates, they also minimize the asymptotic variance of the conditional average treatment effect estimator, meaning that 
\begin{align*}
  \frac{\kappa_2}{f(x)}\Big(\bomega_{\mathtt{ATE}}(x)^\transpose \bSigma^\dagger(x) \bomega_{\mathtt{ATE}}(x) - 2\bomega_{\mathtt{ATE}}(x)^\transpose \bsigma(x)\Big) \leq 0,
\end{align*}
with equality holding only if $\bsigma(x) = 0$.

\section{Theoretical Investigations}\label{section 4:Theoretical investigations}

We now investigate the theoretical properties of the CALM framework introduced in Sections \ref{section 2:Setup and Methodology} and \ref{section 3:Extensions}, which integrates few-shot LLM predictions into treatment effect estimation. We first formalize the core assumptions underpinning our approach. We then establish the asymptotic properties of the CALM estimator for various causal parameters, demonstrating its consistency, asymptotic normality, and the validity of the resulting statistical inference.

Our first assumption concerns unconfounded treatment assignment and strong overlap \cite{ma2025testing,ma2020robust}, a condition inherently satisfied in randomized controlled trials.

\begin{assumption}[Treatment assignment]\label{assumption: Randomized treatment assignment}
(i) Treatment assignment $T \in \{1,\dots,k\}$ is independent of the potential outcomes and the unstructured data, conditional on the structured covariates: $T \perp\!\!\!\perp \{Y(1), \ldots, Y(k), Z\} \mid X$. 
(ii) The propensity score $e_t(x) = \Probability[T=t|X=x]$ is bounded away from $0$ for all $t\in\{1,\dots,k\}$.
\end{assumption}

Next, we assume that the data consist of independent and identically distributed (iid) observations.

\begin{assumption}[Random sampling]\label{assumption: I.I.D. sampling}
The observed data $ \{(Y_i, T_i, X_i, Z_i)\}_{i=1}^n$ are independent and identically distributed.
\end{assumption} 

Recall that a unique feature of our method is that it averages over LLM-based predictions generated using small and randomly sampled demonstrative examples. While this averaging helps reduce the randomness of the LLM-predicted potential outcomes and eliminates dependence on the ordering of the examples, it also complicates the theoretical analysis. To facilitate the discussion, we first introduce a notation that can be viewed as a large-sample analogue of the few-shot predicted outcome $Y_{i}^\dagger(t;\mathcal{I})$. Let $\mathcal{S}^* = (O_{1}^*, O_{2}^*, \dots, O_{m}^*)$ denote a random sample drawn from the arm $T=t$, independent of $(X_i,Z_i)$. In line with Step~\ref{step2:few-shot}, let $Y^\dagger_i(t; \mathcal{S}^*)$ denote the few-shot prediction. We then define
\[
Y^\dagger_i(t) = \Expectation[Y^\dagger_i(t; \mathcal{S}^*) \mid X_i, Z_i],
\]
where the expectation is taken with respect to the random demonstrative sample $\mathcal{S}^*$. To be fully rigorous, a more precise notation would be $Y^\dagger_{i,m,\theta}(t)$ to explicitly denote the dependence on both the number of demonstrative examples, $m$, and the parameters of the underlying language model, $\theta$. However, we suppress these subscripts for notational simplicity. 

Our next assumption requires finite moments and nonvanishing variance for the outcome variables, along with bounded counterfactual predictions. In applications, boundedness is standard since prompts naturally constrain outcome ranges. From a technical standpoint, this condition helps establish the convergence of aggregated predictions, though it can be relaxed to finite higher-order moments.

\begin{assumption}[Finite moments]\label{assumption: finite moments}
There exists some constant $c > 0$ such that for all $t=1,2,\dots,k$, $|\mu_t(X_i)|$, $\Variance[Y_i(t)|X_i]$, $|Y_i^\dagger(t;\mathcal{S}^*)|\leq c$. In addition, $\Variance[Y_i(t)|X_i]$, $\Variance[Y_i^\dagger(t)|X_i]\geq c^{-1}$.
\end{assumption}

We now introduce a high-level assumption stating that the estimated nuisance functions, $\hat{\mu}_{t}(\cdot)$, $\hat{\mu}_{t}^{\dagger}(\cdot)$, and $\hat{\omega}_{t}(\cdot)$, are consistent. Depending on the specific form of the estimators, such as $k$-nearest neighbors, kernel- or series-based methods, regression trees, or random forests, more primitive conditions are available in the literature. Such high-level consistency assumptions are commonly employed in the analysis of semiparametric estimators with Neyman-orthogonal estimating equations. Together with sample splitting in Step~\ref{step1:sample split} and cross-fitting in Step~\ref{step4:estimation and inference}, they help facilitate the establishment of consistency and asymptotic normality. Let $\|\cdot\|_{L_2}$ denote the $L_2$-norm with respect to the distribution of $X$. 

\begin{assumption}[Consistent nuisance function estimation]\label{assumption:Consistent nuisance function estimation}
The cross-fitted function estimates obtained are consistent: $\|\hat{\mu}_{t}(\cdot;\mathcal{I})-\mu_{t}(\cdot)\|_{L_2}$, $\|\widehat{\mu}^{\dagger}_{t}(\cdot;\mathcal{I})-\mu^{\dagger}_{t}(\cdot)\|_{L_2}$ and $\|\widehat{\omega}_{t}(\cdot;\mathcal{I})-\omega_{t}(\cdot)\|_{L_2} \xrightarrow{\mathrm{p}} 0$ for each data fold $\mathcal{I}$ and treatment arm $t\in\{1,\dots,k\}$.
\end{assumption}

We are now ready to state the main theoretical result: the CALM estimator $\hat{\mu}_{t,\mathtt{CALM}}$, which incorporates few-shot predicted potential outcomes using an LLM, is consistent, asymptotically normally distributed, and achieves a smaller asymptotic variance than the standard AIPW estimator whenever the LLM-based predictions are informative about the true potential outcomes. 

\begin{theorem}[Asymptotic properties of $\hat{\mu}_{t,\mathtt{CALM}}$]\label{theorem: Asymptotic properties of CALM with few-shot learning}
Let Assumptions~\ref{assumption: Randomized treatment assignment}--\ref{assumption:Consistent nuisance function estimation} hold, and $B\to\infty$. Then 
\begin{align*}
  \sqrt{n}\big(\hat{\mu}_{t,\mathtt{CALM}} - \overline{\mu}_{t,\mathtt{CALM}}\big)\xrightarrow{\mathrm{p}} 0, \quad \text{which implies}\quad 
  \sqrt{n}\big(\hat{\mu}_{t,\mathtt{CALM}} - \mu_t\big) \xrightarrow{\mathrm{d}} \mathcal{N}(0,\mathsf{V}_{t,\mathtt{CALM}}),
\end{align*}
where $\overline{\mu}_{t,\mathtt{CALM}}$ and $\mathsf{V}_{t,\mathtt{CALM}}$ are defined in \eqref{eq:CALM asymptotic rep} and \eqref{eq:CALM asy Variance}, respectively. 
\end{theorem}

The above theorem establishes that the CALM estimator remains consistent and asymptotically normal when coupled with the resampling-based few-shot strategy introduced in Section~\ref{section 2:Setup and Methodology}. It is also clear that CALM yields strictly smaller asymptotic variance, and thus significant efficiency gains, whenever the few-shot LLM-based prediction is informative about the true potential outcome $Y_i(t)$, i.e., when $\rho_{t}(x)$, or equivalently $\omega_{t}(x)$, is nonzero. This is also borne out in our empirical analysis. 

Next, we examine to what extent the conclusions of Theorem \ref{theorem: Asymptotic properties of CALM with few-shot learning} continue to hold without Assumption \ref{assumption:Consistent nuisance function estimation}, i.e., under outcome model misspecification and constrained calibration, a scenario we discuss in Section \ref{subsection 2-3:Misspecified Outcome Models and Calibration}. We make the following ``stochastic equicontinuity'' assumption on the estimated nuisance functions and calibration weights, which helps establish an asymptotically linear representation and hence, the asymptotic normality of the CALM estimator. Such conditions allow removing one data fold, reducing the algorithm to a two-way sample split, thus significantly reducing computation. They are also relatively straightforward to verify whenever nuisance function estimation is conducted in classes that are simple or parsimonious (i.e., Donsker), a main motivational scenario behind Section \ref{subsection 2-3:Misspecified Outcome Models and Calibration}.

\begin{assumption}[Misspecified outcome models with constrained calibration]\label{assumption: Misspecified outcome models with constrained calibration} 
Let $\tilde{\mu}_t(\cdot)$ and $\tilde{\mu}^\dagger_t(\cdot)$ be two functions with finite variances, $\Variance[\tilde{\mu}_t(X_i)]$, $\Variance[\tilde{\mu}_t^\dagger(X_i)] < \infty$, and $\omega_t(\cdot)\in\mathcal{W}$ be some calibration weight. Then 
\begin{align*}
  &\frac{1}{\sqrt{|\mathcal{I}'|}}\sum_{i \in \mathcal{I}'} \Big(1 - \frac{\Indicator(T_i=t)}{e_t(X_i)}\Big)\big( \hat{\mu}_t(X_i;\mathcal{I}') - \tilde{\mu}_t(X_i) \big)\xrightarrow{\mathrm{p}} 0,\\
  &\frac{1}{\sqrt{|\mathcal{I}'|}}\sum_{i \in \mathcal{I}'} \Big(1 - \frac{\Indicator(T_i=t)}{e_t(X_i)}\Big) Y_i^\dagger(t;\mathcal{I})\big(\hat{\omega}_t(X_i;\mathcal{I}') - \omega_t(X_i)\big)\xrightarrow{\mathrm{p}} 0,\\
  &\frac{1}{\sqrt{|\mathcal{I}'|}}\sum_{i \in \mathcal{I}'} \Big(1 - \frac{\Indicator(T_i=t)}{e_t(X_i)}\Big)\big( \hat{\mu}^\dagger_t(X_i;\mathcal{I}') \hat{\omega}_t(X_i;\mathcal{I}')- \tilde{\mu}_t^\dagger(X_i)\omega_t(X_i)\big)\xrightarrow{\mathrm{p}} 0,
\end{align*}
for data folds $\mathcal{I}\neq\mathcal{I}'$. 
\end{assumption}

We are now ready to present the following theorem. 

\begin{theorem}[$\hat{\mu}_{t,\mathtt{CALM}}$ under misspecification and constrained calibration]\label{theorem: Asymptotic misspecification}
Let Assumptions~\ref{assumption: Randomized treatment assignment}--\ref{assumption: finite moments} and \ref{assumption: Misspecified outcome models with constrained calibration} hold, and $B\to\infty$. Also assume that $\hat{\mu}_{t,\mathtt{CALM}}$ is constructed via Step \ref{step 4 (reduced cross-fit)} in Section \ref{subsection 2-3:Misspecified Outcome Models and Calibration}. Then
\begin{align*}
  \sqrt{n}\big(\hat{\mu}_{t,\mathtt{CALM}} - \overline{\mu}_{t,\mathtt{CALM}}\big)\xrightarrow{\mathrm{p}} 0, \quad \text{which implies}\quad 
  \sqrt{n}\big(\hat{\mu}_{t,\mathtt{CALM}} - \mu_t\big) \xrightarrow{\mathrm{d}} \mathcal{N}(0,\mathsf{V}_{t,\mathtt{CALM}}),
\end{align*}
where $\overline{\mu}_{t,\mathtt{CALM}}$ and $\mathsf{V}_{t,\mathtt{CALM}}$ are defined in \eqref{eq:CALM with misspecification} and \eqref{eq:CALM asy Variance misspecified}, respectively. If, in addition, $\omega_t(\cdot)$ is the optimal constrained calibration weight and $0\in\mathcal{W}$, then $\mathsf{V}_{t,\mathtt{CALM}} \leq \mathsf{V}_{t,\mathtt{AIPW}}$.
\end{theorem}

As we discussed above and in Section \ref{subsection 2-3:Misspecified Outcome Models and Calibration}, the Donsker or stochastic equicontinuity type condition in Assumption \ref{assumption: Misspecified outcome models with constrained calibration} is most appropriate when the function classes used to estimate the nuisance functions are not ``too large.'' This encompasses both parametric models and classical nonparametric techniques, such as kernel- or sieve-based methods under additional regularity conditions on the bandwidth or sieve dimension. Given space constraints, however, we present below a special but important case of covariate discretization. 

\begin{corollary}[Covariate discretization]\label{coro:Covariate discretization}
  Consider the discretized covariates approach introduced in Section~\ref{subsection 2-3:Misspecified Outcome Models and Calibration}, and assume that $\mathbb{P}[X_i \in \mathcal{X}_j] > 0$ for all $j=1,2,\dots,J$. If Assumptions~\ref{assumption: Randomized treatment assignment}--\ref{assumption: finite moments} hold, then Assumption~\ref{assumption: Misspecified outcome models with constrained calibration} is satisfied, and the conclusions of Theorem~\ref{theorem: Asymptotic misspecification} follow.
\end{corollary}

We now turn to the average treatment effect estimator detailed in Section \ref{subsection 3-2:CALM for estimating ATE}. Given the different structure of the calibration weights, we introduce the following assumption. 

\begin{assumption}[Consistent nuisance function estimation]\label{assumption: ATE Consistent nuisance function estimation}
The cross-fitted function estimates are consistent: $\|\hat{\mu}_{t}(\cdot;\mathcal{I})-\mu_{t}(\cdot)\|_{L_2}$, $\|\widehat{\mu}^{\dagger}_{t}(\cdot;\mathcal{I})-\mu^{\dagger}_{t}(\cdot)\|_{L_2}$ and $\|\widehat{\bomega}_{\mathtt{ATE}}(\cdot;\mathcal{I})-\bomega_{\mathtt{ATE}}(\cdot)\|_{L_2} \xrightarrow{\mathrm{p}} 0$ for each data fold $\mathcal{I}$ and treatment arm $t,t'\in\{1,\dots,k\}$.
\end{assumption}

Building on the insights of Section \ref{subsection 3-2:CALM for estimating ATE}, the estimated ATE using CALM is consistent, asymptotically normally distributed, and achieves higher efficiency compared to the standard augmented inverse probability weighting approach. 

\begin{theorem}[Asymptotic properties of $\hat{\tau}_{t,t',\mathtt{CALM}}$]\label{theorem: Asymptotic properties of CALM for ATE}
  Let Assumptions~\ref{assumption: Randomized treatment assignment}--\ref{assumption: finite moments} and \ref{assumption: ATE Consistent nuisance function estimation} hold. In addition, assume the minimum eigenvalue of $\bSigma^\dagger(X_i)$ is uniformly bounded away from 0, and $B\to\infty$. Then 
\begin{align*}
  \sqrt{n}\big(\hat{\tau}_{t,t',\mathtt{CALM}} - \overline{\tau}_{t,t',\mathtt{CALM}}\big)\xrightarrow{\mathrm{p}} 0, \quad \text{which implies}\quad 
  \sqrt{n}\big(\hat{\tau}_{t,t',\mathtt{CALM}} - \tau_{t,t'}\big) \xrightarrow{\mathrm{d}} \mathcal{N}(0,\mathsf{V}_{t,t',\mathtt{CALM}}),
\end{align*}
where $\overline{\tau}_{t,t',\mathtt{CALM}}$ and $\mathsf{V}_{t,t',\mathtt{CALM}}$ are defined in \eqref{eq:CALM ATE asymptotic rep} and \eqref{eq:ATE asym variance}, respectively. 
\end{theorem}

Our final theoretical result establishes an asymptotic efficiency gain of the CALM-based conditional average treatment effect estimator, which we discussed in Section \ref{subsection:CALM for estimating CATE}. To conserve space, additional regularity conditions are postponed to the Supplementary Materials. 
 
\begin{theorem}[Asymptotic properties of $\hat{\tau}_{t,t',\mathtt{CALM}}(x)$]\label{theorem: Asymptotic properties of CALM for CATE}
  Let Assumptions~\ref{assumption: Randomized treatment assignment}--\ref{assumption: finite moments} hold, $B\to\infty$, $nh^p \to \infty$, and $nh^{p+2s} \precsim 1$, where $s\geq 1$ denotes the smoothness of $\mu_t(\cdot)$ and $\mu_{t'}(\cdot)$. Then under the additional regularity conditions laid out in the Supplementary Materials,
\begin{align*}
  \sqrt{nh^p}\big(\hat{\tau}_{t,t',\mathtt{CALM}}(x) - \overline{\tau}_{t,t',\mathtt{CALM}}(x)\big)\xrightarrow{\mathrm{p}} 0,
\end{align*}
where $\overline{\tau}_{t,t',\mathtt{CALM}}(x)$ is defined in \eqref{eq:CALM CATE asymptotic rep}, and therefore
\begin{align*}
  \sqrt{nh^p}  \big(\hat{\tau}_{t,t',\mathtt{CALM}}(x) - \tau_{t,t'}(x)\big)\xrightarrow{\mathrm{d}} \mathcal{N}(0, \mathsf{V}_{t,t',\mathtt{CALM}}(x)),
\end{align*}
and $\mathsf{V}_{t,t',\mathtt{CALM}}(x)$ is defined in \eqref{eq:CATE asym variance}.
\end{theorem}
 
\section{Simulation Studies}\label{section 5:Simulation studies}

We now evaluate the performance of CALM using synthetic data calibrated from the BRIGHTEN study \citep{pratap2022real}. Before detailing our complete simulation design and full results, we highlight four key findings. First, CALM consistently delivers estimates with higher efficiency while maintaining valid statistical inference (Figure~\ref{fig:sim-fig-2-benchmark}). Second, augmented inverse probability weighting may exhibit substantial bias when few-shot predictions are naively incorporated as additional covariates, leading to systematic under-coverage (Figure~\ref{fig:sim-fig-2-benchmark}, panels D--F). Third, surprisingly, we observe efficiency gains even in the absence of unstructured covariates, suggesting that CALM also benefits from intrinsic information encoded in pre-trained language models (Figure~\ref{fig:sim-fig-4-role-of-Z}). Finally, CALM's heterogeneous calibration weighting mechanism selectively leverages counterfactual predictions when they are most informative, leading to substantial variance reduction (Figure~\ref{fig:sim-fig-3-calibration-weight}).

\subsection{Overview of BRIGHTEN Study}\label{subsection:Overview of BRIGHTEN study}

The BRIGHTEN study \citep{pratap2022real} comprises fully remote, smartphone-based randomized controlled trials conducted between 2016 and 2018 in the U.S., with the goal of evaluating the feasibility and effectiveness of delivering mental health care at scale through mobile applications. Recruitment targeted adults ($\geq$18 years old) exhibiting clinically significant depressive symptoms. In total, 2,193 subjects consented and enrolled. They were then randomized to one of the app-delivered interventions and followed for 12 weeks. To match our earlier notation, $T_i\in\{0,1\}$ indicates treatment assignment, to either the intervention group receiving internet-based Problem-Solving Therapy (iPST), or to a control group with alternative interventions. Structured covariates $X_i$ capture demographic and clinical characteristics: sex, education, employment status, marital status, race, age, etc. The unstructured textual covariates $Z_i$ include participants' self-reported reasons for enrollment and their initial satisfaction with the mobile app. The outcome of interest, $Y_i$, is the PHQ-9 score measured after the trial concludes.

\begin{figure}[!ht]
\centering
\includegraphics[width=1\linewidth]{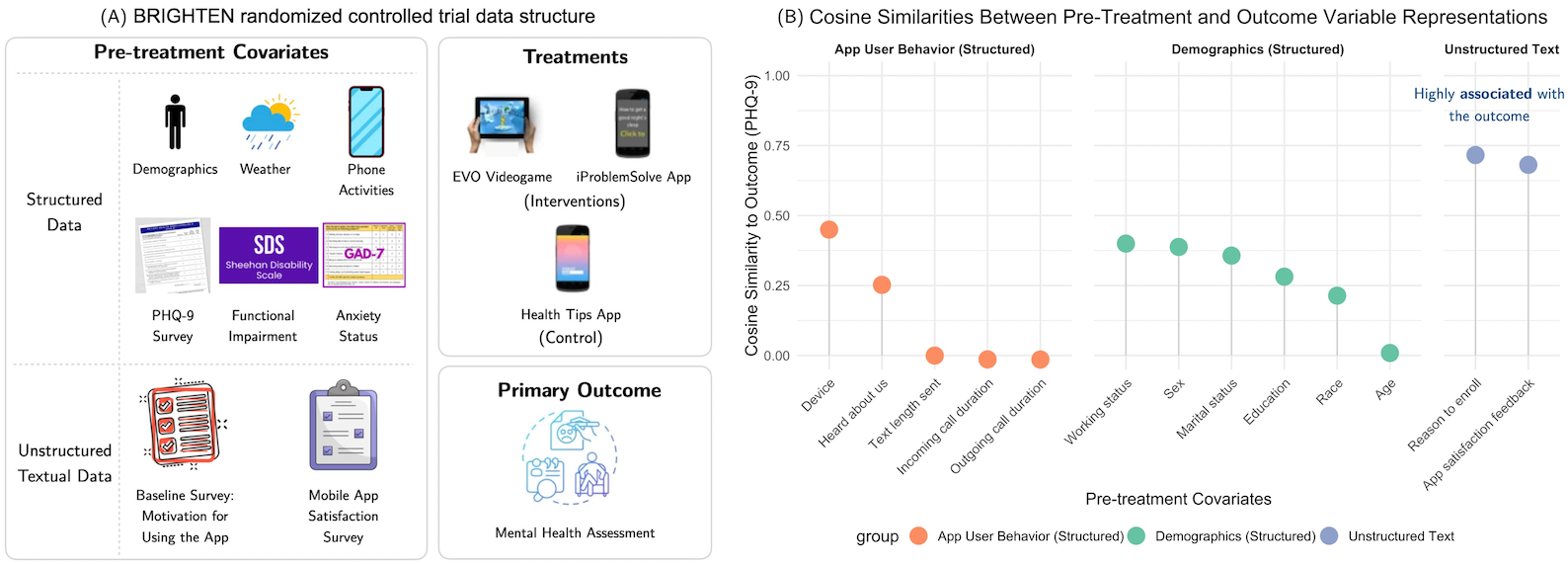}
\caption{(A) BRIGHTEN study data structure. (B) Cosine similarity with the primary outcome. }
\label{fig:BRIGHTEN study}
\end{figure}

The BRIGHTEN study provides a unique data structure that integrates structured covariates with rich, unstructured baseline textual surveys, making it particularly well-suited for evaluating CALM (Figure~\ref{fig:BRIGHTEN study}, Panel A). Furthermore, baseline free-text responses regarding app-use motivation and pre-treatment app satisfaction exhibit the highest cosine similarity with discretized PHQ-9 scores, the primary outcome of the trial, demonstrating the potential of unstructured textual data as valuable prognostic factors (Figure~\ref{fig:BRIGHTEN study}, Panel B).

\subsection{Simulating BRIGHTEN Study with Structured/Unstructured Data}\label{subsection:Simulating BRIGHTEN study with structured/unstructured data}

\begin{figure}[!ht]
\centering
\includegraphics[width=\linewidth]{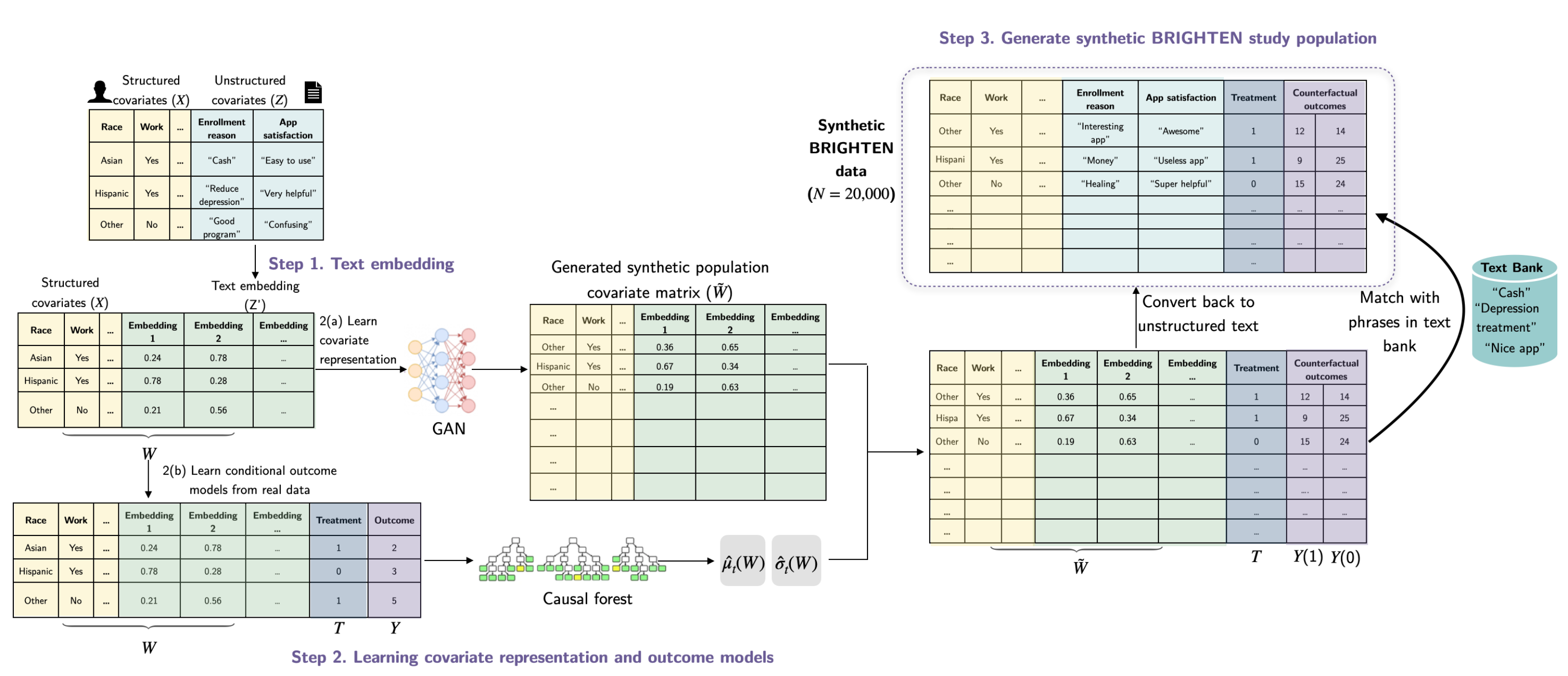}
\captionof{figure}{Illustration of the synthetic BRIGHTEN data generation.}
\label{fig:synthetic-data-generation-illustration}
\end{figure}

In this subsection, we outline our approach to generating a synthetic superpopulation that closely mirrors the BRIGHTEN study (Figure \ref{fig:synthetic-data-generation-illustration}). This endeavor presents three main methodological challenges. To begin with, modeling outcome-covariate relationships is complex when the feature set spans both structured variables (numerical and categorical) and unstructured text (free-text survey responses). Second, preserving dependencies across modalities requires embedding textual responses into continuous vector spaces, a step that introduces high-dimensionality concerns given the limited sample size of the original trial data. Third, as downstream tasks require natural language inputs, reconstructing text from these vector representations demands faithfully capturing the style, diversity, and semantic nuance of the original survey responses. Together, these hurdles make constructing a large-scale yet realistic synthetic population a non-trivial task. 

The initial step involves preprocessing the covariates in the original BRIGHTEN study dataset. Structured numeric covariates are median-imputed and standardized, while categorical covariates are one-hot encoded. Unstructured covariates, such as survey responses, are first normalized so that missing entries are mapped to the string ``No response.'' We then embed each unstructured covariate separately using the pretrained sentence transformer \texttt{all-MiniLM-L6-v2} \citep{wang2020minilm}, applying padding, truncation, and $L_2$-normalization to obtain dense text embeddings. We adopt \texttt{all-MiniLM-L6-v2} because it is a compact, contrastively trained transformer model that produces high-quality sentence embeddings with well-preserved semantic structure. Finally, embeddings from the two unstructured covariate columns are combined with the processed structured covariates to form the final covariate representation, denoted as $W_i = (X_i, Z_i')$, where $Z_i'$ represents the embedding vectors. 

Second, we learn both a generative model for the covariates and a predictive model for the outcomes. Specifically, the joint covariate distribution $p(W)$ is modeled using a conditional Generative Adversarial Network (CTGAN) \citep{xu2019ctgan}. To improve sample efficiency and training stability, each text-embedding block is compressed via principal component analysis (PCA) to 32 components. The structured covariates and PCA-compressed embeddings are then concatenated to form a unified training table, whose schema is validated with \texttt{Metadata} before fitting a \texttt{CTGANSynthesizer} \citep{SDV}. This yields a generator $G$ that models $\hat{p}(W)$, which we use to draw synthetic covariates for downstream simulations.  Next, using the embedded BRIGHTEN dataset $\{W_i, T_i, Y_i\}_{i=1}^{n}$, we fit causal forests to learn the conditional mean and conditional variance, $ m_t(W) = \mathbb{E}[Y|W, T=t]$ and $s_t^2(W) = \Variance[Y|W, T=t]$ for $t \in\{0,1\}$, using the \texttt{R} package \texttt{grf} \citep{tibshirani2024grf}. We denote the fitted models as $\hat{m}_t(W)$ and $\hat{s}_t^2(W)$. Finally, we categorize the outcome variable into five clinically standard levels of depression severity: minimal, mild, moderate, high, and severe, to simplify prediction tasks for LLMs, as they are more effective when predicting discrete, semantically meaningful categories rather than continuous numeric scores. 

The next step is to generate a synthetic super-population of size $N=20,000$. Using the generator learned in the second step, we first form the synthetic covariate matrix, denoted as $ \tilde{W}_j = (\tilde{X}_j, \tilde{Z}_j')$ for $j = 1,\ldots, N$. We then generate the treatment variable as  $ \tilde{T}_j \sim \text{Bernoulli}(0.5)$, $j=1,\ldots,N$, and the potential outcomes as $\tilde{Y}_j(t) = \hat{m}_t(\tilde{W}_j) + \hat{s}_t(\tilde{W}_j) \varepsilon_{j}$, $\varepsilon_j \sim N(0,1)$, $t \in\{0,1\}$. The observed synthetic outcomes are therefore $\tilde{Y}_j = \tilde{T}_j \cdot \tilde{Y}_j(1) + (1-\tilde{T}_j)\cdot \tilde{Y}_j(0)$, $j = 1,\ldots,N$. The true ATE based on the synthetic population is computed as $\tau \equiv \tau_{1,0} = \frac{1}{N} \sum_{j=1}^N \{\tilde{Y}_j(1)- \tilde{Y}_j(0)\}$.

Lastly, because the generated covariates $\tilde{W}_j$ consist of $(\tilde{X}_j, \tilde{Z}_j')$, where $\tilde{Z}_j'$ is represented in the numerical embedding space, we recover $\tilde{Z}_j'$ back into unstructured text. To this end, we first establish a phrase bank for each free-text field (impression of the mobile app, reasons for enrolling) using responses from the original BRIGHTEN dataset after normalizing the raw texts. The surviving unique strings form the seed set $\mathcal{S}_k$ for unstructured covariate column $k$, where $k\in\{1,2\}$. We then paraphrase each seed using pretrained LLMs, Pre-training with Extracted Gap-sentences for Abstractive Summarization (PEGASUS) and Text-to-Text Transfer Transformer (T5), to increase lexical and semantic diversity. PEGASUS is a transformer model specialized for text summarization and paraphrase generation, trained with a gap-sentence generation objective that makes it highly effective for producing semantically faithful variations of input text \citep{zhang2020pegasus}. T5 (Text-to-Text Transfer Transformer) is a general-purpose encoder-decoder transformer that frames all NLP tasks in a unified text-to-text format, enabling effective paraphrase generation among other tasks \citep{raffel2020exploring}. Candidate paraphrases are filtered by enforcing length constraints and requiring cosine similarity in the PCA-compressed embedding space to fall within a semantic band-pass window $[0.65,0.98]$, relative to the original seed. This procedure yields the final phrase bank $\mathcal{B}_k$ for each unstructured covariate column $k\in\{1,2\}$, with their size fixed at 50,000. All phrases in $\mathcal{B}_k$ are then re-embedded using the same encoder applied in the first step and projected through the saved PCA transformation. For each row of $\tilde{Z}_i'$, we perform top-1 nearest neighbor retrieval to identify the closest matching string, thereby recovering natural-language representations $\tilde{Z}_i$ corresponding to the embedding vectors $\tilde{Z}_i'$.

\subsection{Benchmarking CALM with Other Methods}\label{subsection:Benchmarking CALM with other methods}

Using the synthetic BRIGHTEN super population generated in the previous section, we conduct simulation studies using 300 Monte Carlo iterations. Each Monte Carlo sample is randomly drawn from this synthetic population with sample sizes ranging from $n=400$ to 2,000. This subsection focuses on comparing CALM to alternative methods. The next two subsections offer additional insights into CALM's performance.

We consider several variants of CALM: zero-shot learning ($m=0$, as discussed in Section~\ref{subsection 3-1:CALM with Zero-Shot Predictions}) and few-shot learning with $m \in \{6,10,14\}$ demonstrative examples and $B=200$ resampling aggregation (Section~\ref{subsection 2-2:CALM with Few-Shot Learning: Implementation Details}). \texttt{GPT-4o-mini} is employed to generate counterfactual outcome predictions. For further context and robustness, the following case study in Section \ref{section 6:case Study} includes additional results using \texttt{gemini-2.5-flash} and \texttt{GPT-3.5-turbo} as alternative LLMs for generating counterfactual outcomes. An illustrative example of zero- and few-shot learning prompts is provided in the Supplementary Materials.

We compare CALM to several variants of AIPW: (1) \texttt{AIPW}: the standard AIPW estimator using only structured covariates; (2) \texttt{AIPW (zero-shot covariate)}: the AIPW estimator with zero-shot predictions as additional covariates; (3) \texttt{AIPW (few-shot covariate)}: the AIPW estimator with few-shot predictions as additional covariates. For both the CALM and AIPW methods, we use random forests to estimate conditional outcome models. A comparison of performance using random forests versus gradient boosting is provided in Supplementary Materials. We assess performance based on absolute bias, the $\sqrt{n}$-scaled standard deviation (SD) of the ATE estimates, and coverage probability of 95\% confidence intervals. Results are summarized in Figure \ref{fig:sim-fig-2-benchmark}.

\begin{figure}[!ht]
	\centering
	\includegraphics[width=0.9\linewidth]{Figures/sim-fig-2-1-comparison-with-calm.pdf}
	\includegraphics[width=0.9\linewidth]{Figures/sim-fig-2-2-comparison-with-benchmark.pdf}
	\captionof{figure}{Panel (A)--(C): Comparison of CALM with different number of demonstrative examples: $m=0$ (zero-shot) and $m\in\{6,10,14\}$ (few-shot). Panel (D)--(F): Comparison with AIPW-based methods.
	}
	\label{fig:sim-fig-2-benchmark}
\end{figure} 

Panel (A)--(C) of Figure~\ref{fig:sim-fig-2-benchmark} compares zero-shot and few-shot CALM methods. The results show that increasing the number of few-shot examples in the prompt leads to progressively greater efficiency gains, with $m=14$ achieving the lowest standard deviation. We conjecture that this is because few-shot learning provides additional information and guidance to the LLM in generating outcomes, leading to stronger correlation with the observed outcomes and consequently improving efficiency in the estimated treatment effects. The validity of statistical inference of the CALM method is further supported by the coverage probabilities, where CALM with zero-shot or few-shot learning achieves a nearly nominal level of coverage. 

Panels (D)--(F) of Figure~\ref{fig:sim-fig-2-benchmark} show that CALM achieves substantial efficiency gains compared to all evaluated AIPW-based methods. Furthermore, while standard AIPW and zero-shot AIPW yield consistent estimates of the ATE, naively incorporating few-shot predictions as covariates into AIPW introduces bias that scales with the number of prompt examples. This growing bias leads to poor coverage performance. We hypothesize that this bias stems from sample reuse when generating LLM-based counterfactual predictions via few-shot prompting. In contrast, using zero-shot LLM predictions in AIPW avoids this bias and maintains nominal coverage.

\subsection{Does Pre-trained LLM Contain Relevant Information?}
\label{subsubsec:Does pre-trained LLM contain relevant information?}

As outlined in the motivation for our proposed method, leveraging LLMs to generate outcome predictions offers two main advantages: the ability to exploit the knowledge embedded in pre-trained language models, and the seamless integration of unstructured covariates. These considerations naturally raise two questions: How much of the gain can be attributed to the prior knowledge encoded in pre-trained LLMs themselves? And to what extent do unstructured covariates enhance the informativeness of LLM predictions, and consequently improve the efficiency of treatment effect estimation? 

In this section, we consider four variations of the CALM methods with different counterfactual-outcome generation mechanisms: (i) \texttt{CALM(zero-shot,$X+Z$)} that incorporates zero-shot predictions formed with both $X_i$ and $Z_i$; (ii) \texttt{CALM(zero-shot,$X$)} where the zero-shot predictions are generated using only structured information $X_i$; (iii) \texttt{CALM(RF,$X$)} for which counterfactual predictions are from a random forest using only structured $X_i$, instead of a language model; (iv) \texttt{CALM(GB,$X$)}, which is similar to (iii) with the only difference that gradient boosting is employed to construct counterfactual predictions. Monte Carlo standard deviations of the estimates are reported in Panel (A) of Figure \ref{fig:sim-fig-4-role-of-Z}, and we also show the calibration weights for a specific data stratum in Panel (B) of the same figure.

\begin{figure}[!ht]
\centering
\includegraphics[width=0.7\linewidth]{Figures/sim-fig-4-role-of-Z.pdf}
\caption{Panel (A): Monte Carlo standard deviation under different counterfactual outcome prediction mechanisms. \texttt{CALM(zero-shot,$X$)} refers to the CALM zero-shot learning method where only the structured covariates $X_i$ are included in the prompt. \texttt{CALM(zero-shot,$X+Z$)} incorporates both structured $X_i$ and unstructured $Z_i$ information. \texttt{CALM(RF,$X$)} and \texttt{CALM(GB,$X$)} do not employ any language model for counterfactual prediction. They instead use random forest and gradient boosting methods. Panel (B): An example of the calibration weight under different counterfactual outcome prediction mechanisms in the female stratum. }
\label{fig:sim-fig-4-role-of-Z}
\end{figure}

{Figure~\ref{fig:sim-fig-4-role-of-Z} demonstrates that our proposed method with zero-shot learning, whether or not unstructured covariates are included in the prompt, yields higher estimation efficiency compared to using standard machine learning methods alone. Moreover, incorporating unstructured covariates into the zero-shot prompt provides an additional efficiency gain. These results suggest that the pre-trained LLM (\texttt{GPT-4o-mini} in this setting) encodes knowledge that improves the efficiency of ATE estimation. Even without informative unstructured inputs, the LLM's exposure to large and diverse training corpora may enable it to capture latent associations between covariates and outcomes, which CALM can then exploit to enhance efficiency. Incorporating unstructured covariates into the prompt further amplifies this benefit.}

\subsection{How Informative are LLM-based Predictions Through $\omega_t(x)$?}\label{subsubsec:How informative are LLM-based predictions}

Next, we provide some insights on the informativeness of LLM-based predictions through the lens of the calibration weights $\omega_t(x)$. 
Specifically, we examine the relationship between the LLM-predicted outcomes and the observed outcomes across different covariate strata. This analysis is motivated by the inherent structure of the CALM method, which selectively borrows information from distinct regions of the covariate space, depending on where the LLM predictions are most informative. 

To better demonstrate the mechanism underlying the efficiency gain, we 
show the calibration weight $\omega_t(x)$ between the LLM-predicted outcomes and observed outcomes in Figure \ref{fig:sim-fig-3-calibration-weight} (A) across strata defined by gender and race. To quantify how these calibration weights translate into improved efficiency, we report the percentage reduction in estimator variance between the standard AIPW method and the CALM method with zero-shot learning in Figure \ref{fig:sim-fig-3-calibration-weight} (B). 
Specifically, we compute this percentage variance reduction by subtracting the empirical variance under CALM from that under AIPW, dividing the difference by the empirical variance under AIPW, and multiplying by 100. 

Figure \ref{fig:sim-fig-3-calibration-weight}  suggests substantial heterogeneity in the strength of these correlations, across strata and between treatment arms. In some strata, LLM-predicted outcomes exhibit strong alignment with the observed outcomes (e.g., American Indian/Alaskan Native), while in others the alignment is weaker (e.g., Other race). This pattern suggests that the informativeness of the LLM-based predictions is not uniform across the covariate space. This heterogeneity plays a significant role in our proposed CALM method. Rather than uniformly leveraging the LLM predictions, CALM adaptively borrows strength in regions of the covariate space where the predictions are most reliable. To evaluate the practical implications of this selective borrowing, we compare the variance of the CALM estimator to that of the AIPW estimator across strata. Figure \ref{fig:sim-fig-3-calibration-weight} (B) reports the percentage reduction in variance achieved by CALM within each stratum. The results show that the degree of variance reduction achieved by CALM aligns closely with the strength of the calibration weight shown in Figure \ref{fig:sim-fig-3-calibration-weight} (A). In strata where LLM predictions are highly aligned with observed outcomes, CALM yields greater variance reductions. Together, these findings suggest that the calibration weight of selectively borrowing information from LLM-generated predictions contributes to the improved estimation efficiency of our proposed method. 

\begin{figure}[!ht]
\centering
\includegraphics[width=0.9\linewidth]{Figures/sim-fig-3-weight-variance_reduction.pdf}
\caption{Panel (A): Calibration weight $\omega_t(x)$ for the ``race'' and ``gender'' strata using the CALM zero-shot learning method ($n=2,000$). Panel (B): the variance reduction in comparison with the AIPW estimator.} \label{fig:sim-fig-3-calibration-weight}
\end{figure}

\section{Empirical Illustration: The BRIGHTEN Study}\label{section 6:case Study}

In this section, we apply our proposed method to the real BRIGHTEN study dataset. We summarize the takeaways from the case study section as follows: First, the CALM-based methods achieve higher estimation efficiency and are able to detect significant treatment effects in the female Hispanic subgroup compared to the benchmark approaches. Second, the CALM-based methods attain slightly greater efficiency when generating the counterfactual prediction under \texttt{GPT-4o-mini} or \texttt{gemini-2.5-flash}, compared to \texttt{GPT-3.5-turbo}. Third, CALM's performance is robust to variations in prompt design, demonstrating invariance to different prompting strategies. 

We begin by estimating the ATE using CALM and the benchmark methods. The resulting point estimates and 95\% confidence intervals are shown in Figure~\ref{fig:case-study-fig-5-ate}   (A). Consistent with the simulation findings, CALM with few-shot learning yields the narrowest confidence intervals, reflecting higher estimation efficiency. In contrast, AIPW augmented with few-shot learning produces a significantly negative treatment effect estimate, which is likely attributable to the estimation bias documented in Section~\ref{subsection:Benchmarking CALM with other methods}. To assess the impact of different LLMs on CALM's performance, we compare results using three models: \texttt{GPT-4o-mini}, \texttt{GPT-3.5-turbo}, and \texttt{gemini-2.5-flash}, as shown in Figure~\ref{fig:case-study-fig-5-ate}  (B). The figure shows that \texttt{GPT-4o-mini} and \texttt{gemini-2.5-flash} achieve comparable efficiency, whereas \texttt{GPT-3.5-turbo} yields slightly wider confidence intervals, suggesting reduced estimation efficiency due to less informative LLM-based predictions relative to the other two models.

\begin{figure}[!ht]
\centering
\includegraphics[width=0.8\linewidth]{Figures/case-study-fig-5-ate.pdf}
\caption{Panel (A): Comparison of the ATE estimates and the associated 95\% confidence intervals from the BRIGHTEN study under AIPW and CALM-based methods. Panel (B): Comparison across different language models for generating counterfactual outcomes under CALM zero-shot learning method.}
\label{fig:case-study-fig-5-ate}
\end{figure}

Furthermore, we investigate strata-level treatment effects, with results presented in Figure~\ref{fig:case-study-fig-7-subgroup-ate}. The figure shows that CALM with zero-shot or few-shot learning is able to detect a significant treatment effect among female Hispanic participants, which is not identified by the benchmark AIPW methods. This finding highlights how the efficiency gains achieved by CALM translate into improved power for subgroup analyses in the BRIGHTEN study, enabling the detection of significant treatment effect heterogeneity of the mobile app intervention. In this case, the enhanced efficiency also facilitates scientific discovery. 

\begin{figure}[!ht]
\centering
\includegraphics[width=0.6\linewidth]{Figures/case-study-fig-7-subgroup-ate.pdf}
\caption{Comparison of the ATE estimates and the associated 95\% confidence intervals from the BRIGHTEN study under AIPW and CALM-based methods in female Hispanic and male Hispanic strata.}
\label{fig:case-study-fig-7-subgroup-ate}
\end{figure}

\begin{figure}[!ht]
\centering
\begin{minipage}{0.4\linewidth}
\centering
\includegraphics[width=\linewidth]{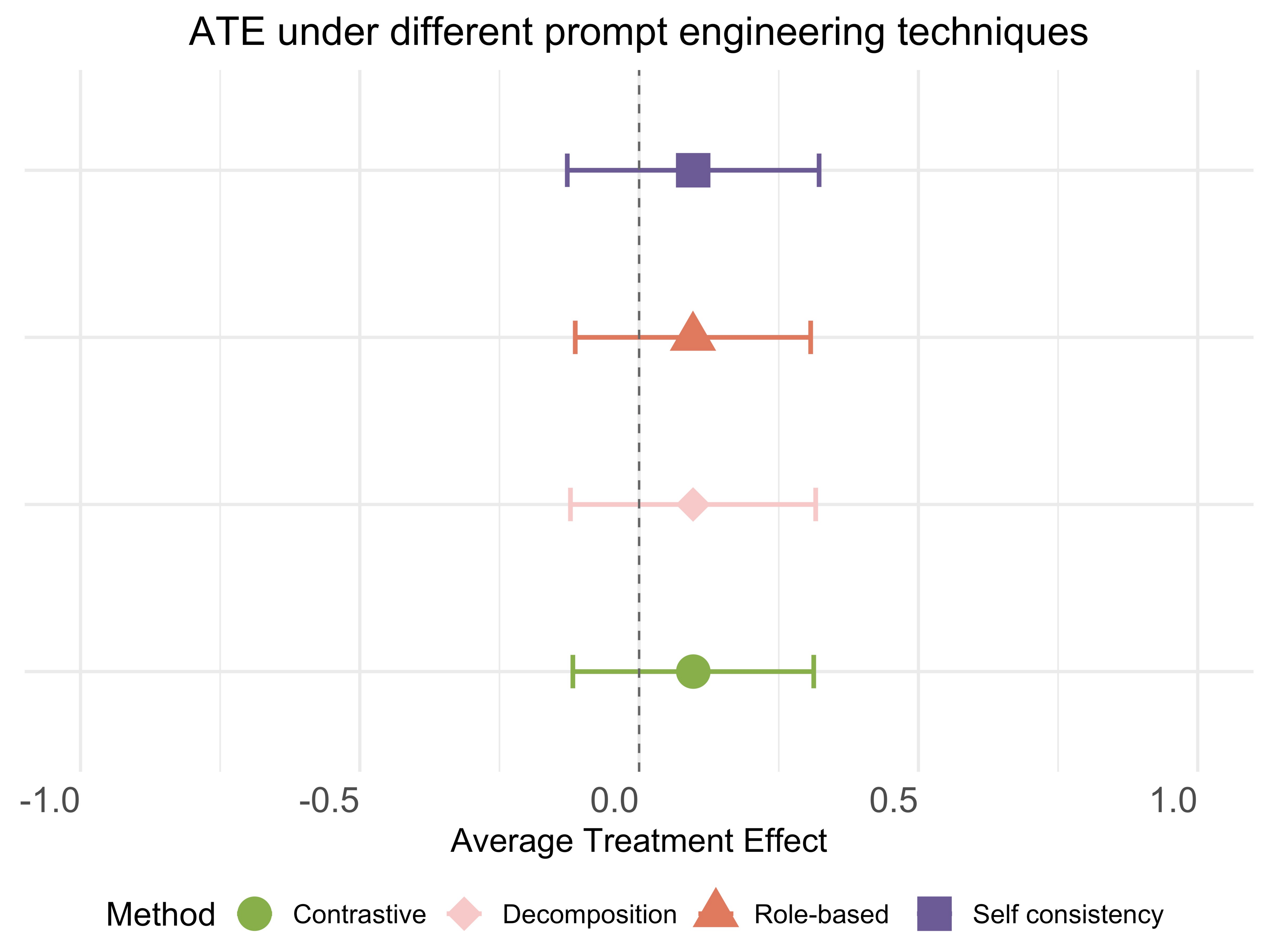}
\end{minipage}%
\hfill
\begin{minipage}{0.6\linewidth}
\centering
\includegraphics[width=\linewidth]{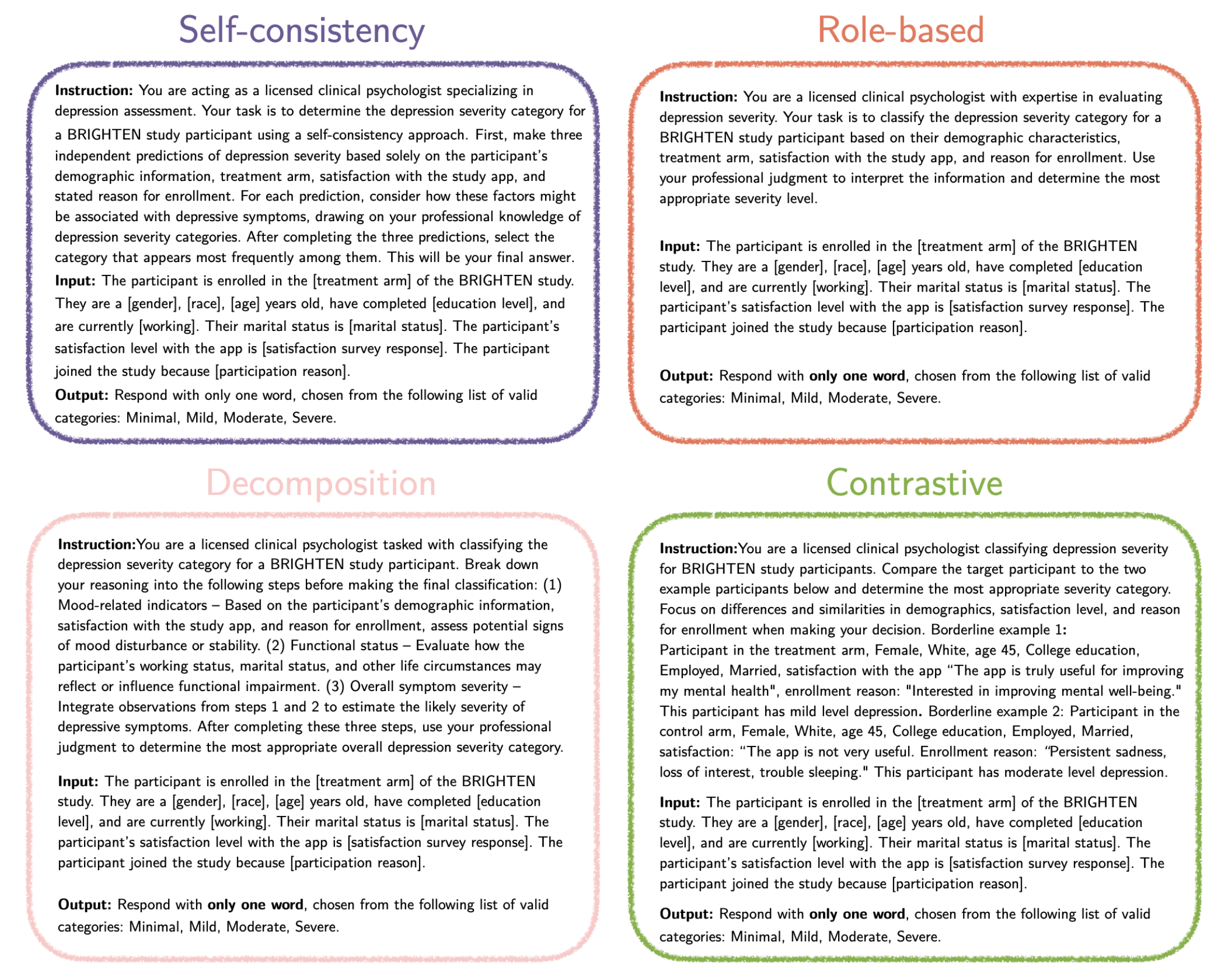}
\end{minipage}
\caption{Comparison of ATE and the associated 95\% confidence intervals under four different prompt engineering techniques: self-consistency, role-based, decomposition, and contrastive methods.}
\label{fig:case-study-fig-6-prompt-designs}
\end{figure}

Additionally, in the preceding simulation studies, we fix a single prompt engineering technique for generating LLM-based predictions. A natural question is whether these predictions are sensitive to prompt design, and whether alternative formulations might alter estimation performance within the CALM framework. To assess this, we conduct a robustness analysis using four prompt engineering strategies: (1) self-consistency, which generates multiple predictions under different sampling seeds and aggregates them to reduce variance; (2) role-based prompting, which assigns the LLM a specific role (e.g., ``You are an experienced clinical researcher'') to provide domain-specific context; (3) decomposition prompting, which breaks the task into sequential sub-questions to capture complex covariate dependencies; and (4) contrastive prompting, which presents systematically varied hypothetical cases to highlight differences in predicted outcomes. Examples of these prompt designs and the ATE estimates with associated 95\% confidence intervals are shown in Figure~\ref{fig:case-study-fig-6-prompt-designs}.

Figure~\ref{fig:case-study-fig-6-prompt-designs} shows that all four prompting strategies yield comparable point estimates and confidence intervals, regardless of the specific prompt design. Among them, role-based prompting produces slightly narrower intervals, indicating a modest efficiency gain, though the difference is small and not statistically significant relative to the other methods. Overall, the results suggest that CALM is robust to variations in prompt engineering, even when the structure and contextual framing differ substantially. This stability under different prompt engineering implies that CALM's performance does not critically depend on prompt format, enhancing its practical applicability in settings where prompt design may vary. 

\section{Conclusion}
In this manuscript, we introduced \textbf{CALM}, a statistically principled framework for leveraging LLM-generated predictions in randomized experiments. By residualizing and heterogeneously calibrating LLM predictions, CALM yields consistent estimators with valid inference even when zero- and few-shot predictions are biased, and it empirically improves efficiency relative to covariate-adjusted AIPW for mean potential outcomes; extensions to average and heterogeneous treatment effects follow directly. In simulations calibrated to the BRIGHTEN RCT, CALM reduces variance and increases power for detecting treatment-effect heterogeneity while remaining robust to prompt design. 

\subsection*{Data availability}
The BRIGHTEN study data are publicly accessible upon request via Synapse: \\
\texttt{https://www.synapse.org/Synapse:syn10848316/wiki/548727} (Synapse Project:syn10848316). 

\bibliography{reference}

@misc{arbour2026ai,
  title={AI-Assisted Variance Reduction in Randomized Experiments},
  author={Arbour, David and Ben-Michael, Eli and Feller, Avi and Lal, Apoorva and Yuan, Lo-Hua},
  howpublished={arXiv:2606.08853},
  year={2026}
}

@article{angelopoulos2023prediction,
 title={Prediction-Powered Inference},
 author={Angelopoulos, Anastasios N and Bates, Stephen and Fannjiang, Clara and Jordan, Michael I and Zrnic, Tijana},
 journal={Science},
 volume={382}, number={6671}, pages={669--674}, year={2023}
}

@article{bryan2025developing,
 title={Developing a Supportive Virtual Human to Deliver Clinical Trial Education for Older Women and Other Populations Historically Excluded From Research},
 author={Bryan, Emma G and Chen, Huan and Vilaro, Melissa and Chu, Haoran and Grillo, Gabriella and Te, Palani and others},
 journal={Patient Education and Counseling},
 volume={130}, number={}, pages={108485}, year={2025}
}

@misc{carlson2025unifying,
  title={A Unifying Framework for Robust and Efficient Inference with Unstructured Data},
  author={Carlson, Jacob and Dell, Melissa},
  howpublished={arXiv:2505.00282},
  year={2026}
}

@article{cattaneo2010efficient,
 title={Efficient Semiparametric Estimation of Multi-Valued Treatment Effects Under Ignorability},
 author={Cattaneo, Matias D},
 journal={Journal of Econometrics},
 volume={155}, number={2}, pages={138--154}, year={2010}
}

@article{cattaneo2019two,
 title={Two-Step Estimation and Inference with Possibly Many Included Covariates},
 author={Cattaneo, Matias D and Jansson, Michael and Ma, Xinwei},
 journal={The Review of Economic Studies},
 volume={86}, number={3}, pages={1095--1122}, year={2019}
}

@article{chen2008semiparametric,
  title={Semiparametric Efficiency in GMM Models with Auxiliary Data},
  author={Chen, Xiaohong and Hong, Han and Tarozzi, Alessandro},
  journal={The Annals of Statistics},
  volume={36}, number={2}, 
  pages={808--843},
  year={2008}
}

@article{chernozhukov2018double,
 title={Double/Debiased Machine Learning for Treatment and Structural Parameters},
 author={Chernozhukov, Victor and Chetverikov, Denis and Demirer, Mert and Duflo, Esther and Hansen, Christian and Newey, Whitney and others},
 journal={The Econometrics Journal},
 volume={21}, number={1}, pages={C1--C68}, year={2018}
}

@article{coravos2019modernizing,
 title={Modernizing and Designing Evaluation Frameworks for Connected Sensor Technologies in Medicine},
 author={Coravos, Andrea and Goldsack, Jennifer C and Karlin, Daniel R and Nebeker, Camille and Perakslis, Eric and Zimmerman, Naomi},
 journal={npj Digital Medicine},
 volume={3}, number={37}, pages={1--10}, year={2020}
}

@article{freedman2008regression,
 title={On Regression Adjustments to Experimental Data},
 author={Freedman, David A},
 journal={Advances in Applied Mathematics},
 volume={40}, number={2}, pages={180--193}, year={2008}
}

@article{guterman2023care,
 title={Care Ecosystem Collaborative Model and Health Care Costs in Medicare Beneficiaries with Dementia: A Secondary Analysis of a Randomized Clinical Trial},
 author={Guterman, Elan L and Kiekhofer, Rachel E and Wood, Andrew J and Allen, I Elaine and Kahn, James G and Dulaney, Sarah and others},
 journal={JAMA Internal Medicine},
 volume={183}, number={11}, pages={1222--1228}, year={2023}
}

@article{hahn1998role,
 title={On the Role of the Propensity Score in Efficient Semiparametric Estimation of Average Treatment Effects},
 author={Hahn, Jinyong},
 journal={Econometrica},
 volume={66}, number={2}, pages={315--331}, year={1998}
}

@article{hanvey2024investigating,
 title={Investigating Inequities in Cancer Clinical Trial Participation: Understanding Demographic and Socioeconomic Determinants of Behavioral Cancer Research Participation},
 author={Hanvey, Grace and Dede, Duane and Krieger, Janice and Ross, Kathryn and Amdur, Robert and Pereira, Deidre},
 journal={Psychosomatic Medicine},
 volume={86}, number={5}, pages={A50}, year={2024}
}

@article{lin2013agnostic,
 title={Agnostic Notes on Regression Adjustments to Experimental Data: Reexamining Freedman's Critique},
 author={Lin, Winston},
 journal={The Annals of Applied Statistics},
 volume={7}, number={1}, pages={295--318}, year={2013}
}

@article{luo2022biogpt,
 title={BioGPT: Generative Pre-Trained Transformer for Biomedical Text Generation and Mining},
 author={Luo, Renqian and Sun, Liai and Xia, Yingce and Qin, Tao and Zhang, Sheng and Poon, Hoifung and others},
 journal={Briefings in Bioinformatics},
 volume={23}, number={6}, pages={bbac409}, year={2022}
}

@article{ma2025testing,
title={Testing Limited Overlap},
author={Ma, Xinwei and Sasaki, Yuya and Wang, Yulong},
journal={Econometric Theory},
volume ={41}, number ={5}, pages ={1129--1162}, year ={2025}
}

@article{ma2020robust,
 title={Robust Inference Using Inverse Probability Weighting},
 author={Ma, Xinwei and Wang, Jingshen},
 journal={Journal of the American Statistical Association},
 volume={115}, number={532}, pages={1851--1860}, year={2020}
}

@article{pratap2022real,
 title={Real-World Behavioral Dataset from Two Fully Remote Smartphone-Based Randomized Clinical Trials for Depression},
 author={Pratap, Abhishek and Homiar, Ava and Waninger, Luke and Herd, Calvin and Suver, Christine and Volponi, Joshua and others},
 journal={Scientific Data},
 volume={9}, number={1}, pages={522}, year={2022}
}

@article{raffel2020exploring,
 title={Exploring the Limits of Transfer Learning with a Unified Text-to-Text Transformer},
 author={Raffel, Colin and Shazeer, Noam and Roberts, Adam and Lee, Katherine and Narang, Sharan and Matena, Michael and others},
 journal={Journal of Machine Learning Research},
 volume={21}, number={140}, pages={1--67}, year={2020}
}

@article{ramos2024assessing,
 title={Assessing Predictive Factors of Attitudes Toward Peer-Supported Mental Health Interventions in the Metaverse: Mixed Methods Study},
 author={Ramos, Francisco Nicolas and Bernstein, Rachel A and Ezawa, Iony D},
 journal={JMIR XR and Spatial Computing (JMXR)},
 volume={1}, number={1}, pages={e57990}, year={2024}
}

@article{zrnic2024cross,
 title={Cross-Prediction-Powered Inference},
 author={Zrnic, Tijana and Cand{\`e}s, Emmanuel J},
 journal={Proceedings of the National Academy of Sciences},
 volume={121}, number={15}, pages={e2322083121}, year={2024}
}

@incollection{alayrac2022flamingo,
  title={Flamingo: A Visual Language Model for Few-Shot Learning},
  author={Alayrac, Jean-Baptiste and Donahue, Jeff and Luc, Pauline and Miech, Antoine and Barr, Iain and Hasson, Yana and others},
  booktitle={Proceedings of the 36th International Conference on Neural Information Processing Systems},
  pages={23716--23730}, year={2022}
}

@misc{angelopoulos2023ppi++,
 title={Ppi++: Efficient Prediction-Powered Inference},
 author={Angelopoulos, Anastasios N and Duchi, John C and Zrnic, Tijana},
 howpublished={arXiv:2311.01453},
 year={2024}
}

@incollection{brown2020language,
  title={Language Models Are Few-Shot Learners},
  author={Brown, Tom and Mann, Benjamin and Ryder, Nick and Subbiah, Melanie and Kaplan, Jared D and Dhariwal, Prafulla and others},
  booktitle={Proceedings of the 34th International Conference on Neural Information Processing Systems},
  pages={1877--1901}, year={2020}
}

@misc{DeBartolomeis2025Efficient,
 title={Efficient Randomized Experiments Using Foundation Models},
 author={De Bartolomeis, Piersilvio and Abad, Javier and Wang, Guanbo and Donhauser, Konstantin and Duch, Raymond M. and Yang, Fanny and others},
 howpublished={arXiv:2502.04262},
 year={2025}
}

@incollection{driess2023palme,
  title  = {PaLM{-}E: An Embodied Multimodal Language Model},
  author = {Driess, Danny and Xia, Fei and Sajjadi, Mehdi~S.~M. and Lynch, Corey and Chowdhery, Aakanksha and Ichter, Brian and others},
  booktitle= {Proceedings of the 40th International Conference on Machine Learning
  },
  pages = {8469--8488}, year   = {2023}
}

@incollection{egami2023using,
  title={Using Imperfect Surrogates for Downstream Inference: Design-Based Supervised Learning for Social Science Applications of Large Language Models},
  author={Egami, Naoki and Hinck, Musashi and Stewart, Brandon and Wei, Hanying},
  booktitle={Proceedings of the 37th International Conference on Neural Information Processing Systems},
  pages={68589--68601}, year={2023}
}

@incollection{fisch2024stratified,
  title={Stratified Prediction-Powered Inference for Effective Hybrid Evaluation of Language Models},
  author={Fisch, Adam and Maynez, Joshua and Hofer, R and Dhingra, Bhuwan and Globerson, Amir and Cohen, William W},
  booktitle={Proceedings of the 38th International Conference on Neural Information Processing Systems},
  pages={111489--111514}, year={2024}
}

@incollection{gao2020making,
  title={Making Pre-Trained Language Models Better Few-Shot Learners},
  author={Gao, Tianyu and Fisch, Adam and Chen, Danqi},
  booktitle={Proceedings of the 59th Annual Meeting of the Association for Computational Linguistics
  and the 11th International Joint Conference on Natural Language Processing},
  pages = {3816--3830}, year={2021}
}

@incollection{li2023blip,
  title={{BLIP-2}: Bootstrapping Language-Image Pre-Training with Frozen Image Encoders and Large Language Models},
  author={Li, Junnan and Li, Dongxu and Savarese, Silvio and Hoi, Steven},
  booktitle={Proceedings of the 40th International Conference on Machine Learning},
  pages={19730--19742}, year={2023}
}

@incollection{liu2023llava,
  title={Visual Instruction Tuning},
  author={Liu, Haotian and Li, Chunyuan and Wu, Qingyang and Lee, Yong Jae},
  booktitle={Proceedings of the 37th International Conference on Neural Information Processing Systems},
  pages={34892--34916}, year={2023}
}

@misc{ludwig2025large,
 title={Large Language Models: An Applied Econometric Framework},
 author={Ludwig, Jens and Mullainathan, Sendhil and Rambachan, Ashesh},
 howpublished={arXiv:2412.07031},
 year={2025}
}

@incollection{min2022rethinking,
  title={Rethinking the Role of Demonstrations: What Makes In-Context Learning Work?},
  author={Min, Sewon and Lyu, Xinxi and Holtzman, Ari and Artetxe, Mikel and Lewis, Mike and Hajishirzi, Hannaneh and Zettlemoyer, Luke},
  booktitle={Proceedings of the 2022 Conference on Empirical Methods in Natural Language Processing},
  pages = {11048--11064}, year={2022}
}

@misc{openai2024gpt4,
 title={GPT-4 Technical Report},
 author={{OpenAI} and others},
 howpublished={arXiv preprint arXiv:2303.08774},
 year={2023}
}

@incollection{SDV,
  title        = {The Synthetic Data Vault},
  author       = {Neha Patki and Roy Wedge and Kalyan Veeramachaneni},
  booktitle    = {IEEE International Conference on Data Science and Advanced Analytics (DSAA)},
  year         = {2016},
  pages        = {399--410}
}

@misc{poulet2025prediction,
 title={Prediction-Powered Inference for Clinical Trials},
 author={Poulet, Pierre-Emmanuel and Tran, Maylis and du Montcel, Sophie Tezenas and Dubois, Bruno and Durrleman, Stanley and Jedynak, Bruno},
 howpublished={medRxiv},
 year={2025}
}

@incollection{radford2021clip,
  title  = {Learning Transferable Visual Models From Natural Language Supervision},
  author = {Radford, Alec and Kim, Jong Wook and Hallacy, Chris and Ramesh, Aditya and Goh, Gabriel and Agarwal, Sandhini and others},
  booktitle= {Proceedings of the 38th International Conference on Machine Learning},
   pages  = {8748--8763}, year   = {2021}
}

@misc{schick2020exploiting,
 title={Exploiting Cloze Questions for Few Shot Text Classification and Natural Language Inference},
 author={Schick, Timo and Sch{\"u}tze, Hinrich},
 howpublished={arXiv:2001.07676},
 year={2021}
}

@misc{taylor2022galactica,
 title={Galactica: A Large Language Model for Science},
 author={Taylor, Ross and Kardas, Marcin and Cucurull, Guillem and Scialom, Thomas and Hartshorn, Anthony and Saravia, Elvis and others},
 howpublished={arXiv:2211.09085},
 year={2022}
}

@howpublished{tibshirani2024grf,
  title={{{GRF}: Generalized Random Forests}},
  author={Tibshirani, Julie and Athey, Susan and Wager, Stefan and Friedberg, Rina and Hadad, Vitor and Miner, Luke and others},
  year={2024},
  howpublished={R package version 2.3.2}
}

@misc{touvron2023llama,
 title={Llama: Open and Efficient Foundation Language Models},
 author={Touvron, Hugo and Lavril, Thibaut and Izacard, Gautier and Martinet, Xavier and Lachaux, Marie-Anne and Lacroix, Timoth{\'e}e and others},
 howpublished={arXiv:2302.13971},
 year={2023}
}

@incollection{wang2020minilm,
  title={Minilm: Deep self-attention distillation for task-agnostic compression of pre-trained transformers},
  author={Wang, Wenhui and Wei, Furu and Dong, Li and Bao, Hangbo and Yang, Nan and Zhou, Ming},
  booktitle={Proceedings of the 34th International Conference on Neural Information Processing Systems},
  pages={5776--5788}, year={2020}
}

@incollection{xu2019ctgan,
  title={Modeling Tabular Data Using Conditional {GAN}},
  author={Xu, Lei and Skoularidou, Maria and Cuesta-Infante, Alfredo and Veeramachaneni, Kalyan},
  booktitle={Proceedings of the 33rd International Conference on Neural Information Processing Systems},
  pages={7335--7345}, year={2019}
}

@incollection{zhang2020pegasus,
  title={Pegasus: Pre-Training with Extracted Gap-Sentences for Abstractive Summarization},
  author={Zhang, Jingqing and Zhao, Yao and Saleh, Mohammad and Liu, Peter},
  booktitle={Proceedings of the 37th International Conference on Machine Learning},
  pages={11328--11339}, year={2020}
}

@misc{zhang2025agentic,
 title={Agentic Economic Modeling},
 author={Zhang, Bohan and Li, Jiaxuan and Horta{\c{c}}su, Ali and Ye, Xiaoyang and Chernozhukov, Victor and Ni, Angelo and others},
 howpublished={arXiv:2510.25743},
 year={2025}
}

@inproceedings{zhao2021calibrate,
  title={Calibrate Before Use: Improving Few-Shot Performance of Language Models},
  author={Zhao, Zihao and Wallace, Eric and Feng, Shi and Klein, Dan and Singh, Sameer},
  booktitle={Proceedings of the 38th International Conference on Machine Learning},
  pages={12697--12706},
  year={2021},
  volume={139}
}

@article{cattaneo2018kernel,
 title={Kernel-Based Semiparametric Estimators: Small Bandwidth Asymptotics and Bootstrap Consistency},
 author={Cattaneo, Matias D and Jansson, Michael},
 journal={Econometrica},
 volume={86}, number={3}, pages={955--995}, year={2018}
}

@article{hirano2003efficient,
 title={Efficient Estimation of Average Treatment Effects Using the Estimated Propensity Score},
 author={Hirano, Keisuke and Imbens, Guido W and Ridder, Geert},
 journal={Econometrica},
 volume={71}, number={4}, pages={1161--1189}, year={2003}
}

@article{robinson1988root,
 title={Root-N-Consistent Semiparametric Regression},
 author={Robinson, P.M.},
 journal={Econometrica},
 volume={56}, number={4}, pages={931--954}, year={1988}
}

@misc{ruan2026hero,
  title={HERO: Improving the Reliability and Sensitivity of Generative Model Evaluation Using Historical Data},
  author={Ruan, Xinrui and Zhao, Zhenyu and Wei, Waverly and Zhang, Yueshan and Zheng, Zeyu and Huang, Sui and Wang, Jingshen},
  howpublished={arXiv:2606.29784},
  year={2026}
}
\bibliographystyle{jasa}

\end{document}


\title{\Large Supplementary Materials for ``Can Language Models Boost the Power of Randomized Experiments Without Statistical Bias?'' }

\author{Xinrui Ruan$^1$ \and Xinwei Ma$^2$ \and Yingfei Wang$^3$ \and Waverly Wei$^4$ \and Jingshen Wang$^1$}

\date{
	$^1${\small Division of Biostatistics, University of California, Berkeley}\\
	$^2${\small Department of Economics, University of California, San Diego}\\
	$^3${\small Michael G. Foster School of Business, University of Washington}\\
	$^4${\small Department of Data Sciences and Operations, University of Southern California}\\
	\ \\
	{\small October 7, 2025 \\ Revised: \today}
}

\maketitle
\tableofcontents

\clearpage

\section{Proof of Results in the Main Paper}

We provide proofs of the results established in the main paper. Throughout the Supplementary Materials, limits are taken with respect to the sample size tending to infinity (i.e., $n\to \infty$). Convergence in probability and in distribution are denoted by $\toProbability$ and $\toDist$, respectively. For two possibly random sequences $a_n$ and $b_n$ with $b_n\geq 0$, $a_n = O(b_n)$ or $a_n \precsim b_n$ indicates that $|a_n|/b_n$ is bounded. Analogously, we employ $a_n = \Op(b_n)$ or $a_n \precsimp b_n$ if $|a_n|/b_n$ is bounded in probability. Asymptotic equivalence between two nonrandom sequences is $a_n \asymp b_n$. Finally, $a_n = \op(1)$ if $a_n \toProbability 0$. 

\subsection{Technical Lemmas}

We first establish the asymptotic properties for the oracle estimator. Recall that $Y^\dagger_i(t)$ denotes the expected few-shot prediction obtained by averaging over the randomness of the demonstrative examples, and its conditional mean is
\begin{align*}
\mu_{t}^{\dagger}(x) &= \Expectation[Y^\dagger_i(t)|X_i=x].
\end{align*}
Also recall that
\begin{align*} 
\omega_{t}(x) &= \frac{\cov[Y_i(t),Y^\dagger_i(t)|X_i=x]}{\Variance[Y^\dagger_i(t)|X_i=x]}.
\end{align*}
The oracle CALM estimator for the parameter $\mu_t$ is defined as
\begin{align*}
\overline{\mu}_{t,\mathtt{CALM}} &= \frac{1}{n}\sum_{i=1}^n \varphi_{t}(i),
\end{align*}
where
\begin{align*}
\varphi_{t}(i) &= \frac{\Indicator(T_i=t)}{e_t(X_i)}\Big(Y_i - \mu_{t}(X_i)\Big) + \mu_{t}(X_i) 
+ \Big(1-\frac{\Indicator(T_i=t)}{e_t(X_i)}\Big)\Big(Y^{\dagger}_{i}(t) - \mu^{\dagger}_{t}(X_i)\Big)\omega_{t}(X_i).
\end{align*}
To streamline the presentation, let 
\begin{align*}
  \nu_{i,t} = 1-\frac{\Indicator(T_i=t)}{e_t(X_i)},
\end{align*} 
which satisfies
\begin{align*}
  \Expectation[\nu_{i,t}|X_i] = \Expectation[\nu_{i,t}|X_i,Z_i] = 0,\quad \text{and}\quad \Variance[\nu_{i,t}|X_i] = \Variance[\nu_{i,t}|X_i,Z_i] = \frac{1 - e_t(X_i)}{e_t(X_i)} = \lambda_t(X_i).
\end{align*}

\begin{lemma}[Asymptotic properties of the oracle CALM estimator]\label{lem:oracle-fs}
Under Assumptions~1--3, the oracle estimator $\overline{\mu}_{t,\mathtt{CALM}}$ is consistent and asymptotically normally distributed. Specifically, we have $\overline{\mu}_{t,\mathtt{CALM}} \toProbability \mu_t$, and $\sqrt n \left(\overline{\mu}_{t,\mathtt{CALM}}-\mu_t\right) \xrightarrow{\mathrm{d}} \mathcal N\left(0, \mathsf V_{t,\mathtt{CALM}}\right)$, where
\begin{align*}
  &\mathsf{V}_{t,\mathtt{AIPW}} =  \Expectation\Big[ \frac{\Variance[Y_i(t)|X_i]}{e_t(X_i)} + (\mu_t(X_i) - \mu_t)^2 \Big],
  \quad\mathsf{V}_{t,\mathtt{CALM}}  = \mathsf{V}_{t,\mathtt{AIPW}} -  \Expectation\Big[ \frac{\Variance[Y_i(t)|X_i]}{e_t(X_i)}(1-e_t(X_i))\rho^2_t(X_i) \Big],
\end{align*}
and
\[
\rho_{t}(x) = \mathbb{C}\mathrm{orr}[Y_i(t),Y_i^\dagger(t)|X_i=x].
\] 
\end{lemma}

\begin{proof}
We prove the result in two steps: (i) consistency and asymptotic normality, and (ii) asymptotic variance computation.

\medskip
\noindent\textit{Consistency and asymptotic normality.}
For any integrable function $h(X_i,Z_i)$, we have
\begin{align*}
\Expectation\left[\nu_{i,t}\cdot h(X_i,Z_i)\right]
&= \Expectation\left[\Expectation\left[\left.\nu_{i,t}\cdot h(X_i,Z_i)\right|X_i,Z_i\right]\right] \\
&= \Expectation\left[h(X_i,Z_i)\left(1-\frac{\Expectation[\Indicator(T_i=t)|X_i,Z_i]}{e_t(X_i)}\right)\right] 
=0,
\end{align*}
where the last equality uses Assumption~1.
Since $Y_i^\dagger(t)$ is a function of $(X_i,Z_i)$, the term
\[
\mu_t(X_i)+\omega_{t}(X_i)\{Y_i^\dagger(t)-\mu_{t}^{\dagger}(X_i)\}
\]
is also a function of $(X_i,Z_i)$. The above argument then implies
\begin{align*}
\Expectation[\varphi_{t}(i)] &= \Expectation\left[\frac{\Indicator(T_i=t)Y_i}{e_t(X_i)}\right] = \Expectation[Y_i(t)] = \mu_t.
\end{align*}
Thus, $\varphi_{t}(i)$ is unbiased for $\mu_t$. By Assumption~2 and 3, the sequence
$\{\varphi_{t}(i)\}_{i=1}^n$ is i.i.d. with finite and nonvanishing variance. Therefore, a standard application of the CLT implies
\[
\overline{\mu}_{t,\mathtt{CALM}} \toProbability \mu_t, \quad \sqrt n \left(\overline{\mu}_{t,\mathtt{CALM}}-\mu_t\right) \xrightarrow{\mathrm{d}} \mathcal N\left(0, \Variance[\varphi_{t}(i)]\right).
\]

\medskip
\noindent\textit{Asymptotic variance computation.}
We now compute $\Variance[\varphi_{t}(i)]$. By definition,
\[
\begin{aligned}
\Variance[\varphi_{t}(i)] &= \Expectation\left[\left\{\varphi_{t}(i)-\mu_t\right\}^2\right] \\
&= \Expectation\Bigg[\bigg\{\mu_t(X_i)-\mu_t + \frac{\Indicator(T_i=t)}{e_t(X_i)}\{Y_i-\mu_t(X_i)\} + \nu_{i,t}\omega_{t}(X_i)\{Y_i^\dagger(t)-\mu_{t}^{\dagger}(X_i)\}\bigg\}^2\Bigg].
\end{aligned}
\]
Let
\[
\Delta_{t}(i) := \frac{\Indicator(T_i=t)}{e_t(X_i)}\{Y_i-\mu_t(X_i)\} + \nu_{i,t}\omega_{t}(X_i)\{Y_i^\dagger(t)-\mu_{t}^{\dagger}(X_i)\}.
\]
Then
\[
\Variance[\varphi_{t}(i)] = \Expectation\left[\{\mu_t(X_i)-\mu_t\}^2\right] + \underbrace{\Expectation[\Delta_{t}(i)^2]}_{\text{(I)}} + 2\underbrace{\Expectation\left[\{\mu_t(X_i)-\mu_t\}\Delta_{t}(i)\right]}_{\text{(II)}}.
\]
Since $\Expectation[\Delta_{t}(i)|X_i]=0$, we have
\[
\text{(II)} = \Expectation\left[\{\mu_t(X_i)-\mu_t\}\Expectation[\Delta_{t}(i)|X_i]\right] = 0.
\]
A direct variance-covariance expansion gives
\begin{align*}
\text{(I)} &= \Expectation\Big[\frac{1}{e_t(X_i)}\Variance[Y_i(t)|X_i] +\frac{1-e_t(X_i)}{e_t(X_i)}\left\{\omega_{t}(X_i)^2\Variance[Y_i^\dagger(t)|X_i] - 2\omega_{t}(X_i)\cov[Y_i(t),Y_i^\dagger(t)|X_i]\right\}\Big].
\end{align*}
Therefore,
\[
\begin{aligned}
\Variance[\varphi_{t}(i)] &= \Expectation\left[\{\mu_t(X_i)-\mu_t\}^2 + \Variance[Y_i(t)|X_i] + \frac{1-e_t(X_i)}{e_t(X_i)}\Variance[Y_i(t)-\omega_{t}(X_i)Y_i^\dagger(t)|X_i]\right].
\end{aligned}
\]
By the definition of $\omega_{t}(x)$,
\[
\Variance[Y_i(t)-\omega_{t}(X_i)Y_i^\dagger(t)|X_i] = \Variance[Y_i(t)|X_i]\{1-\rho_{t}^2(X_i)\}.
\]
This completes the proof.
\end{proof}

For the next lemma, we write $Y_i^\dagger(t) =  Y^\dagger(X_i,Z_i,t)$ to emphasize its dependence on the unit's pre-intervention covariates. 

\begin{lemma}[Convergence of resampling-based few-shot predictions]
\label{lem:fs-resampling-convergence}
Under Assumptions 2--3,
\begin{align*}
  \Expectation_{X_i,Z_i}\big|Y_{i}^\dagger(t;\mathcal I) - Y_{i}^\dagger(t)\big|^2 \precsimp \frac{1}{B} + \frac{1}{n},
\end{align*}
where the expectation is taken with respect to the marginal distribution of $(X_i,Z_i)$.
\end{lemma}

\begin{proof}
Throughout this proof, we assume that $(X_i,Z_i)$ is statistically independent of the demonstrative examples. Recall that our resampling and aggregation step gives
\[
Y_{i}^\dagger(t;\mathcal I) = \frac{1}{B}\sum_{b=1}^B Y^\dagger\big(X_i,Z_i,t,\mathcal{S}_b^*\big).
\]
Given an unordered sample $\mathcal{I}$, we denote its size by $n'$. Note that $n'\asymp n$ by our sample splitting procedure. We will also define the symmetrized version: for an unordered sample $\mathcal{S}$ of size $m$, 
\[
Y^\dagger_{\mathtt{sym}}\big(X_i,Z_i,t,\mathcal{S}\big) = \frac{1}{m!}\sum_b Y^\dagger\big(X_i,Z_i,t,\mathcal{S}_b\big),
\]
where the above takes summation over all ordered versions of $\mathcal{S}$: $\mathcal{S}_b$ for $b=1,2,\dots,m!$. Now consider the auxiliary quantities
\[
\overline{Y}_{i}^\dagger(t;\mathcal I) = \frac{1}{\binom{n'}{m}}\sum_b Y^\dagger_{\mathtt{sym}}\big(X_i,Z_i,t,\mathcal{S}_b\big),
\]
where $\mathcal{S}_b$ in the above enumerates over all unordered subsamples of size $m$, and
\[
\tilde{Y}_{i}^\dagger(t;\mathcal I) = \frac{1}{\lfloor n'/m \rfloor}\sum_b Y^\dagger_{\mathtt{sym}}\big(X_i,Z_i,t,\mathcal{S}_b\big),
\]
where the summation takes over some nonoverlapping subsets of $m$. (This can be done by arbitrarily ordering observations in $\mathcal{I}$ into $O_1,O_2,\dots,O_{n'}$, and let $\mathcal{S}_1 = \{O_1,\dots,O_m\}$, and $\mathcal{S}_2 = \{ O_{m+1},\dots,O_{2m} \}$, $\dots$) The notation $\lfloor n'/m \rfloor$ represents the floor function, which is the largest integer below $n'/m$. Given the unordered sample $\mathcal{I}$, one immediate observation is that 
\[
\Expectation[Y_{i}^\dagger(t;\mathcal I)|X_i,Z_i,\mathcal{I}] 
= \Expectation[\tilde{Y}_{i}^\dagger(t;\mathcal I)|X_i,Z_i,\mathcal{I}] 
= \overline{Y}_{i}^\dagger(t;\mathcal I).
\]
Given the above observation, we immediately have 
\begin{align*}
  \Variance[\overline{Y}_{i}^\dagger(t;\mathcal I)|X_i,Z_i] \leq \Variance[\tilde{Y}_{i}^\dagger(t;\mathcal I)|X_i,Z_i] = \frac{m}{n'}\Variance[Y^\dagger_{\mathtt{sym}}\big(X_i,Z_i,t,\mathcal{S}\big) | X_i,Z_i] \precsim \frac{1}{n},
\end{align*}
where the final bound in the above is due to our Assumption 3. 

Now consider $Y_{i}^\dagger(t;\mathcal I)$. Conditioning on the unordered sample $\mathcal{I}$ and $(X_i,Z_i)$, its variance is 
\begin{align*}
  \Variance[Y_{i}^\dagger(t;\mathcal I)|X_i,Z_i,\mathcal{I}] = \frac{1}{B}\frac{1}{\binom{n'}{m}}\frac{1}{m!}\sum_b \Big( Y^\dagger\big(X_i,Z_i,t,\mathcal{S}_b\big) - \overline{Y}_{i}^\dagger(t;\mathcal I)\Big)^2,
\end{align*}
and the summation above is taken over all ordered subsets $\mathcal{S}_b$ of size $m$. Again, by Assumption 3, we have 
\begin{align*}
  \Variance[Y_{i}^\dagger(t;\mathcal I)|X_i,Z_i,\mathcal I] \precsim \frac{1}{B}.
\end{align*}

Combining the previous results with triangle inequality, 
\begin{align*}
  \big|Y_{i}^\dagger(t;\mathcal I) - Y_{i}^\dagger(t)\big|^2 
  \leq 2\big|Y_{i}^\dagger(t;\mathcal I) - \overline{Y}_{i}^\dagger(t;\mathcal I)\big|^2 
  + 2\big|\overline{Y}_{i}^\dagger(t;\mathcal I) - Y_{i}^\dagger(t)\big|^2 \precsimp \frac{1}{B} + \frac{1}{n}, 
\end{align*}
Finally, dominated convergence implies 
\begin{align*}
  \Expectation_{X_i,Z_i}\big|Y_{i}^\dagger(t;\mathcal I) - Y_{i}^\dagger(t)\big|^2 
   \precsimp \frac{1}{B} + \frac{1}{n}, 
\end{align*}
where $\Expectation_{X_i,Z_i}[\cdot]$ denotes marginal expectation with respect to $X_i$ and $Z_i$. 
\end{proof}

\begin{lemma}[Asymptotic equivalence]
\label{lem:fs-equivalence-oracle}
Under Assumptions~1--4 and $B\to \infty$,
\[
\sqrt n \left(\hat{\mu}_{t,\mathtt{CALM}} - \overline{\mu}_{t,\mathtt{CALM}}\right) \toProbability 0.
\]
\end{lemma}

\begin{proof}
With slight abuse of notation, we define
\[
\hat{\mu}_{t,\mathtt{CALM}}^{\mathcal I_3} = \frac{1}{|\mathcal{I}_3|}\sum_{i\in\mathcal{I}_3} \hat{\varphi}_{t}(i;\mathcal{I}_{1},\mathcal{I}_{2}),\qquad \overline{\mu}_{t,\mathtt{CALM}}^{\mathcal I_3} = \frac{1}{|\mathcal{I}_3|}\sum_{i\in\mathcal{I}_3} \varphi_{t}(i),
\]
and $|\mathcal{I}_3|$ is the sample size of the fold $\mathcal{I}_3$. Note that $|\mathcal{I}_3| \asymp n$. It will suffice to show that
\[
\sqrt{n}\left(\hat{\mu}_{t,\mathtt{CALM}}^{\mathcal I_3} - \overline{\mu}_{t,\mathtt{CALM}}^{\mathcal I_3}\right) \toProbability 0,
\]
and the desired conclusion will follow from the cross-fitting step. We recall the difference between these quantities is that the oracle estimator uses $Y_i^\dagger(t)$ and the true nuisance functions
$\mu_t(\cdot)$, $\mu^\dagger_{t}(\cdot)$, and $\omega_{t}(\cdot)$, whereas the feasible estimator uses the aggregated few-shot prediction
$Y_{i}^\dagger(t;\mathcal I_1)$ and the cross-fitted estimates
$\hat\mu_t(\cdot;\mathcal I_2)$, $\hat\mu_{t}^{\dagger}(\cdot;\mathcal I_2)$, and
$\hat\omega_{t}(\cdot;\mathcal I_2)$.

We first decompose the difference as
\begin{align*}
\tag{I}\sqrt{n}\left(\hat{\mu}_{t,\mathtt{CALM}}^{\mathcal I_3} - \overline{\mu}_{t,\mathtt{CALM}}^{\mathcal I_3}\right) 
&= \frac{\sqrt n}{|\mathcal I_3|}\sum_{i\in\mathcal I_3} \nu_{i,t} \cdot\Big(\hat\mu_t(X_i;\mathcal I_2)-\mu_t(X_i)\Big) \\
\tag{II}&\quad - \frac{\sqrt n}{|\mathcal I_3|}\sum_{i\in\mathcal I_3} \nu_{i,t}\cdot\hat{\omega}_{t}(X_i;\mathcal{I}_2)\Big(\hat\mu_{t}^{\dagger}(X_i;\mathcal I_2)-\mu_{t}^\dagger(X_i)\Big) \\
\tag{III}&\quad - \frac{\sqrt n}{|\mathcal I_3|}\sum_{i\in\mathcal I_3} \nu_{i,t}\cdot\mu_{t}^\dagger(X_i)\Big(\hat\omega_{t}(X_i;\mathcal I_2)-\omega_{t}(X_i)\Big) \\
\tag{IV}&\quad + \frac{\sqrt n}{|\mathcal I_3|}\sum_{i\in\mathcal I_3} \nu_{i,t}\cdot Y_i^\dagger(t)\Big(\hat\omega_{t}(X_i;\mathcal I_2)-\omega_{t}(X_i)\Big) \\
\tag{V}&\quad + \frac{\sqrt n}{|\mathcal I_3|}\sum_{i\in\mathcal I_3} \nu_{i,t}\cdot\hat\omega_{t}(X_i;\mathcal I_2)\Big(Y_{i}^\dagger(t;\mathcal I_1) - Y_i^\dagger(t)\Big),
\end{align*}
where we recall that $\nu_{i,t} = 1 - {\Indicator(T_i=t)}/{e_t(X_i)}$.

We now bound these terms. For the first term, conditional on $\mathcal I_1$, $\mathcal I_2$ and the observed covariates $\{X_i,Z_i:i\in\mathcal I_3\}$, its variance is
\begin{align*}
\Variance\big[\text{(I)} \big| \mathcal I_1, \mathcal I_2,\{X_i,Z_i\}_{i\in\mathcal I_3}\big] 
&= \frac{n}{|\mathcal I_3|^2}\sum_{i\in\mathcal I_3} \frac{1-e_t(X_i)}{e_t(X_i)}\big(\hat\mu_t(X_i;\mathcal I_2)-\mu_t(X_i)\big)^2 \\
&\precsim  \frac{1}{|\mathcal I_3|}\sum_{i\in\mathcal I_3} \big(\hat\mu_t(X_i;\mathcal I_2)-\mu_t(X_i)\big)^2 \toProbability 0
\end{align*}
by our Assumption 4. Similar arguments apply to (II)--(IV):
\begin{align*}
\Variance\big[\text{(II)} \big| \mathcal I_1, \mathcal I_2,\{X_i,Z_i\}_{i\in\mathcal I_3}\big] 
&= \frac{n}{|\mathcal I_3|^2}\sum_{i\in\mathcal I_3} \frac{1-e_t(X_i)}{e_t(X_i)}\hat{\omega}_{t}(X_i;\mathcal{I}_2)^2\big(\hat\mu_t^\dagger(X_i;\mathcal I_2)-\mu_t^\dagger(X_i)\big)^2 \\
&\precsim  \frac{1}{|\mathcal I_3|}\sum_{i\in\mathcal I_3} \big(\hat\mu_t^\dagger(X_i;\mathcal I_2)-\mu_t^\dagger(X_i)\big)^2 \toProbability 0,
\end{align*}
where we have used the assumption that the estimated calibration weight $\hat{\omega}_{t}(X_i;\mathcal{I}_2)$ is bounded. Next for (III),
\begin{align*}
\Variance\big[\text{(III)} \big| \mathcal I_1, \mathcal I_2,\{X_i,Z_i\}_{i\in\mathcal I_3}\big] 
&= \frac{n}{|\mathcal I_3|^2}\sum_{i\in\mathcal I_3} \frac{1-e_t(X_i)}{e_t(X_i)} \mu_{t}^\dagger(X_i)^2\big(\hat\omega_{t}(X_i;\mathcal I_2)-\omega_{t}(X_i)\big)^2 \\
&\precsim  \frac{1}{|\mathcal I_3|}\sum_{i\in\mathcal I_3} \big(\hat\omega_{t}(X_i;\mathcal I_2)-\omega_{t}(X_i)\big)^2 \toProbability 0.
\end{align*}
Finally for (IV),
\begin{align*}
\Variance\big[\text{(IV)} \big| \mathcal I_1, \mathcal I_2,\{X_i,Z_i\}_{i\in\mathcal I_3}\big] 
&= \frac{n}{|\mathcal I_3|^2}\sum_{i\in\mathcal I_3} \frac{1-e_t(X_i)}{e_t(X_i)} Y_i^\dagger(t)^2\big(\hat\omega_{t}(X_i;\mathcal I_2)-\omega_{t}(X_i)\big)^2 \\
&\precsim  \frac{1}{|\mathcal I_3|}\sum_{i\in\mathcal I_3} \big(\hat\omega_{t}(X_i;\mathcal I_2)-\omega_{t}(X_i)\big)^2 \toProbability 0.
\end{align*}
It remains to bound the few-shot aggregation error in term (V). By the same conditioning argument, 
\begin{align*}
\Variance\big[\text{(V)} \big| \mathcal I_1, \mathcal I_2,\{X_i,Z_i\}_{i\in\mathcal I_3}\big] 
&= \frac{n}{|\mathcal I_3|^2}\sum_{i\in\mathcal I_3} \frac{1-e_t(X_i)}{e_t(X_i)} \hat\omega_{t}(X_i;\mathcal I_2)^2\big(Y_{i}^\dagger(t;\mathcal I_1) - Y_i^\dagger(t)\big)^2 \\
&\precsim  \frac{1}{|\mathcal I_3|}\sum_{i\in\mathcal I_3} \big(Y_{i}^\dagger(t;\mathcal I_1) - Y_i^\dagger(t)\big)^2 \\
&\precsimp \Expectation_{X_i,Z_i}\big(Y_{i}^\dagger(t;\mathcal I_1) - Y_i^\dagger(t)\big)^2 \precsimp \frac{1}{B} + \frac{1}{n}.
\end{align*}
The last line is due to Markov inequality applied to nonnegative random variables and our Lemma \ref{lem:fs-resampling-convergence}.
\end{proof}

\subsection{Proof of Theorem~1}
This theorem follows from Lemmas \ref{lem:oracle-fs}, \ref{lem:fs-resampling-convergence}, and \ref{lem:fs-equivalence-oracle}.

\subsection{Proof of Theorem~2}
The proof consists of two parts. The first part establishes asymptotic equivalence, and the second computes the asymptotic variance. 

\medskip\noindent\textit{Asymptotic equivalence.} The main departure of this theorem from Theorem 1 is that it replaces the $\mathcal{I}_2/\mathcal{I}_3$ split with Assumption 5. We start from the decomposition:
\begin{align*}
\tag{I}\sqrt{n}\left(\hat{\mu}_{t,\mathtt{CALM}}^{\mathcal I_2} - \overline{\mu}_{t,\mathtt{CALM}}^{\mathcal I_2}\right) 
&= \frac{\sqrt n}{|\mathcal I_2|}\sum_{i\in\mathcal I_2} \nu_{i,t} \cdot \Big(\hat\mu_t(X_i;\mathcal I_2)-\tilde{\mu}_t(X_i)\Big) \\
\tag{II}&\quad + \frac{\sqrt n}{|\mathcal I_2|}\sum_{i\in\mathcal I_2} \nu_{i,t} \cdot Y_i^\dagger(t;\mathcal{I}_1)\Big(\hat\omega_{t}(X_i;\mathcal I_2)-\omega_{t}(X_i)\Big) \\
\tag{III}&\quad - \frac{\sqrt n}{|\mathcal I_2|}\sum_{i\in\mathcal I_2} \nu_{i,t} \cdot\Big(\hat\mu_{t}^{\dagger}(X_i;\mathcal I_2)\hat\omega_{t}(X_i;\mathcal I_2)-\tilde{\mu}_{t}^\dagger(X_i)\omega_{t}(X_i)\Big)\\
\tag{IV}&\quad + \frac{\sqrt n}{|\mathcal I_2|}\sum_{i\in\mathcal I_2} \nu_{i,t} \cdot\omega_{t}(X_i)\Big(Y_i^\dagger(t;\mathcal{I}_1) - Y_i^\dagger(t)\Big).
\end{align*}
The first three terms in the decomposition vanish thanks to Assumption 5. The last term can be handled by conditional variance calculation: 
\begin{align*}
\Variance\big[ \text{(IV)} \big| \mathcal I_1, \{X_i,Z_i\}_{i\in\mathcal I_2} \big] 
&\precsim \frac{ 1}{|\mathcal I_2|}\sum_{i\in\mathcal I_2} \frac{1-e_t(X_i)}{e_t(X_i)}\omega_t(X_i)^2 \Big(Y_i^\dagger(t;\mathcal{I}_1) - Y_i^\dagger(t)\Big)^2 \\
&\precsim \frac{ 1}{|\mathcal I_2|}\sum_{i\in\mathcal I_2}  \Big(Y_i^\dagger(t;\mathcal{I}_1) - Y_i^\dagger(t)\Big)^2 
\precsimp \Expectation_{X_i,Z_i}\big(Y_{i}^\dagger(t;\mathcal I_1) - Y_i^\dagger(t)\big)^2 \precsimp \frac{1}{B} + \frac{1}{n}.
\end{align*}
The last line is due to Markov inequality applied to nonnegative random variables and our Lemma \ref{lem:fs-resampling-convergence}.

\medskip\noindent\textit{Asymptotic variance calculation.} Recall the asymptotic representation:
\begin{align*}
  \overline{\mu}_{t,\mathtt{CALM}} = \frac{1}{n}\sum_{i=1}^n \underbrace{\frac{\Indicator(T_i=t)}{e_t(X_i)}\Big(Y_i - \tilde{\mu}_t(X_i)\Big) + \tilde{\mu}_t(X_i)}_{\textstyle = A_i} + \underbrace{\Big(1-\frac{\Indicator(T_i=t)}{e_t(X_i)}\Big)\Big(Y_i^\dagger(t) - \tilde{\mu}^\dagger_t(X_i)\Big)\omega_t(X_i)}_{\textstyle = B_i}.
\end{align*}  
Clearly variance of $A_i$ corresponds to the asymptotic variance of the AIPW approach under misspecification, which is
\begin{align*}
  \Variance[A_i] &= \Variance\Big[ \frac{\Indicator(T_i=t)}{e_t(X_i)}\Big(Y_i - \tilde{\mu}_t(X_i)\Big) + \tilde{\mu}_t(X_i) \Big]\\
  & = \Variance\Big[ \frac{\Indicator(T_i=t)}{e_t(X_i)}\Big(Y_i - \mu_t(X_i)\Big) + \nu_{i,t}\big(\tilde{\mu}_t(X_i) - \mu_t(X_i)\big) + \big(\mu_t(X_i)- \mu_t\big) \Big]\\
  &= \Expectation\Big[\frac{\Variance[Y_i(t)|X_i]}{e_t(X_i)} + \lambda_t(X_i)\big(\tilde{\mu}_t(X_i) - \mu_t(X_i)\big)^2 + \big(\mu_t(X_i)- \mu_t\big)^2\big].
\end{align*}
Notice that the three terms in the second line are conditionally uncorrelated. 

To find the asymptotic variance of the CALM approach, we need the variance of $B_i$ as well as the covariance between $A_i$ and $B_i$. Specifically, 
\begin{align*}
  \Variance[B_i] &= \Expectation\big[ \lambda_t(X_i)\big(Y_i^\dagger(t) - \tilde{\mu}^\dagger_t(X_i)\big)^2\omega_t(X_i)^2   \big],
\end{align*}
and the covariance is 
\begin{align*}
  2\cov\big[A_i,B_i\big] &= 2\Expectation\Big[\frac{\Indicator(T_i=t)}{e_t(X_i)}\Big(Y_i - \tilde{\mu}_t(X_i)\Big)\Big(1-\frac{\Indicator(T_i=t)}{e_t(X_i)}\Big)\Big(Y_i^\dagger(t) - \tilde{\mu}^\dagger_t(X_i)\Big)\omega_t(X_i) \Big] \\
  &\qquad + 2\Expectation\Big[\tilde{\mu}_t(X_i)\Big(1-\frac{\Indicator(T_i=t)}{e_t(X_i)}\Big)\Big(Y_i^\dagger(t) - \tilde{\mu}^\dagger_t(X_i)\Big)\omega_t(X_i) \Big]\\
  &= 2\Expectation\Big[\frac{\Indicator(T_i=t)}{e_t(X_i)}\Big(Y_i - \tilde{\mu}_t(X_i)\Big)\Big(1-\frac{\Indicator(T_i=t)}{e_t(X_i)}\Big)\Big(Y_i^\dagger(t) - \tilde{\mu}^\dagger_t(X_i)\Big)\omega_t(X_i) \Big]\\
  &= -2\Expectation\Big[ \lambda_t(X_i)\omega_t(X_i) \Expectation\big[(Y_i - \tilde{\mu}_t(X_i))(Y_i^\dagger(t) - \tilde{\mu}^\dagger_t(X_i))|X_i\big]\Big],
\end{align*}
which completes the calculation.

\subsection{Proof of Corollary~1}
  To show this result, we will verify Assumption 5 for the special case of covariate discretization. We define the local constant basis expansion:
  \begin{align*}
    \mathsf{p}(X_i) = \Big( \Indicator(X_i\in\mathcal{X}_1),\ \Indicator(X_i\in\mathcal{X}_2),\ \dots,\ \Indicator(X_i\in\mathcal{X}_J) \Big) \in \{0,1\}^J.
  \end{align*}
  Therefore, the nuisance functions and their estimates all have the representation: 
  \begin{alignat*}{2}
    \tilde{\mu}_t(X_i) &= \mathsf{p}(X_i)^\transpose \bbeta_{t,\mu},\qquad   &&\hat{\mu}_t(X_i;\mathcal{I}_2) = \mathsf{p}(X_i)^\transpose \hat{\bbeta}_{t,\mu}(\mathcal{I}_2), \\
    \tilde{\mu}_t^\dagger(X_i) &= \mathsf{p}(X_i)^\transpose \bbeta_{t,\mu}^\dagger,\qquad   &&\hat{\mu}^\dagger_t(X_i;\mathcal{I}_2) = \mathsf{p}(X_i)^\transpose \hat{\bbeta}^\dagger_{t,\mu}(\mathcal{I}_2), \\
    {\omega}_t(X_i) &= \mathsf{p}(X_i)^\transpose \bbeta_{t,\omega},\qquad   &&\hat{\omega}_t(X_i;\mathcal{I}_2) = \mathsf{p}(X_i)^\transpose \hat{\bbeta}_{t,\omega}(\mathcal{I}_2).
  \end{alignat*}
  A direct consequence is that
  \begin{align*}
    \| \bbeta_{t,\mu} - \hat{\bbeta}_{t,\mu}(\mathcal{I}_2) \| \precsimp \frac{1}{\sqrt{n}},\quad 
    \| \bbeta^\dagger_{t,\mu} - \hat{\bbeta}^\dagger_{t,\mu}(\mathcal{I}_2) \| \precsimp \frac{1}{\sqrt{n}},\quad 
    \| \bbeta_{t,\omega} - \hat{\bbeta}_{t,\omega}(\mathcal{I}_2) \| \precsimp \frac{1}{\sqrt{n}},
  \end{align*}
  where we note that $|\mathcal{I}_2| \asymp n$. To verify Assumption 5, we again check each term in the expansion. First,
  \begin{align*}
  &\frac{1}{\sqrt{|\mathcal{I}_2|}}\sum_{i \in \mathcal{I}_2} \nu_{i,t}\cdot\big( \hat{\mu}_t(X_i;\mathcal{I}_2) - \tilde{\mu}_t(X_i) \big)  
  = \frac{1}{\sqrt{|\mathcal{I}_2|}}\sum_{i \in \mathcal{I}_2} \nu_{i,t}\cdot\mathsf{p}(X_i)^\transpose \Big(\hat{\bbeta}_{t,\mu}(\mathcal{I}_2) - \bbeta_{t,\mu}\Big)\\
  &\leq \Big\|\frac{1}{{|\mathcal{I}_2|}}\sum_{i \in \mathcal{I}_2} \nu_{i,t}\cdot\mathsf{p}(X_i)^\transpose \Big\| \cdot \sqrt{|\mathcal{I}_2|} \Big\|\hat{\bbeta}_{t,\mu}(\mathcal{I}_2) - \bbeta_{t,\mu} \Big\|\\
  &\precsimp \Big\|\frac{1}{{|\mathcal{I}_2|}}\sum_{i \in \mathcal{I}_2} \nu_{i,t}\cdot\mathsf{p}(X_i)^\transpose \Big\| \toProbability 0,
  \end{align*}
  by the standard law of large numbers. 

  Similarly, 
  \begin{align*}
  &\frac{1}{\sqrt{|\mathcal{I}_2|}}\sum_{i \in \mathcal{I}_2} \nu_{i,t}\cdot Y_i^\dagger(t;\mathcal{I}_1)\big(\hat{\omega}_t(X_i;\mathcal{I}_2) - \omega_t(X_i)\big) \\
  &\leq \Big\|\frac{1}{{|\mathcal{I}_2|}}\sum_{i \in \mathcal{I}_2} \nu_{i,t}\cdot Y_i^\dagger(t)\mathsf{p}(X_i)^\transpose \Big\| \cdot \sqrt{|\mathcal{I}_2|} \Big\|\hat{\bbeta}_{t,\omega}(\mathcal{I}_2) - \bbeta_{t,\omega} \Big\|\\
  &\qquad+ \frac{1}{{|\mathcal{I}_2|}}\sum_{i \in \mathcal{I}_2} \Big\|\nu_{i,t}\cdot\mathsf{p}(X_i)^\transpose \Big\| \cdot \Big|Y_i^\dagger(t;\mathcal{I}_1) - Y_i^\dagger(t) \Big| \cdot \sqrt{|\mathcal{I}_2|} \Big\|\hat{\bbeta}_{t,\omega}(\mathcal{I}_2) - \bbeta_{t,\omega} \Big\| \\
  &\precsimp \Big\|\frac{1}{{|\mathcal{I}_2|}}\sum_{i \in \mathcal{I}_2} \nu_{i,t}\cdot Y_i^\dagger(t)\mathsf{p}(X_i)^\transpose \Big\|  + \frac{1}{{|\mathcal{I}_2|}}\sum_{i \in \mathcal{I}_2} \Big\|\nu_{i,t}\cdot\mathsf{p}(X_i)^\transpose \Big\| \cdot \Big|Y_i^\dagger(t;\mathcal{I}_1) - Y_i^\dagger(t) \Big|.
  \end{align*}
  The first term in the last line clearly converges to zero. The second one can be bounded as
  \begin{align*}
    \frac{1}{{|\mathcal{I}_2|}}\sum_{i \in \mathcal{I}_2} \Big\|\nu_{i,t}\cdot\mathsf{p}(X_i)^\transpose \Big\| \cdot \Big|Y_i^\dagger(t;\mathcal{I}_1) - Y_i^\dagger(t) \Big| &\leq \sqrt{\frac{1}{{|\mathcal{I}_2|}}\sum_{i \in \mathcal{I}_2}\Big\|\nu_{i,t}\cdot\mathsf{p}(X_i)^\transpose \Big\|^2} \sqrt{\frac{1}{{|\mathcal{I}_2|}}\sum_{i \in \mathcal{I}_2}\Big|Y_i^\dagger(t;\mathcal{I}_1) - Y_i^\dagger(t) \Big|^2} \\
    &\precsimp \sqrt{\frac{1}{{|\mathcal{I}_2|}}\sum_{i \in \mathcal{I}_2}\Big|Y_i^\dagger(t;\mathcal{I}_1) - Y_i^\dagger(t) \Big|^2}\\
    &\precsimp \sqrt{\Expectation_{X_i,Z_i}\Big|Y_i^\dagger(t;\mathcal{I}_1) - Y_i^\dagger(t) \Big|^2},
  \end{align*}
  which again tends to zero due to Lemma \ref{lem:fs-resampling-convergence}.

  Finally, 
  \begin{align*}
    &\frac{1}{\sqrt{|\mathcal{I}_2|}}\sum_{i \in \mathcal{I}_2} \nu_{i,t}\cdot\big( \hat{\mu}^\dagger_t(X_i;\mathcal{I}_2) \hat{\omega}_t(X_i;\mathcal{I}_2)- \tilde{\mu}_t^\dagger(X_i)\omega_t(X_i)\big)\\
    &= \frac{1}{\sqrt{|\mathcal{I}_2|}}\sum_{i \in \mathcal{I}_2} \nu_{i,t}\cdot \hat{\omega}_t(X_i;\mathcal{I}_2)\big( \hat{\mu}^\dagger_t(X_i;\mathcal{I}_2) - \tilde{\mu}_t^\dagger(X_i)\big) \\ 
    &+ \frac{1}{\sqrt{|\mathcal{I}_2|}}\sum_{i \in \mathcal{I}_2} \nu_{i,t}\cdot\tilde{\mu}_t^\dagger(X_i)\big( \hat{\omega}_t(X_i;\mathcal{I}_2)- \omega_t(X_i)\big).
  \end{align*}
  Each of the terms can be bounded via 
  \begin{align*}
  &\frac{1}{\sqrt{|\mathcal{I}_2|}}\sum_{i \in \mathcal{I}_2} \nu_{i,t}\cdot \hat{\omega}_t(X_i;\mathcal{I}_2)\big( \hat{\mu}^\dagger_t(X_i;\mathcal{I}_2) - \tilde{\mu}_t^\dagger(X_i)\big) \\
  &\leq \Big\|\frac{1}{{|\mathcal{I}_2|}}\sum_{i \in \mathcal{I}_2} \nu_{i,t}\cdot \omega_t(X_i)\mathsf{p}(X_i)^\transpose \Big\| \cdot \sqrt{|\mathcal{I}_2|} \Big\|\hat{\bbeta}^\dagger_{t,\mu}(\mathcal{I}_2) - \bbeta^\dagger_{t,\mu} \Big\|\\
  &\qquad+ \frac{1}{{|\mathcal{I}_2|}}\sum_{i \in \mathcal{I}_2} \Big\|\nu_{i,t}\cdot\mathsf{p}(X_i)^\transpose \Big\| \cdot \Big\|\mathsf{p}(X_i)^\transpose \Big\| \cdot \Big\|\hat{\bbeta}_{t,\omega}(\mathcal{I}_2) - \bbeta_{t,\omega} \Big\| \cdot  \sqrt{|\mathcal{I}_2|} \Big\|\hat{\bbeta}^\dagger_{t,\mu}(\mathcal{I}_2) - \bbeta^\dagger_{t,\mu} \Big\| \\
  &\precsimp \Big\|\frac{1}{{|\mathcal{I}_2|}}\sum_{i \in \mathcal{I}_2} \nu_{i,t}\cdot \omega_t(X_i)\mathsf{p}(X_i)^\transpose \Big\| + \frac{1}{{|\mathcal{I}_2|}^2}\sum_{i \in \mathcal{I}_2} \Big\|\nu_{i,t}\cdot\mathsf{p}(X_i)^\transpose \Big\| \cdot \Big\|\mathsf{p}(X_i)^\transpose \Big\| \toProbability 0,
  \end{align*}
  and
  \begin{align*}
  &\frac{1}{\sqrt{|\mathcal{I}_2|}}\sum_{i \in \mathcal{I}_2} \nu_{i,t}\cdot\tilde{\mu}_t^\dagger(X_i)\big( \hat{\omega}_t(X_i;\mathcal{I}_2)- \omega_t(X_i)\big)  
  = \frac{1}{\sqrt{|\mathcal{I}_2|}}\sum_{i \in \mathcal{I}_2} \nu_{i,t}\cdot\tilde{\mu}_t^\dagger(X_i)\cdot \mathsf{p}(X_i)^\transpose \Big(\hat{\bbeta}_{t,\omega}(\mathcal{I}_2) - \bbeta_{t,\omega}\Big)\\
  &\leq \Big\|\frac{1}{{|\mathcal{I}_2|}}\sum_{i \in \mathcal{I}_2} \nu_{i,t}\cdot\tilde{\mu}_t^\dagger(X_i)\cdot \mathsf{p}(X_i)^\transpose \Big\| \cdot \sqrt{|\mathcal{I}_2|} \Big\|\hat{\bbeta}_{t,\omega}(\mathcal{I}_2) - \bbeta_{t,\omega} \Big\|\\
  &\precsimp \Big\|\frac{1}{{|\mathcal{I}_2|}}\sum_{i \in \mathcal{I}_2} \nu_{i,t}\cdot\tilde{\mu}_t^\dagger(X_i)\cdot \mathsf{p}(X_i)^\transpose \Big\|  \toProbability 0,
  \end{align*}
  which closes the proof.

\subsection{Proof of Theorem~3}

The first conclusion of the theorem, asymptotic equivalence, follows from Lemmas \ref{lem:oracle-fs}, \ref{lem:fs-resampling-convergence}, and \ref{lem:fs-equivalence-oracle}. We therefore only calculate the asymptotic variance. Recall that the oracle estimator takes the form 
\begin{align*}
&\overline{\tau}_{t,t',\mathtt{CALM}} = \frac1n\sum_{i=1}^n\ \underbrace{\bigg[\frac{\Indicator(T_i=t)}{e_t(X_i)}\big(Y_i-\mu_t(X_i)\big)+\mu_t(X_i)\bigg] - \bigg[\frac{\Indicator(T_i=t')}{e_{t'}(X_i)}\big(Y_i-\mu_{t'}(X_i)\big)+\mu_{t'}(X_i)\bigg]}_{\textstyle = A_i}\\
&\qquad+\underbrace{\nu_{i,t}\cdot \big(Y_i^\dagger(t)-\mu_t^\dagger(X_i)\big)}_{\textstyle = B_i}\omega_t(X_i) + \underbrace{- \nu_{i,t'}\cdot\big(Y_i^\dagger(t')-\mu_{t'}^\dagger(X_i)\big)}_{\textstyle = C_i}\omega_{t'}(X_i).
\end{align*}
Clearly the asymptotic variance of the AIPW approach is simply the variance of $A_i$. Also recall that 
\begin{align*}
  \nu_{i,t} = 1-\frac{\Indicator(T_i=t)}{e_t(X_i)},
\end{align*} 
with
\begin{align*}
  \Expectation[\nu_{i,t}|X_i] = \Expectation[\nu_{i,t}|X_i,Z_i] = 0,\quad \text{and}\quad \Variance[\nu_{i,t}|X_i] = \Variance[\nu_{i,t}|X_i,Z_i] = \frac{1 - e_t(X_i)}{e_t(X_i)} = \lambda_t(X_i).
\end{align*}
In addition, 
\begin{align*}
  \Expectation\Big[\nu_{i,t} \frac{\Indicator(T_i=t)}{e_t(X_i)}\Big|X_i\Big] = \Expectation\Big[\nu_{i,t} \frac{\Indicator(T_i=t)}{e_t(X_i)}\Big|X_i,Z_i\Big] = -\lambda_t(X_i),
\end{align*}
and for $t\neq t'$,
\begin{align*}
  \Expectation\Big[\nu_{i,t'} \frac{\Indicator(T_i=t)}{e_t(X_i)}\Big|X_i\Big] = \Expectation\Big[\nu_{i,t'} \frac{\Indicator(T_i=t)}{e_t(X_i)}\Big|X_i,Z_i\Big] = 1,
\end{align*}
as the indicators will be mutually exclusive. Finally, we also have for $t\neq t'$
\begin{align*}
  \Expectation\big[\nu_{i,t}\nu_{i,t'}\big|X_i\big] = \Expectation\big[\nu_{i,t}\nu_{i,t'}\big|X_i,Z_i\big] = -1.
\end{align*}

We start with the variances of $B_i$ and $C_i$, which  take the form  
\begin{align*}
  \Variance[B_i|X_i] &= \lambda_t(X_i)\Variance[Y_i^\dagger(t)|X_i],\qquad \Variance[C_i|X_i] = \lambda_{t'}(X_i)\Variance[Y_i^\dagger(t')|X_i].
\end{align*}
Their covariance is
\begin{align*}
  \cov[B_i,C_i|X_i] = \cov[Y_i^\dagger(t),Y_i^\dagger(t')|X_i].
\end{align*}
The above calculation gives the form of the matrix 
\begin{align*}
  \bSigma^\dagger(X_i) &= \begin{bmatrix}
    \lambda_t(X_i)\Variance[Y_i^\dagger(t)|X_i] & &   \cov[Y_i^\dagger(t),Y_i^\dagger(t')|X_i] \\
    \cov[Y_i^\dagger(t),Y_i^\dagger(t')|X_i] & & \lambda_{t'}(X_i)\Variance[Y_i^\dagger(t')|X_i],
  \end{bmatrix}.
\end{align*}

We next compute the additional covariances, where we add an extra negative sign to match the expressions in the main paper:
\begin{align*}
  -\cov[A_i,B_i|X_i] &= \lambda_t(X_i)\cov[Y_i(t),Y_i^\dagger(t)|X_i] + \cov[Y_i(t'),Y_i^\dagger(t)|X_i]\\
  -\cov[A_i,C_i|X_i] &= \cov[Y_i(t),Y_i^\dagger(t')|X_i] + \lambda_{t'}(X_i)\cov[Y_i(t'),Y_i^\dagger(t')|X_i],
\end{align*}
which corresponds to the matrix 
\begin{align*}
  \bsigma(X_i) &= \begin{bmatrix}
    \lambda_t(X_i)\cov[Y_i(t),Y_i^\dagger(t)|X_i] \ \ +\ \  \cov[Y_i(t'),Y_i^\dagger(t)|X_i]\\
    \cov[Y_i(t),Y_i^\dagger(t')|X_i] \ \ +\ \  \lambda_{t'}(X_i)\cov[Y_i(t'),Y_i^\dagger(t')|X_i]
  \end{bmatrix}.
\end{align*}
This completes the variance calculation.

\subsection{Proof of Theorem~4}

We start by providing the regularity conditions needed for CATE estimation, which are omitted from the main paper to conserve space. The first assumption below imposes smoothness for certain nuisance functions. Let $f(\cdot)$ denote the density function of the structured covariates. We fix an evaluation point $x$. 

\begin{assumption}[Smoothness of nuisance functions]\label{SA-assu: continuity}
The functions $f(\cdot)$, $\mu_t^\dagger(\cdot)$, $\bSigma^\dagger(\cdot)$ and $\bsigma(\cdot)$ are continuous in a neighborhood of $x$. $\mu_t(\cdot)$ is $s$-times continuously differentiable in a neighborhood of $x$ with $s\geq 1$. In addition, $f(x) > 0$ and $\bSigma^\dagger(x)$ is nonsingular. 
\end{assumption}

We also require the true potential outcomes to have bounded fourth conditional moments, which helps establish the asymptotic normality of our CATE estimator. 

\begin{assumption}[Bounded fourth moments]\label{SA-assu: boundedness}
There exists some $c>0$ such that $\Expectation[Y_i(t)^4|X_i]\leq c$ for all $t=1,2,\dots,k$.
\end{assumption}

The next assumption concerns the kernel function. In particular, we allow the use of both second- and higher-order kernels, where the latter can be appealing for bias reduction. 

\begin{assumption}[Kernel function]\label{SA-assu: kernel}
The kernel function $k(\cdot)$ is continuous, compactly supported, and integrates to 1. In addition, its order is at least $s+1$.
\end{assumption}

As the CATE estimator is local in nature, we assume the nonparametric estimates are uniformly consistent in a local neighborhood of the evaluation point of interest. 

\begin{assumption}[Consistent nuisance function estimation]\label{assumption: Uniform ATE Consistent nuisance function estimation}
The cross-fitted function estimates satisfy: for some $\epsilon >0$, 
\begin{align*}
  &\sup_{|x'-x|\leq \epsilon}|\hat{\mu}_{t}(x';\mathcal{I})-\mu_{t}(x')| \toProbability 0,\quad \sup_{|x'-x|\leq \epsilon}|\widehat{\mu}^{\dagger}_{t}(x';\mathcal{I})-\mu^{\dagger}_{t}(x')| \toProbability 0,\\
  &\sup_{|x'-x|\leq \epsilon}|\widehat{\bomega}_{\mathtt{ATE}}(x';\mathcal{I})-\bomega_{\mathtt{ATE}}(x')| \toProbability 0,
\end{align*}
for each data fold $\mathcal{I}$ and treatment arm $t,t'\in\{1,2,\dots,k\}$.
\end{assumption}

We are ready to prove the theorem. We first show the formula of the variance comparison. 

\medskip\noindent\textit{Variance calculation.} From the calculation in Section 3.2 of the main paper, one has 
\begin{align*}
  \label{eq:CATE asym variance}&\mathsf{V}_{t,t',\mathtt{CALM}}(x) = nh^p\cdot\Variance[\overline{\tau}_{t,t',\mathtt{CALM}}(x)|\mathbf{X}] \\
  \nonumber&\quad = nh^p\cdot\Variance[\overline{\tau}_{t,t',\mathtt{AIPW}}(x)|\mathbf{X}] + nh^p\sum_{i=1}^n \kappa_i(x)^2\Big(\bomega(X_i)^\transpose \bSigma^\dagger(X_i) \bomega(X_i) - 2\bomega(X_i)^\transpose \bsigma(X_i)\Big).
\end{align*}
We now further expand on the last term:
\begin{align*}
  &nh^p\sum_{i=1}^n \kappa_i(x)^2\Big(\bomega(X_i)^\transpose \bSigma^\dagger(X_i) \bomega(X_i) - 2\bomega(X_i)^\transpose \bsigma(X_i)\Big)\\
  &= nh^p\sum_{i=1}^n \frac{1}{h^{2p}}k\Big(\frac{X_i-x}{h}\Big)^2\Big(\bomega(X_i)^\transpose \bSigma^\dagger(X_i) \bomega(X_i) - 2\bomega(X_i)^\transpose \bsigma(X_i)\Big) \Bigg/ \Big(\sum_{i=1}^n \frac{1}{h^{p}}k\Big(\frac{X_i-x}{h}\Big)\Big)^2\\
  &= \sum_{i=1}^n \frac{1}{nh^{p}}k\Big(\frac{X_i-x}{h}\Big)^2\Big(\bomega(X_i)^\transpose \bSigma^\dagger(X_i) \bomega(X_i) - 2\bomega(X_i)^\transpose \bsigma(X_i)\Big) \Bigg/ \Big(\sum_{i=1}^n \frac{1}{nh^{p}}k\Big(\frac{X_i-x}{h}\Big)\Big)^2.
\end{align*}
Convergence of both the numerator and the denominator follows from standard mean and variance calculations. To illustrate, consider the numerator. Its mean is
\begin{align*}
  &\Expectation\Big[\sum_{i=1}^n \frac{1}{nh^{p}}k\Big(\frac{X_i-x}{h}\Big)^2\Big(\bomega(X_i)^\transpose \bSigma^\dagger(X_i) \bomega(X_i) - 2\bomega(X_i)^\transpose \bsigma(X_i)\Big)\Big]\\
  &=\int_{\mathcal{X}} \frac{1}{h^{p}}k\Big(\frac{u-x}{h}\Big)^2\Big(\bomega(u)^\transpose \bSigma^\dagger(u) \bomega(u) - 2\bomega(u)^\transpose \bsigma(u) \Big)f(u)\diff u \\ 
  &=\int_{(\mathcal{X}-x)/h} k(v)^2\Big(\bomega(x+hv)^\transpose \bSigma^\dagger(x+hv) \bomega(x+hv) - 2\bomega(x+hv)^\transpose \bsigma(x+hv)\Big) f(x+hv)\diff v \\
  &= \kappa_2 f(x) \Big(\bomega(x)^\transpose \bSigma^\dagger(x) \bomega(x) - 2\bomega(x)^\transpose \bsigma(x)\Big) + o(1).
\end{align*}
On the other hand, the variance satisfies
\begin{align*}
  &\Variance\Big[\sum_{i=1}^n \frac{1}{nh^{p}}k\Big(\frac{X_i-x}{h}\Big)^2\Big(\bomega(X_i)^\transpose \bSigma^\dagger(X_i) \bomega(X_i) - 2\bomega(X_i)^\transpose \bsigma(X_i)\Big)\Big]\\
  &\precsim \frac{1}{n}\Variance\Big[ \frac{1}{h^{p}}k\Big(\frac{X_i-x}{h}\Big)^2\Big(\bomega(X_i)^\transpose \bSigma^\dagger(X_i) \bomega(X_i) - 2\bomega(X_i)^\transpose \bsigma(X_i)\Big)\Big] \precsim \frac{1}{nh^p} \to 0.
\end{align*}
As such, 
\begin{align*}
  \sum_{i=1}^n \frac{1}{nh^{p}}k\Big(\frac{X_i-x}{h}\Big)^2\Big(\bomega(X_i)^\transpose \bSigma^\dagger(X_i) \bomega(X_i) - 2\bomega(X_i)^\transpose \bsigma(X_i)\Big) \toProbability \kappa_2 f(x) \Big(\bomega(x)^\transpose \bSigma^\dagger(x) \bomega(x) - 2\bomega(x)^\transpose \bsigma(x)\Big).
\end{align*}
The same analysis applies to the denominator, so that
\begin{align*}
  \sum_{i=1}^n \frac{1}{nh^{p}}k\Big(\frac{X_i-x}{h}\Big) \toProbability f(x).
\end{align*}

\medskip\noindent\textit{Asymptotic equivalence.} We now show that the CALM estimator of the CATE is asymptotically equivalent to the oracle one. To reduce notational burden, we focus on one treatment arm, leading to the expansion:
\begin{align*}
\tag{I}\sqrt{nh^p}\left(\hat{\mu}_{t,\mathtt{CALM}}^{\mathcal I_3} - \overline{\mu}_{t,\mathtt{CALM}}^{\mathcal I_3}\right) 
&= \sqrt{nh^p}\sum_{i\in\mathcal I_3} \kappa_i(x;\mathcal{I}_3)\nu_{i,t} \Big(\hat\mu_t(X_i;\mathcal I_2)-\mu_t(X_i)\Big) \\
\tag{II}&\quad - \sqrt{nh^p}\sum_{i\in\mathcal I_3} \kappa_i(x;\mathcal{I}_3)\nu_{i,t}\Big(\hat\mu_{t}^{\dagger}(X_i;\mathcal I_2)-\mu_{t}^\dagger(X_i)\Big)\hat{\omega}_{t}(X_i;\mathcal{I}_2) \\
\tag{III}&\quad - \sqrt{nh^p}\sum_{i\in\mathcal I_3} \kappa_i(x;\mathcal{I}_3)\nu_{i,t}\mu_{t}^\dagger(X_i)\Big(\hat\omega_{t}(X_i;\mathcal I_2)-\omega_{t}(X_i)\Big) \\
\tag{IV}&\quad + \sqrt{nh^p}\sum_{i\in\mathcal I_3} \kappa_i(x;\mathcal{I}_3)\nu_{i,t}\Big(\hat\omega_{t}(X_i;\mathcal I_2)-\omega_{t}(X_i)\Big)Y_i^\dagger(t) \\
\tag{V}&\quad + \sqrt{nh^p}\sum_{i\in\mathcal I_3} \kappa_i(x;\mathcal{I}_3)\nu_{i,t}\hat\omega_{t}(X_i;\mathcal I_2)\Big(Y_{i}^\dagger(t;\mathcal I_1) - Y_i^\dagger(t)\Big).
\end{align*}
Due to sample splitting, it is easy to see that they all are conditional mean 0. We therefore consider their conditional variance. Let $\mathbf{X}(\mathcal{I}_3)$ denote all structured covariates from fold $\mathcal{I}_3$. Also define the local sup-norm $\| \cdot \|_{\infty,\epsilon} = \sup_{|\cdot - x|\leq \epsilon}|\cdot|$. Starting from (I), 
\begin{align*}
  \Variance[\text{(I)}|\mathcal{I}_1,\mathcal{I}_2,\mathbf{X}(\mathcal{I}_3)] &= nh^p\sum_{i\in\mathcal I_3} \kappa_i(x;\mathcal{I}_3)^2 \lambda_t(X_i)\Big(\hat\mu_t(X_i;\mathcal I_2)-\mu_t(X_i)\Big)^2 \\
  &\precsim \| \hat{\mu}_t(\cdot;\mathcal I_2)-\mu_t(\cdot) \|_{\infty,\epsilon} \Big(nh^p\sum_{i\in\mathcal I_3} \kappa_i(x;\mathcal{I}_3)^2 \Big).
\end{align*}
By our assumption, the first term in the last line vanishes asymptotically. With the same mean and variance calculation, the second term in the above converges to $\kappa_2/f(x)$. As a result, (I) is asymptotically vanishing. The same technique applies to (II), (III) and (IV):
\begin{align*}
  \Variance[\text{(II)}|\mathcal{I}_1,\mathcal{I}_2,\mathbf{X}(\mathcal{I}_3)] &= nh^p\sum_{i\in\mathcal I_3} \kappa_i(x;\mathcal{I}_3)^2 \lambda_t(X_i)\hat{\omega}_{t}(X_i;\mathcal{I}_2)^2\Big(\hat\mu_{t}^{\dagger}(X_i;\mathcal I_2)-\mu_{t}^\dagger(X_i)\Big)^2 \\
  &\precsim \| \hat\mu_{t}^{\dagger}(X_i;\mathcal I_2)-\mu_{t}^\dagger(X_i) \|_{\infty,\epsilon} \Big(nh^p\sum_{i\in\mathcal I_3} \kappa_i(x;\mathcal{I}_3)^2 \Big)\toProbability 0.
\end{align*}
\begin{align*}
  \Variance[\text{(III)}|\mathcal{I}_1,\mathcal{I}_2,\mathbf{X}(\mathcal{I}_3)] &= nh^p\sum_{i\in\mathcal I_3} \kappa_i(x;\mathcal{I}_3)^2 \lambda_t(X_i)\mu_{t}^\dagger(X_i)^2\Big(\hat\omega_{t}(X_i;\mathcal I_2)-\omega_{t}(X_i)\Big)^2 \\
  &\precsim \| \hat\omega_{t}(X_i;\mathcal I_2)-\omega_{t}(X_i) \|_{\infty,\epsilon} \Big(nh^p\sum_{i\in\mathcal I_3} \kappa_i(x;\mathcal{I}_3)^2 \Big)\toProbability 0,
\end{align*}
and 
\begin{align*}
  \Variance[\text{(IV)}|\mathcal{I}_1,\mathcal{I}_2,\mathbf{X}(\mathcal{I}_3)] &= nh^p\sum_{i\in\mathcal I_3} \kappa_i(x;\mathcal{I}_3)^2 \lambda_t(X_i)Y_i^\dagger(t)^2\Big(\hat\omega_{t}(X_i;\mathcal I_2)-\omega_{t}(X_i)\Big)^2 \\
  &\precsim \| \hat\omega_{t}(X_i;\mathcal I_2)-\omega_{t}(X_i) \|_{\infty,\epsilon} \Big(nh^p\sum_{i\in\mathcal I_3} \kappa_i(x;\mathcal{I}_3)^2 \Big)\toProbability 0.
\end{align*}

Now consider (V). Variance computation yields
\begin{align*}
\Variance[\text{(V)}|\mathcal{I}_1,\mathcal{I}_2,\mathbf{X}(\mathcal{I}_3)] &=nh^p\sum_{i\in\mathcal I_3} \kappa_i(x;\mathcal{I}_3)^2 \hat\omega_{t}(X_i;\mathcal I_2)^2 \Big(Y_{i}^\dagger(t;\mathcal I_1) - Y_i^\dagger(t)\Big)^2\\
&\precsimp \frac{1}{nh^p}\sum_{i\in\mathcal I_3} k\Big(\frac{X_i-x}{h}\Big)^2 \Big(Y_{i}^\dagger(t;\mathcal I_1) - Y_i^\dagger(t)\Big)^2.
\end{align*}
As the kernel function is bounded and supported on a compact set, the above has the bound 
\begin{align*}
&\frac{1}{nh^p}\sum_{i\in\mathcal I_3} k\Big(\frac{X_i-x}{h}\Big)^2 \Big(Y_{i}^\dagger(t;\mathcal I_1) - Y_i^\dagger(t)\Big)^2 \\
&\precsimp \Expectation\Big[\frac{1}{nh^p}\sum_{i\in\mathcal I_3} k\Big(\frac{X_i-x}{h}\Big)^2 \Big(Y_{i}^\dagger(t;\mathcal I_1) - Y_i^\dagger(t)\Big)^2\Big| (X_i,Z_i):i\in\mathcal{I}_3\Big] \\
&= \frac{1}{nh^p}\sum_{i\in\mathcal I_3} k\Big(\frac{X_i-x}{h}\Big)^2 \Expectation\Big[\Big(Y_{i}^\dagger(t;\mathcal I_1) - Y_i^\dagger(t)\Big)^2\Big| X_i,Z_i\Big],
\end{align*}
where from the proof of Lemma \ref{lem:fs-resampling-convergence}, the conditional expectation in the last line is of order $B^{-1} + n^{-1}$ uniformly in $X_i$ and $Z_i$. 

\medskip\noindent\textit{Asymptotic bias order.} Given the asymptotically linear representation, we note that the asymptotic bias is determined by 
\begin{align*}
  \sqrt{nh^p}\Expectation\Big[\frac{1}{h^p}k\Big(\frac{X_i-x}{h}\Big)\big(\mu_t(X_i) - \mu_{t'}(X_i) - \mu_t(x) + \mu_{t'}(x)\big) \Big],
\end{align*}
which vanishes as long as $nh^{p+2s}\precsim 1$. 

\medskip\noindent\textit{Asymptotic normality.} To establish asymptotic normality, it suffices to verify a conditional Lindeberg condition with fourth moments, which follows from our assumption on the bandwidth sequence and the finite fourth moments of the potential outcomes.

\clearpage

\section{Motivating Real-world RCTs}
In this section, we present five motivating RCTs that collect unstructured data such as survey responses, video, and audio recordings from participants. These trials span cancer, mental health, and dementia care, as summarized in Table~\ref{table:motivating RCTs}, and demonstrate the broad applicability of our proposed methods in real-world settings.

\begin{table}[ht]
\centering
\footnotesize
\setlength{\tabcolsep}{5pt}
\renewcommand{\arraystretch}{1.15}

\begin{tabularx}{0.95\linewidth}{%
  >{\RaggedRight\arraybackslash}p{3.6cm} 
  Y 
  Y 
  Y 
  Y 
}
\toprule
\makecell[tl]{\textbf{Trial}\\\textbf{Name}} &
\textbf{Brief Summary} &
\makecell[tl]{\textbf{Patient-Centered}\\\textbf{Outcomes}} &
\textbf{Effect Modifiers} &
\makecell[tl]{\textbf{Unstructured}\\\textbf{Pre-treatment Data}} \\
\midrule

A. Precision Clinical Trial Recruitment to Promote Cancer Health Equity \citep{hanvey2024investigating} &
Cluster RCTs on the ALEX Portal testing a virtual Community Health Educator; compares consent and engagement tactics to increase trust, participation, and referrals in Florida. &
Referral to NCI-supported trials; engagement with trial education. &
Race/ethnicity; language preference; geographic region. &
Interaction logs with virtual Community Health Educator. \\

\addlinespace
B. Tailoring Recruitment Communication using Virtual Human Technology for Older Minority Adults \citep{bryan2025developing} &
Cluster RCT on the ALEX platform testing a Virtual Human intervention to improve enrollment of 1{,}363 underrepresented older adults in NIH trials using remote, patient-informed consent and communication. &
Enrollment in active NIH-funded trials; engagement with culturally sensitive recruitment materials. &
Race/ethnicity; rural vs.\ urban residence; age; modality of VHT delivery. &
Survey responses; interaction with Virtual Human Technology. \\

\addlinespace
C. Researching/Improving Psychotherapy Techniques in Interventions DEpression Trial \citep{ramos2024assessing} &
RCT comparing Cognitive Behavioral Therapy vs.\ Acceptance and Commitment Therapy for major depressive disorder in 100 adults (California, USA). &
Improvement in depressive symptoms; improvement in quality of life. &
Treatment modality; age; sex; race/ethnicity; socioeconomic status; pretreatment depressive severity. &
Video, audio, and transcripts of digital therapy interactions. \\

\addlinespace
D. BRIGHTEN: Bridging Research Innovations for Greater Health in Technology, Emotion, and Neuroscience \citep{pratap2022real} &
RCT evaluating mobile app-based therapy for depression management among 2{,}193 U.S. patients; supports symptom tracking and management. &
Depression severity via mobile surveys; app engagement; behavioral changes. &
Baseline depression severity; demographics; behavioral data. &
Open-ended survey responses; user satisfaction; motivation for app use. \\

\addlinespace
E. Care Ecosystem Dementia Randomized Controlled Trial \citep{guterman2023care} &
RCT of a telehealth intervention for 460 patients with dementia and their caregivers. &
Cost of care reimbursed by Medicare. &
Age; sex; race; ethnicity; dementia severity. &
Unstructured text from telephone surveys. \\

\bottomrule
\end{tabularx}

\caption{Five motivating trials that collect unstructured pre-treatment data.}
\label{table:motivating RCTs}
\end{table}

\clearpage
\section{Additional Simulation Study Results}\label{supplement sec:additional simulation study details}

\begin{figure}[h]
\centering
  \centering  \includegraphics[width=0.5\linewidth]{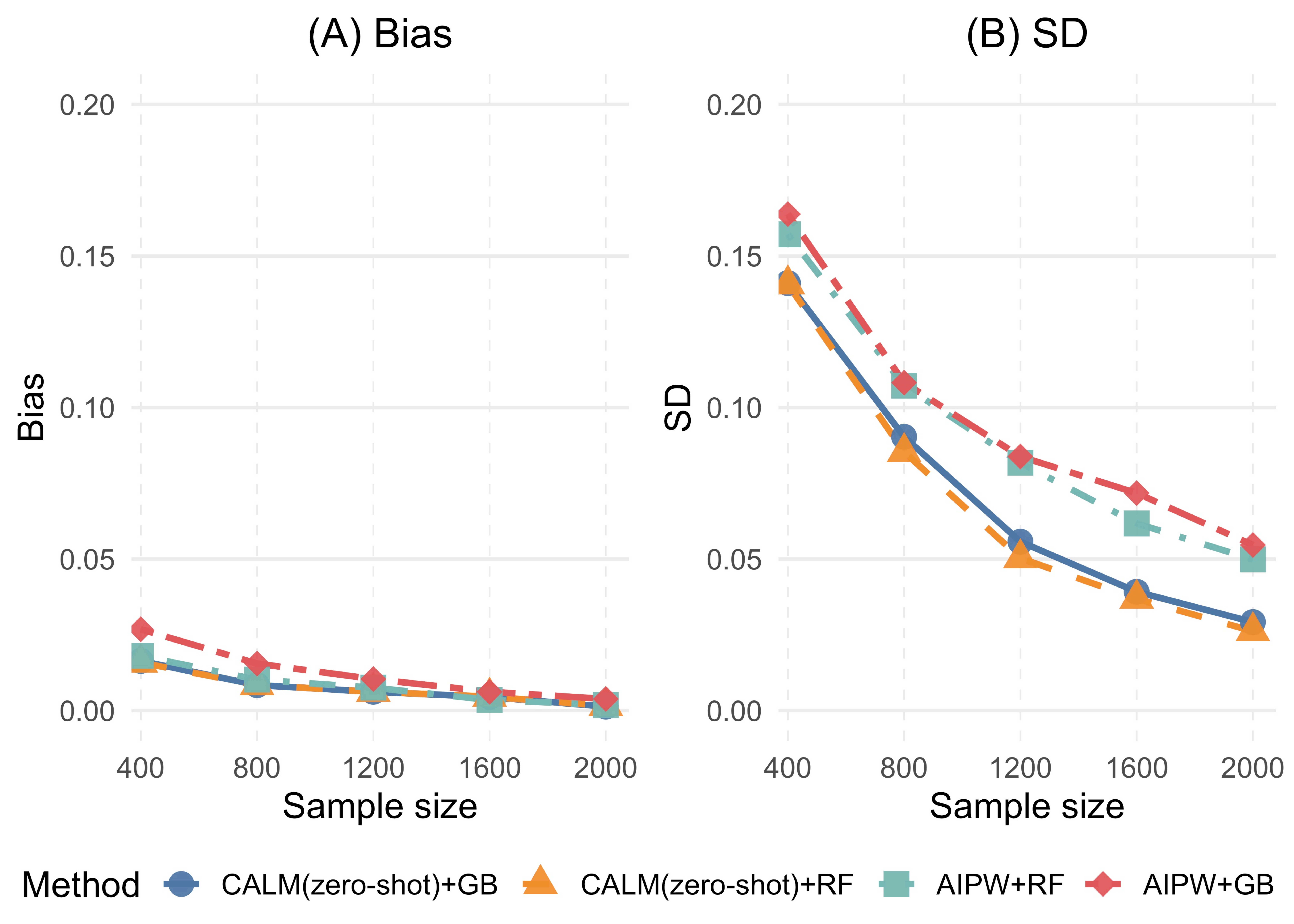}
  \caption{Absolute bias and standard deviation of ATE estimates for CALM (zero-shot) and AIPW-based methods across 300 Monte Carlo simulations. ``CALM(zero-shot)+GB'' and ``CALM(zero-shot)+RF'' refer to using gradient boosting and random forest for estimating the conditional mean model, respectively. ``AIPW+GB'' and ``AIPW+RF'' refer to the standard AIPW estimator where the conditional mean model is estimated using gradient boosting and random forest, respectively. }
  \label{supp fig:sim-fig-1-CALM-zero-shot-and-AIPW}
\end{figure}

In this section, we present additional simulation study results. First, we provide a simulation study comparing different machine learning methods to estimate the conditional outcome model for CALM and AIPW in Figure \ref{supp fig:sim-fig-1-CALM-zero-shot-and-AIPW}. Second, we provide a table of the coverage probabilities (Table \ref{table:coverage-probabilities}) for all the methods discussed in the main manuscript. Third, we provide example prompts under zero-shot prediction, few-shot prediction, and different prompt engineering techniques.

For both CALM and AIPW, when estimating the conditional mean model, we consider two machine learning methods: random forest (RF) and gradient boosting (GB). We refer to the CALM with zero-shot learning as $\texttt{CALM(zero-shot)+RF}$ and $\texttt{CALM(zero-shot)+GB}$ and the AIPW-based method as $\texttt{AIPW+RF}$ and $\texttt{AIPW+GB}$. We present the absolute bias and the standard deviation in Figure \ref{supp fig:sim-fig-1-CALM-zero-shot-and-AIPW}. Figure \ref{supp fig:sim-fig-1-CALM-zero-shot-and-AIPW} shows that for both CALM and AIPW, using random forest for conditional outcome estimation yields slightly lower standard deviation and thus higher estimation efficiency. Therefore, in the main manuscript, we adopt the random forest to estimate the conditional outcome model for all the methods in comparison.

\begin{table}[htbp]
  \centering
  \caption{Comparison of coverage probabilities. }
  \label{table:coverage-probabilities}
  \begin{tabular}{l cc}
    \toprule
    \multirow{2}{*}{\textbf{Method}} & \multicolumn{2}{c}{\textbf{Coverage probability (SE)}} \\
    \cmidrule(lr){2-3}
     & $n=400$ & $n=2,000$ \\
    \midrule
    AIPW+RF                 & 0.94(0.01) & 0.96(0.01) \\
    AIPW+GB                 & 0.96(0.01) & 0.97(0.01) \\
    CALM(zero-shot)+GB      & 0.95(0.02) & 0.95(0.01) \\
    CALM(zero-shot)+RF      & 0.94(0.02) & 0.96(0.01) \\
    CALM(few-shot, $m{=}6$) & 0.94(0.01) & 0.95(0.01) \\
    CALM(few-shot, $m{=}10$)& 0.96(0.01) & 0.95(0.02) \\
    CALM(few-shot, $m{=}14$)& 0.95(0.02) & 0.94(0.01) \\
    AIPW(zero-shot covariate) & 0.96(0.01) & 0.94(0.01) \\
    AIPW(few-shot covariate, $m{=}6$) & 0.92(0.01) & 0.93(0.01) \\
    AIPW(few-shot covariate, $m{=}10$) & 0.90(0.02) & 0.92(0.02) \\
    AIPW(few-shot covariate, $m{=}14$) & 0.89(0.02) & 0.92(0.01)\\
    \bottomrule
  \end{tabular}
\end{table}

In what follows, we provide an example of the zero-shot learning prompt in Figure \ref{supp fig:zero-shot-prompt}, the few-shot learning prompt in Figure \ref{supp fig:few-shot-prompt}, and example prompts under different prompt engineering techniques in   Figure \ref{supp fig:prompt-engineering-examples}. 

\begin{figure}[ht]
\centering
\begin{tcolorbox}[colback=white, colframe=black, width=0.8\textwidth, boxrule=0.7pt, arc=2pt, title=Example prompt for zero-shot learning,
colbacktitle=white, coltitle=black,  fonttitle=\bfseries\small]
\footnotesize
\textbf{Instruction}: You are a licensed clinical psychologist with expertise in evaluating depression severity. Your task is to classify the depression severity category for a BRIGHTEN study participant based on their demographic characteristics, treatment arm, satisfaction with the study app, and reason for enrollment. Use your professional judgment to interpret the information and determine the most appropriate severity level.  Predict the participant’s end-of-study depression severity using exactly one word chosen from the following categories: [Minimal, Mild, Moderate, High, Severe].\\

\noindent\textbf{Input:} The participant is enrolled in the [treatment arm] of the BRIGHTEN study. They are a [Male], [Hispanic/Latino], [45] years old, have completed [High school degree], and are currently [working]. Their marital status is [Married/Partner]. The participant joined the study because [``to gain compensation'']. The participant’s satisfaction level with the app is [``i thought it may be like luminosity. but i was hoping it would help my mood swings'']. 
\end{tcolorbox}
\caption{Example zero-shot learning prompt. The input example is synthetic and does not reflect real participant information from the study.}
\label{supp fig:zero-shot-prompt}
\end{figure}

\begin{figure}[ht]
\centering
\begin{tcolorbox}[colback=white, colframe=black, width=0.9\textwidth, boxrule=0.7pt, arc=2pt, title=Example prompt for few-shot learning,
colbacktitle=white, coltitle=black,  fonttitle=\bfseries\small]
\footnotesize
\textbf{Instruction}: You are a licensed clinical psychologist with expertise in evaluating depression severity. Your task is to classify the depression severity category for a BRIGHTEN study participant based on their demographic characteristics, treatment arm, satisfaction with the study app, and reason for enrollment. Use your professional judgment to interpret the information and determine the most appropriate severity level. Based on the following example participant profiles and their depression level, predict the participant’s end-of-study depression severity using exactly one word chosen from the following categories: [Minimal, Mild, Moderate, High, Severe].\\

\noindent\textbf{Input:} \\
Example 1: The participant is enrolled in the [treatment arm] of the BRIGHTEN study. They are a [Male], [Hispanic/Latino], [45] years old, have completed [High school degree], and are currently [working]. Their marital status is [Married/Partner]. The participant joined the study because [``to gain compensation'']. The participant’s satisfaction level with the app is [``i thought it may be like luminosity. but i was hoping it would help my mood swings'']\\

Example 2: The participant is enrolled in the [control arm] of the BRIGHTEN study. They are a [Male], [African-American], [32] years old, have completed [Graduate degree], and are currently [working]. Their marital status is [Married/Partner]. The participant joined the study because [``Interest'']. The participant’s satisfaction level with the app is [``New ways to deal with stress and mood'']. The participant has [Minimal] level of depression.\\

Example 3: The participant is enrolled in the [treatment arm] of the BRIGHTEN study. They are a [Female], [Hispanic/Latino], [18] years old, have completed [University], and are currently [not working]. Their marital status is [Single]. The participant joined the study because [``To be part of a study'']. The participant’s satisfaction level with the app is [``NA'']. The participant has [Moderate] level of depression.\\

\end{tcolorbox}
\caption{Example few-shot learning prompt. All input examples are synthetic and do not reflect real participant information from the study.}
\label{supp fig:few-shot-prompt}
\end{figure}

\begin{figure}[ht]
\centering
\begin{tcolorbox}[colback=white, colframe=black, width=0.9\textwidth, boxrule=0.7pt, arc=2pt, title=Example prompts under different prompt engineering techniques,
colbacktitle=white, coltitle=black,  fonttitle=\bfseries\small]
\footnotesize
\textbf{(1) Self-consistency.}\\
\textbf{Instruction}: You are acting as a licensed clinical psychologist specializing in depression assessment. Your task is to determine the depression severity category for a BRIGHTEN study participant using a self-consistency approach. First, make three independent predictions of depression severity based solely on the participant’s demographic information, treatment arm, satisfaction with the study app, and stated reason for enrollment. For each prediction, consider how these factors might be associated with depressive symptoms, drawing on your professional knowledge of depression severity categories. After completing the three predictions, select the category that appears most frequently among them. This will be your final answer. Predict the participant’s end-of-study depression severity using exactly one word chosen from the following categories: [Minimal, Mild, Moderate, High, Severe].\\
\textbf{(2) Role-based.}\\
\textbf{Instruction}: You are a licensed clinical psychologist with expertise in evaluating depression severity. Your task is to classify the depression severity category for a BRIGHTEN study participant based on their demographic characteristics, treatment arm, satisfaction with the study app, and reason for enrollment. Use your professional judgment to interpret the information and determine the most appropriate severity level.  Predict the participant’s end-of-study depression severity using exactly one word chosen from the following categories: [Minimal, Mild, Moderate, High, Severe].\\
\textbf{(3) Decomposition.}\\
\textbf{Instruction}:You are a licensed clinical psychologist tasked with classifying the depression severity category for a BRIGHTEN study participant. Break down your reasoning into the following steps before making the final classification: (1) Mood-related indicators – Based on the participant’s demographic information, satisfaction with the study app, and reason for enrollment, assess potential signs of mood disturbance or stability. (2) Functional status – Evaluate how the participant’s working status, marital status, and other life circumstances may reflect or influence functional impairment. (3) Overall symptom severity – Integrate observations from steps 1 and 2 to estimate the likely severity of depressive symptoms. After completing these three steps, use your professional judgment to determine the most appropriate overall depression severity category. Predict the participant’s end-of-study depression severity using exactly one word chosen from the following categories: [Minimal, Mild, Moderate, High, Severe].\\
\textbf{(4) Contrastive}\\
\textbf{Instruction}:You are a licensed clinical psychologist classifying depression severity for BRIGHTEN study participants. Compare the target participant to the two example participants below and determine the most appropriate severity category. Focus on differences and similarities in demographics, satisfaction level, and reason for enrollment when making your decision. Borderline example 1:Participant in the treatment arm, Female, White, age 45, College education, Employed, Married, satisfaction with the app ``Maybe give me a lil bit of pleasure'', enrollment reason: ``Interested in improving mental well-being.'' This participant has mild level depression. Borderline example 2: Participant in the control arm, Female, White, age 45, College education, Employed, Married, satisfaction: ``Had no hopes for this app.'' Enrollment reason: ``Persistent sadness, loss of interest, trouble sleeping.'' This participant has moderate level depression. Predict the participant’s end-of-study depression severity using exactly one word chosen from the following categories: [Minimal, Mild, Moderate, High, Severe].\\

\noindent\textbf{Input:} The participant is enrolled in the [treatment arm] of the BRIGHTEN study. They are a [female], [Asian], [28] years old, have completed [graduate degree], and are currently [working]. Their marital status is [single]. The participant joined the study because [``\$, gift cards'']. The participant’s satisfaction level with the app is [``i thought it may be like luminosity. but i was hoping it would help my mood swings'']. 
\end{tcolorbox}
\caption{Example zero-shot learning prompts under four different prompt engineering techniques. The input example is synthetic and does not reflect real participant information from the study.}
\label{supp fig:prompt-engineering-examples}
\end{figure}

\clearpage

\bibliography{reference}  
\bibliographystyle{jasa}